\documentclass[fleqn,usenatbib]{mnras}
\usepackage{fix-cm}

\usepackage{newtxtext,newtxmath}

\usepackage[T1]{fontenc}

\DeclareRobustCommand{\VAN}[3]{#2}
\let\VANthebibliography\thebibliography
\def\thebibliography{\DeclareRobustCommand{\VAN}[3]{##3}\VANthebibliography}

\usepackage{graphicx}	
\usepackage{amsmath}	
\usepackage{array}

\title[Near-IR and Optical Study of NGC\,5822]{A Near-Infrared and Optical Study of NGC\,5822: An Open Cluster Hosting Barium-stars and Lithium-Enriched Giant Stars}

\author[N. Holanda et al.]{{\parbox{\linewidth}{\flushleft
N. Holanda,$^{1}$\thanks{E-mail: nacizoholanda@on.br (NH)}
V. Loaiza-Tacuri,$^{1,2}$
A. Sonally,$^{1}$
S. Bijavara Seshashayana,$^{3,4}$
M. P. Roriz,$^{1}$
C. F. Martinez,$^{5, 6}$
M. Borges Fernandes,$^{1}$
C. B. Pereira,$^{1}$
O. J. Katime Santrich,$^{7}$
S. Daflon$^{1}$
}}
\\ \\
$^{1}$Observatório Nacional, Rua General José Cristino 77, CEP 20921-400, São Cristóvão, Rio de Janeiro, RJ, Brazil \\
$^{2}$Departamento de F\'isica, Universidade Federal de Sergipe, Av. Marcelo Deda Chagas, S/N, 49107-230 S{\~ a}o Cristov{\~ a}o, SE, Brazil \\
$^{3}$Materials Science and Applied Mathematics, Malm{\"o} University, 205 06 Malm{\"o}, Sweden \\
$^{4}$Nordic Optical Telescope, Rambla José Ana Fernández Pérez, ES-38711 Breña Baja, Spain \\
$^{5}$Universidad Nacional de Córdoba – Observatorio Astronómico de Córdoba, Laprida 854, X5000BGR, Córdoba, Argentina \\
$^{6}$ Consejo Nacional de Investigaciones Científicas y Técnicas (CONICET), Godoy Cruz 2290, CABA, CPC 1425FQB, Argentina \\
$^{7}$Universidade Estadual de Santa Cruz, UESC, Rodovia Jorge Amado km 16, Ilh\'eus 45662000, Bahia, Brazil \\
}

\date{Accepted XXX. Received YYY; in original form ZZZ}

\pubyear{\the\year{}}

\begin{document}
\label{firstpage}
\pagerange{\pageref{firstpage}--\pageref{lastpage}}
\maketitle

\begin{abstract}
We present a chemical abundance study of giant stars in the Galactic open cluster NGC 5822, which hosts two barium stars (\#002 and \#201) and three lithium-enriched giants (\#006, \#102, and \#240). Using high-resolution optical and near-infrared ($H$ and $K$ band) spectra from \texttt{FEROS} and \texttt{IGRINS}, we determine atmospheric parameters and abundances for 23 elements (Li, C, N, O, F, Na, Mg, Al, Si, P, S, K, Ca, Sc, Ti, Cr, Fe, Ni, Y, Ce, Nd, Yb, and Pb). This includes species not yet studied in this cluster, such as F, P, K, Yb, and Pb, as well as oxygen isotopic ratios $^{16}$O/$^{17}$O and $^{16}$O/$^{18}$O. Membership was assessed using astrometry and chemical abundances, providing insight into the evolutionary stages of Li-enriched giants and cluster parameters (age, distance, extinction). However, the identification of Ba-stars remains challenging due to their binary nature and less reliable astrometric solutions. The cluster's abundances are broadly consistent with expectations for the Galactic thin disk. The mean fluorine abundance agrees with chemical evolution models predicting that young clusters (<2\,Gyr) exhibit elevated [F/Fe], with production from SN II, SN Ia, AGB, and Wolf-Rayet stars. No distinct chemical or rotational features were found to explain the lithium enrichment, likely occurring either during the red clump phase or near the RGB tip. For the Ba-stars, nucleosynthesis models combined with the cluster's turn-off mass suggest polluting companion masses of 3.00 and 3.75\,$M_{\odot}$ for stars \#002 and \#201. These results highlight the importance of open clusters as laboratories for chemically peculiar stars.
\end{abstract}

\begin{keywords}
techniques: spectroscopic – stars: abundances – stars: fundamental parameters – (Galaxy:) open clusters and associations: individual: NGC 5822
\end{keywords}



\section{Introduction}\label{sec:intro}

Open clusters (OCs) are important benchmarks for studying stellar evolution and Galactic structure, as their member stars share common distances, initial chemical compositions, and ages. Moreover, age–chemical-clock relations, derived from empirical calibrations of stellar atmospheric abundances in OCs, provide critical insights into stellar ages and Galactocentric distances \citep[e.g.,][]{casali2019,vazquez2022,katime2022,sales2022}. 

\begin{table*}
\centering
\caption{General information about the sample.}
\label{tab:general}
\begin{tabular} {l c c c c c c c c c c c c c}
\hline
Star & RA & DEC & $V$ & $K$ & $G$ & $G_{\rm BP}\,-\,G_{\rm RP}$ & $\pi$ & $\mu_{\alpha}^{*}$ & $\mu_{\delta}$ & \texttt{RUWE} & Prob. \\
 & [deg] & [deg] & [mag] & [mag] & [mag] & [mag] & [mas] & [mas/yr] & [mas/yr] &  &  \\
\hline                                                              
NGC\,5822-001 & 226.0093 & $-$54.3392 & 9.11 & 5.97 & 8.66 & 1.45 & 1.283 & $-$7.348 & $-$4.702 & 3.59 & --- \\
NGC\,5822-002 & 225.9222 & $-$54.3758 & 9.49 & 7.07 & 9.22 & 1.20 & 1.380 & $-$7.751 & $-$5.869 & 1.65 & --- \\
NGC\,5822-006 & 226.0884 & $-$54.3849 & 10.79 & 8.35 & 10.49 & 1.22 & 1.257 & $-$7.555 & $-$5.591 & 0.95 & 1.00 \\
NGC\,5822-008 & 226.1642 & $-$54.3511 & 10.33 & 7.87 & 10.09 & 1.24 & 1.256 & $-$7.641 & $-$5.172 & 1.07 & 0.98 \\
NGC\,5822-102 & 225.9558 & $-$54.3363 & 10.86 & 8.37 & 10.54 & 1.22 & 1.262 & $-$7.495 & $-$5.148 & 1.01 & 0.99 \\
NGC\,5822-201 & 225.9584 & $-$54.2418 & 10.36 & 7.87 & 9.96 & 1.19 & 1.293 & $-$6.809 & $-$6.188 & 1.23 & --- \\
NGC\,5822-224 & 225.8452 & $-$54.4424 & 10.85 & 8.31 & 10.50 & 1.23 & 1.250 & $-$7.886 & $-$5.460 & 1.00 & 0.93 \\
NGC\,5822-240 & 226.1265 & $-$54.5291 & 9.53 & 6.25 & 9.04 & 1.51 & 1.215 & $-$7.605 & $-$5.469 & 1.03 & 1.00 \\
NGC\,5822-316 & 226.1271 & $-$54.5969 & 10.48 & 7.91 & 10.16 & 1.25 & 1.261 & $-$7.540 & $-$5.505 & 0.99 & 1.00 \\
NGC\,5822-348 & 225.7519 & $-$54.4716 & 10.98 & 8.34 & 10.57 & 1.23 & 1.272 & $-$7.572 & $-$5.691 & 0.94 & 1.00 \\
NGC\,5822-375 & 225.8159 & $-$54.2776 & 9.69 & 6.76 & 9.31 & 1.39 & 1.264 & $-$7.542 & $-$5.426 & 1.00 & 1.00 \\
NGC\,5822-443 & 225.7554 & $-$54.2143 & 9.75 & 6.81 & 9.35 & 1.39 & 1.279 & $-$7.465 & $-$5.349 & 0.93 & 1.00 \\
TYC\,8681-389-1 & 226.0142 & $-$54.0950 & 10.93 & 8.23 & 10.42 & 1.22 & 1.239 & $-$7.441 & $-$5.375 & 0.92 & 1.00 \\
\hline
\end{tabular}
\end{table*}

\par Although optical spectroscopy has traditionally been the standard for chemical abundance studies in OCs, near-infrared (NIR) spectroscopy has gained prominence due to its lower sensitivity to interstellar extinction and its access to molecular and atomic lines, which is particularly relevant for red giants and inaccessible in the optical domain. The \texttt{APOGEE} survey \citep{majewski2017} has significantly advanced the chemical characterization of stars in OCs by providing high-quality NIR spectra. For instance, \citet{donor2020} and \citet{myers2022} used \texttt{APOGEE} DR16/17 data to analyze over 120 open clusters, tracing Galactic abundance gradients in 16 elements and investigating their evolution with age, thereby offering new insights into the Galaxy’s chemical evolution. However, \texttt{APOGEE} is limited to the $H$ band and offers a resolution of $R\,\sim\,22,500$. Separately, \citet{topcu2019,topcu2020} analyzed $H$ and $K$ band spectra of red giant stars in the OCs NGC\,6940 and NGC\,752, respectively, deriving abundances for 20 chemical species and offering a more comprehensive chemical characterization. Building on this approach, \citet{holanda2024a} examined four stars in the young OC NGC\,2345, including the determination of fluorine abundances. Beyond these efforts, \citet{bijavara2024a,bijavara2024b} investigated fluorine abundances in OC stars to study chemical gradients and Galactic chemical evolution, also using high-resolution NIR spectroscopy; fluorine is only detectable via HF molecular lines in the $K$ band. \citet{nandakumar2022,nandakumar2023,nandakumar2024} conducted a series of investigations using high-resolution IGRINS spectra of 37 M giant stars, providing discussion on Galactic chemical evolution of phosphorus, fluorine, $\alpha$-elements, and other species. These studies also emphasized the identification and analysis of previously neglected lines, particularly for heavy elements such as Y\,{\sc i} and Ba\,{\sc i}. Moreover, these works establish a key foundation in the literature and contribute to the validation of atomic and molecular line lists.

\par Parallel developments using NIR spectroscopy have focused in the study of the Galaxy’s nuclear star cluster (NSC). Recent findings show that the NSC shares chemical similarities with the inner bulge, particularly in the enhanced $\alpha$-element abundances of metal-rich stars \citep[e.g.,][]{ryde2025}. These results suggest a shared chemical enrichment history and challenge models that invoke a recent dominant star formation burst in the NSC. Also recently, \citet{nandakumar2025} performed a comprehensive chemical census of nine NSCs, identifying abundance trends consistent with those of the inner bulge population.

\begin{figure*}
    \centering
    \includegraphics[width=1\linewidth]{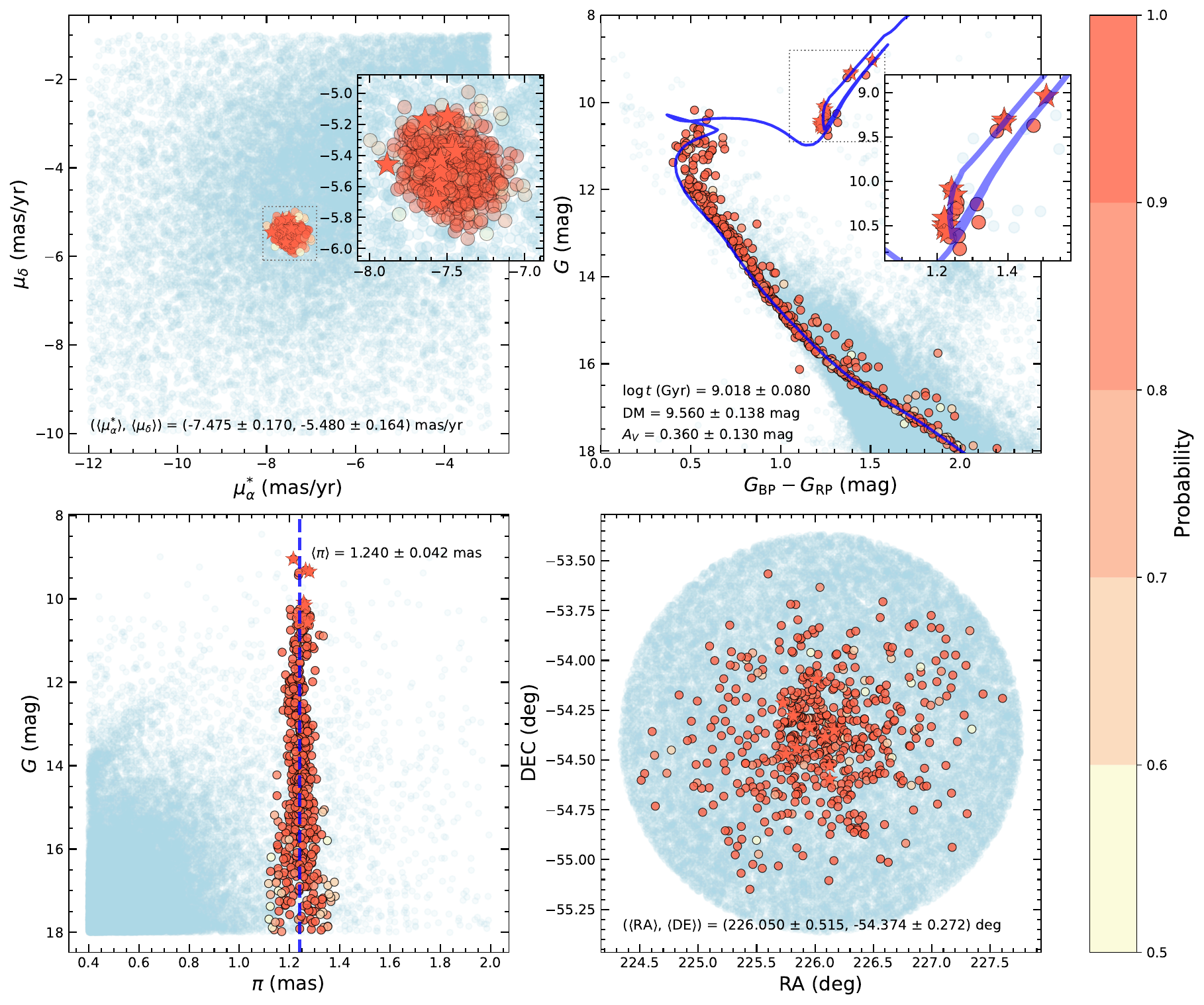}
    \caption{Proper motion distribution for bright stars around of the centre of NGC\,5822 (top left), where the colour coded and light blue points (field contaminants) represent stars with probability to belonging to cluster of P\,$\geq$\,0.5 and 0.0\,$\leq$\,P\,$<$\,0.5, respectively. Colour–magnitude diagram (top right) is shown with an isochrone fitting \citep{bressan2012} for stars with proper motions and parallaxes characteristics of the cluster ($\log\,t$\,=\,9.018).The bottom left panel shows magnitude versus parallax, where blue dashed line represents the mean parallax, while the bottom right panel represents the plane distribution of stars around the cluster. The stars analyzed in this work are marked with ``star'' symbols. Data were taken from \citet{gaia2023}.}
    \label{fig:membership}
\end{figure*}

\par NGC\,5822 is a well-studied OC in the optical domain, with several works investigating its red giant stars and providing chemical abundances for various species \citep[e.g.,][]{pace2010,santrich2013,sales2014,penasuarez2018,randich2022}. Among its main characteristics, we highlight its old age of 1.04\,Gyr, distance of about 0.80\,kpc, mean extinction in $V$ band of 0.20\,mag \citep{hunt2023}, and mean metallicity of $-0.09\,\pm\,0.06$\,dex \citep{penasuarez2018}. However, no study to date has explored its red giants using NIR spectroscopy covering both the $H$ and $K$ bands. As a result, elements such as fluorine and ytterbium have never been investigated in the stars of this cluster. Moreover, the cluster NGC\,5822 is also notable for hosting chemically peculiar stars: for instance, \citet{santrich2013} identified and analyzed two Ba-stars, \#002 and \#201. The authors cited that in the optical domain, these stars show strong Ba\,{\sc ii} lines with equivalent widths greater than 200\,m\AA. These absorption features are found in the saturated regime of the curve-of-growth, preventing the derivation of reliable Ba abundances in these stars. Moreover, barium stars are exceptionally rare in OCs, and recent discussions on chemically peculiar cluster members have often centered on blue stragglers enriched in s-process elements \citep[e.g.,][]{pal2024,nine2024}. In contrast, low-mass Li-rich giants account for only about 1\,\% of all G–K giants \citep[e.g.,][]{martell2021}. Notably, NGC\,5822 hosts three giant stars with a significant degree of lithium enrichment (\#006, \#102, and \#240). Unlike Ba-stars, the mechanism responsible for lithium enrichment in giant atmospheres remains unclear and continues to be a topic of active investigation.

In this work, we present the most comprehensive chemical study to date of giant stars in the OC NGC\,5822, combining high-resolution optical and NIR spectroscopy. We also perform a robust membership analysis to identify the true members of the cluster and better constrain evolutionary stages and the mean abundances of 23 chemical elements (Li, C, N, O, F, Na, Mg, Al, Si, P, S, K, Ca, Sc, Ti, Cr, Fe, Ni, Y, Ce, Nd, Yb, and Pb) and three isotopic ratios ($^{12}$C/$^{13}$C, $^{16}$O/$^{17}$O, and $^{16}$O/$^{18}$O), applied both chemically normal and peculiar stars associated with the cluster. Section\,\ref{sec:obser} provides a general overview of the observational data and describes the membership and spectroscopic analyses. In Section\,\ref{sec:discuss}, we present a detailed discussion of the derived abundances across different groups of elements, mixing models tests, and a particular approach on the nature of peculiar stars and nucleosynthesis models. Finally, Section\,\ref{sec:conc} summarizes our main findings and presents the concluding remarks.

\section{Observations and analysis}\label{sec:obser}

The selected sample for this work is based on stars analyzed by \citet{penasuarez2018} (\#001, \#006, \#008, \#102, \#224, \#240, \#316, \#348, \#375, \#443, and TYC\,8681-389-1) and \citet{katime2013} (\#002 and \#201). Our analysis combines public astrometric and photometric data with spectra obtained through a one observation proposal and an agreement between the Max Planck Institute/European Southern Observatory (ESO) and Observatório Nacional. 

\par The publicly available data \citep{gaia2023} were used for a comprehensive membership analysis of stars in the OC’s direction, while the obtained spectra were used to derive detailed physical parameters and perform a chemical abundance analysis. The observations were carried out using the Immersion GRating INfrared Spectrometer \citep[\texttt{IGRINS};][]{yuk2010} on the 8.1\,m Gemini South telescope at Cerro Pach\'on, Chile, by proposal submitted to Laborat\'orio Nacional de Astrof\'isica (GS-2024A-Q-301; P.I. Nacizo Holanda). The \texttt{IGRINS} is a cross-dispersed NIR spectrograph with a resolving power of R\,$\approx$\,45,000 that covers the $H$ (14,900\,-\,18,000\,\AA) and $K$ bands (19,600\,-\,24,600\,\AA), providing broad spectral coverage and high spectral resolution. The targets were observed using an ABBA nod sequence along the slit, employing exposure times to ensure spectra with a high signal-to-noise ratio, S/N$_{\rm IR}$\,$\geq$\,150, measured at 21,500\,\AA\, region. The data reduction process involved utilizing the \texttt{IGRINS} pipeline package, \texttt{PLP}. As part of the pipeline processing, the spectra were subjected to key steps such as telluric correction, wavelength calibration, and flat-field correction. Telluric features were corrected using observations of an A0V star acquired at a similar airmass to the target.

\par The second part of the observations were carried out using the Fiber-fed Extended Range Optical Spectrograph \citep[\texttt{FEROS};]{kaufer1999} at the 2.2\,m Max Planck Gesellschaft/ESO Telescope in La Silla, Chile. The \texttt{FEROS} spectra have a resolving power of R\,$\approx$\,48,000 within the spectral coverage of 3,600\,-\,9,200\,\AA. The exposure time were estimated to achieve a S/N$_{\rm OPT}$ high enough to resolve atomic and molecular lines. The \texttt{FEROS} Data Reduction System pipeline was used to reduce all optical spectra. The next subsections describes each dataset.

\subsection{Membership analysis}

\par To conduct a detailed study of the stars in NGC\,5822, we identified all cluster members with $G$\,$\leq$\,18.0\,mag and determined key cluster parameters, including mean proper motion components ($\mu^*_\alpha$, $\mu_\delta$), parallax ($\pi$), distance ($d$), distance modulus ($M\,-\,m$), extinction ($A_{\rm V}$), and age ($t$), based on data from \textit{Gaia} DR3 \citep{gaia2023}. To achieve this, a zero-point offset correction was applied following \citet{lindegren2021}, and quality cuts were implemented to exclude unreliable and contaminated data. Stars with negative parallax, discrepant proper motion, and high Renormalized Unit Weight Error (\texttt{RUWE}\,$\geq$\,1.2) were removed to mitigate the impact of poor astrometric solutions and unresolved binaries \citep[e.g.,][]{krolikowski2021,hernandez2024}.

\par Membership was determined using the Unsupervised Photometric Membership Assignment in Stellar Clusters code \citep[\texttt{UPMASK};][]{krone2014}, implemented via the Python-based \texttt{pyUPMASK} \citep{pera2021}, which improves performance and usability over the original R version. Using three principal component analysis (PCA) dimensions, 10 outer loops, 25 maximum inner loops, and a minimum group size of 10 stars, we identified 557 likely members with astrometric probabilities\,$\geq$\,0.50, including 10 giant stars from our spectroscopic sample (\#006, \#008, \#102, \#224, \#240, \#316, \#348, \#375, \#443, and TYC\,8681-389-1).

\par The threshold of \texttt{RUWE}\,$\geq$\,1.2 includes stars such as \#002 and \#201, classified as Ba-stars by \citet{santrich2013}. Ba-stars are chemically peculiar objects believed to form in binary systems \citep{mcclure1983}, typically through mass-transfer processes such as stellar wind accretion or Roche lobe overflow. As a result, their binary nature may have affected their astrometric solutions, potentially compromising the reliability of their membership probabilities. Additionally, \#001 is not classified as a spectroscopic binary by \citet{mermilliod2008} but exhibits a high \texttt{RUWE} value. Therefore, these three stars do not present probability determined. In Sec.\,\ref{sec:discuss}, we present the chemical abundance results that can be linked with membership results. With the exception of neutron-capture elements, stars \#002 and \#201 show abundances consistent with the mean values observed for the OC. In contrast, star \#001 presents discrepant abundance results in some cases when compared to other cluster members. Table\,\ref{tab:general} compile all basic information about the thirteen stars, including photometric and astrometric information from \textit{Gaia} DR3 catalog, \texttt{RUWE}, and final astrometric pobability obtained in the membership analysis.

\par For the isochrone fitting, we employed the routine available in the Automated Stellar Cluster Analysis code \citep[\texttt{ASteCA};][]{perren2015}, which uses a likelihood-based approach to achieve the best match between model isochrones and the observed stellar distribution. The stellar sample was defined using \texttt{pyUPMASK}, considering only stars with an astrometric membership probability $\geq$\,0.50. Figure\,\ref{fig:membership} shows the proper motion distribution for stars in NGC\,5822 (top left), where the color scale represents the membership probability (P\,$\geq$\,0.5). The top right panel presents the color-magnitude diagram (CMD) for stars that exhibit proper motion and parallax values consistent with cluster membership and field stars (in blue), along with the best-fit isochrone from \citet{bressan2012}, corresponding to $\log\,t$\,=\,9.018. 

\par The results obtained from the isochrone fitting are in excellent agreement with previous studies \citep{Cantat2020,dias2021,hunt2023}. The only significant discrepancy arises in the extinction value reported by \citet{hunt2023}, which differs from all other determinations for this OC. Table\,\ref{tab:cluster_par} presents the cluster parameters derived from isochrone fitting and from mean values based on the final list of member stars. For completeness, [Fe/H] results obtained from spectroscopic analysis are also provided.

\subsection{Spectroscopic analysis}

Spectroscopic analysis was performed using the radiative transfer code \texttt{MOOG} \citep[version 2019;][]{sneden1973}, which serves as the primary tool for deriving atmospheric parameters and chemical abundances and assumes local thermodynamical equilibrium (\texttt{LTE}). The atmospheric parameters were determined using high-resolution \texttt{FEROS} spectra, which provide extensive spectral coverage, enabling the measurement of numerous absorption lines from neutral and singly ionized iron. The analysis was conducted using the grid of plane-parallel atmosphere models by \citet{castelli2004}, which also assume \texttt{LTE}. \texttt{IGRINS} spectra were used to derive chemical abundances via spectral synthesis method, as described in the following subsections.

\begin{table}
\centering
\caption{Main cluster parameters. The values obtained in this work were calculated considering the mean values of the cluster members and assuming 8.34\,$\pm$\,0.6\,kpc for the solar galactocentric distance.}
\label{tab:cluster_par}
\begin{tabular} {l c c}
\hline
RA (deg)                        & 226.050\,$\pm$\,0.515 & This Work \\
DEC (deg)                       & $-$\,54.374\,$\pm$\,0.272 & This Work \\
$\mu^*_\alpha$ (mas yr$^{-1}$)  & $-$\,7.475\,$\pm$\,0.170 & This Work \\
                                & $-$\,7.422\,$\pm$\,0.222 & \citet{Cantat2020} \\
                                & $-$\,7.417\,$\pm$\,0.247 & \citet{dias2021} \\
                                & $-$\,7.462\,$\pm$\,0.189 & \citet{hunt2023} \\
$\mu_\delta$ (mas yr$^{-1}$)    & $-$\,5.480\,$\pm$\,0.164 & This Work \\
                                & $-$\,5.534\,$\pm$\,0.205 & \citet{Cantat2020} \\
                                & $-$\,5.534\,$\pm$\,0.209 & \citet{dias2021} \\
                                & $-$\,5.478\,$\pm$\,0.179 & \citet{hunt2023} \\ 
$\pi$ (mas)                     & 1.240\,$\pm$\,0.042 & This Work \\
                                & 1.187\,$\pm$\,0.050 & \citet{Cantat2020} \\
                                & 1.188\,$\pm$\,0.054 & \citet{dias2021} \\
                                & 1.199\,$\pm$\,0.048 & \citet{hunt2023} \\
${R_\text{GC}}$ (kpc)           & 7.72\,$\pm$\,0.16 & This Work \\
                                & 7.69      & \citet{Cantat2020} \\
$[\text{Fe/H}]^*$ (dex)         & $-$\,0.06\,$\pm$\,0.05 & This Work \\
                                & $-$\,0.09\,$\pm$\,0.06 & \citet{penasuarez2018} \\
                                & $+$0.02 & \citet{randich2022} \\
$M_\text{turn-off}$ (M$_\odot$) & 2.05\,$\pm$\,0.10   & This Work \\
distance (pc)                   & 817\,$\pm$\,52 & This Work \\    
                                & 854       & \citet{Cantat2020} \\
                                & 796\,$\pm$\,27      & \citet{dias2021} \\
                                & 808       & \citet{hunt2023} \\
$\log\,t$ (Gyr)                 & 9.018\,$\pm$\,0.080 & This Work \\    
                                & 8.960     & \citet{Cantat2020} \\
                                & 9.039\,$\pm$\,0.060 & \citet{dias2021} \\
                                & 9.017     & \citet{hunt2023} \\
$A_{\rm V}$ (mag)               & 0.448\,$\pm$\,0.056 & This Work \\
                                & 0.390     & \citet{Cantat2020} \\
                                & 0.481\,$\pm$\,0.060 & \citet{dias2021} \\
                                & 0.202     & \citet{hunt2023} \\
$M\,-\,m$ (mag)                 & 9.516\,$\pm$\,0.138 & This Work \\    
                                & 9.660     & \citet{Cantat2020} \\
                                & 9.453     & \citet{hunt2023} \\
\hline
\multicolumn{3}{l}{\footnotesize $^*$The iron abundance was derived from spectroscopy.} \\
\end{tabular}
\end{table}

\subsubsection{Atmospheric parameters} \label{sec:atm}

First, the effective temperature ($T_{\rm eff}$), surface gravity ($\log\,g$), microturbulent velocity ($\xi_{t}$), and metallicity ([Fe/H]), were derived using the standard spectroscopic approach. For that, we employed a carefully measured set of absorption lines from neutral and singly ionized iron (Fe\,{\sc i} and Fe\,{\sc ii}), initially compiled by \citet{lambert1996}. While the original catalog contained 165 Fe\,{\sc i} and 24 Fe\,{\sc ii} lines, a meticulous line-by-line evaluation was performed to select the most reliable measurements, utilizing Automatic Routine for line Equivalent widths in stellar Spectra \citep[\texttt{ARES} v2;][]{sousa2015} and \texttt{IRAF} through the {\sl splot} routine \citep{tody1986}. We imposed selection criteria of equivalent width between 10\,m\AA\, to 150\,m\AA\, with relative error of $\sigma_{\rm EW}$\,$\leq$\,10\,\%.

\par The parameter $T_\text{eff}$ was determined by ensuring that the Fe\,{\sc i} abundance showed no dependence on the excitation potential ($\chi$). On the other hand, $\xi_{\rm t}$ was established by requiring that the Fe\,{\sc i} abundance were independent of the reduced equivalent width ($\text{EW}/\lambda$). Surface gravity was calculated by enforcing ionization equilibrium, ensuring that $\log\,\varepsilon(\text{Fe\,{\sc i}})$ is very similar to $\log\,\varepsilon(\text{Fe\,{\sc ii}})$. Table\,\ref{tab:atm} presents final results for atmospheric parameters based on the standard spectroscopic analysis.

\begin{table*}
\centering
\caption{Derived atmospheric parameters, metallicities, and rotational velocities for stars of the open cluster NGC\,5822. The symbol ``\#'' indicates the number of lines used to determine each set of spectroscopic parameters.}
\label{tab:atm}
\begin{tabular} {l c c c c c c c c c c c}
\hline
Star & $T_{\rm eff}^{\rm V-K}$ & $T_{\rm eff}$ & $\log\,g$ & $\xi_{\rm t}$ & [Fe\,{\sc i}/H] & \# & [Fe\,{\sc ii}/H] & \# & $v\,\sin\,i$ & S/N$_{\rm OPT}$ & S/N$_{\rm IR}$ \\
 & [K] & [K] & [cm\,s$^{-2}$] & [km\,s$^{-1}$] & [dex] &  & [dex] &  & [km\,s$^{-1}$] &  &  \\
\hline                                                              
NGC\,5822-001 & 4415 & 4360 & 1.55 & 1.38 & $-$0.12 & 52 & $-$0.12 & 7 & 4.7 & 116 & 280 \\ 
NGC\,5822-002 & 5104 & 5170 & 2.65 & 1.65 & $-$0.06 & 57 & $-$0.06  & 9 & 5.8 & 200 & 360 \\ 
NGC\,5822-006 & 5079 & 5110 & 2.85 & 1.22 & $+$0.00 & 65 & $+$0.00 & 9 & 6.1 & 157 & 170 \\ 
NGC\,5822-008 & 5055 & 4990 & 2.70 & 1.31 & $-$0.01 & 64 & $-$0.01 & 12 & 5.0 & 167 & 230 \\ 
NGC\,5822-102 & 5021 & 5040 & 2.75 & 1.23 & $-$0.09 & 65 & $-$0.09 & 10 & 5.7 & 185 & 156 \\ 
NGC\,5822-201 & 5022 & 5150 & 2.60 & 1.44 & $-$0.12 & 57 & $-$0.12 & 11 & 4.7 & 140 & 250 \\ 
NGC\,5822-224 & 4983 & 5030 & 2.70 & 1.14 & $-$0.03 & 64 & $-$0.03 & 10 & 4.0 & 160 & 270 \\ 
NGC\,5822-240 & 4315 & 4410 & 1.90 & 1.43 & $-$0.10 & 45 & $-$0.10 & 10 & 3.0 & 210 & 255 \\ 
NGC\,5822-316 & 4929 & 5050 & 2.75 & 1.35 & $+$0.00 & 63 & $+$0.01 & 11 & 3.0 & 149 & 310 \\ 
NGC\,5822-348 & 4855 & 5030 & 2.80 & 1.29 & $-$0.07 & 59 & $-$0.07 & 12 & 2.8 & 132 & 222 \\ 
NGC\,5822-375 & 4581 & 4640 & 2.10 & 1.45 & $-$0.09 & 49 & $-$0.11 & 12 & 2.2 & 106 & 200 \\ 
NGC\,5822-443 & 4574 & 4670 & 2.30 & 1.55 & $-$0.15 & 47 & $-$0.15 & 10 & 2.8 & 128 & 237 \\ 
TYC\,8681-389-1 & 4793& 5100 & 2.90 & 1.38 & $-$0.02 & 43 & $-$0.02 & 10 & 3.1 & 70 & 170 \\ 
\hline 
\end{tabular}
\end{table*}

\par In addition, to derive a second and independent set of effective temperature values, we employed a photometric method based on the ($V-K$) color index, which is considered one of the most reliable temperature indicators for giant stars \citep{alonso1999}. For this calculation, we adopted the extinction ratio $A_{\rm V}/E(V-K)\,=\,1.13$ \citep[][and references therein]{cox2000} and a color excess of $E(B-V)\,=\,0.145$\,mag, obtained from isochrone fitting. The observed mean difference (T$_{\rm{eff}}$\,--\,T$_{\rm{eff}}^{\rm V-K}$) is 39\,K, with a standard deviation of 82\,K. The Pearson correlation coefficient between the two sets is $+$0.98, indicating a strong linear agreement between the spectroscopic and photometric temperatures, despite a small systematic offset.

\begin{figure*}
    \centering
    \includegraphics[scale=0.35]{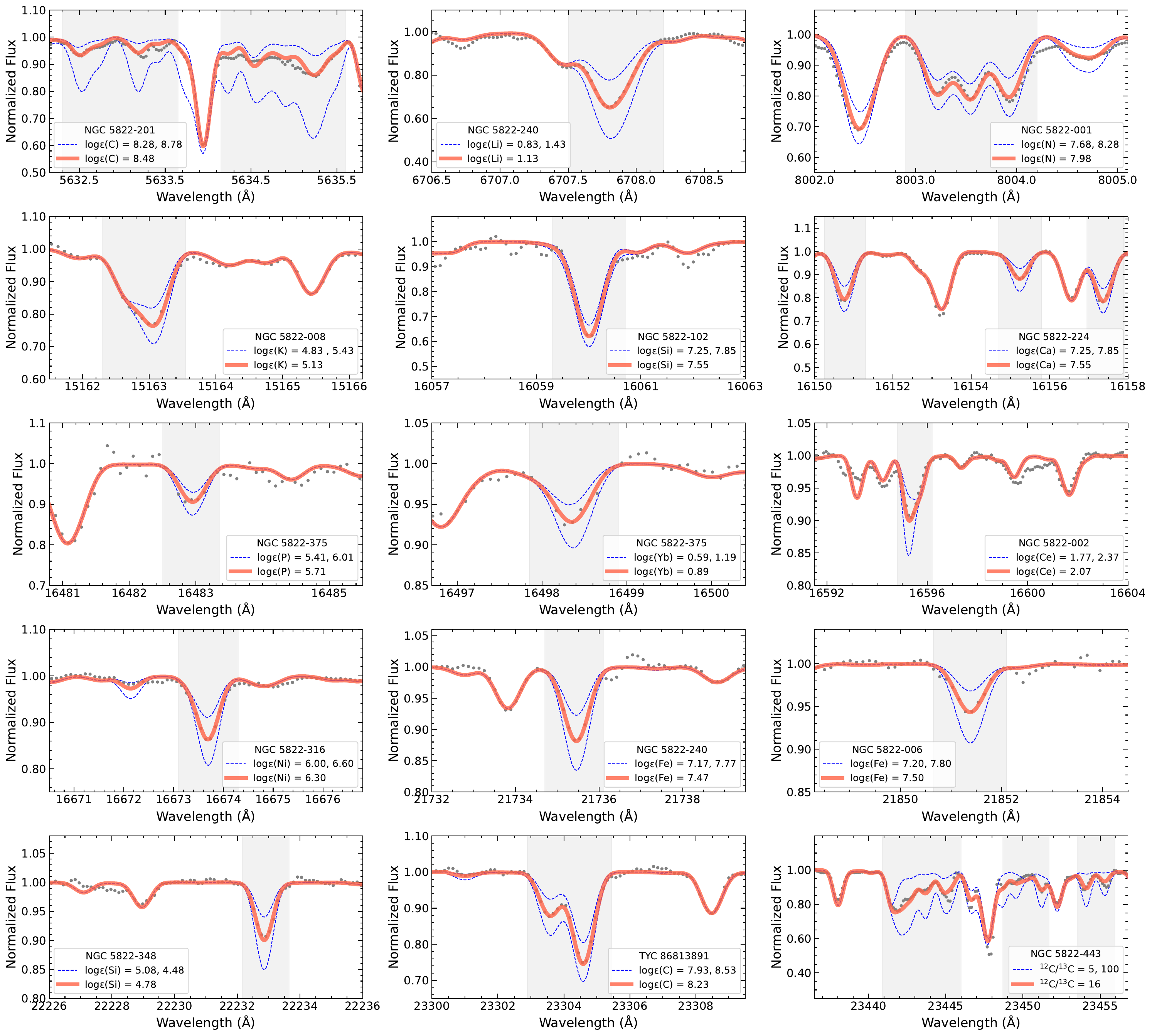}
    \caption{Examples of best-fitting results obtained from synthetic spectra (red lines) and observed IGRINS spectra (grey points) of stars. Additional lines are included to illustrate variations in abundance for the same spectral features (blue dashed-lines; do not represent uncertainties).}
    \label{fig:dev_synth}
\end{figure*}

\subsubsection{Projected rotational velocities}

Stellar rotation can be investigated through the projected rotational velocity ($v\,\sin\,i$) and is a powerful characteristic that can be related to chemical abundances for understanding internal structure. To estimate the possible contributions of rotational velocity in the chemical pattern of the peculiar stars analyzed here, we determined the $v\,\sin\,i$ by using spectral synthesis of the Fe\,{\sc i} 5848\,\AA, 6027\,\AA, 6188\,\AA\, lines. Following the same methodology adopted in \citet{holanda2021}, we fixed macroturbulence velocity at 3.0\,km\,s$^{-1}$, typical for G and K giants \citep{fekel1997}, and considering the instrumental broadening of the \texttt{FEROS} spectral resolution (FWHM\,$\approx$\,0.13). The best value of $v\,\sin\,i$ for each line was obtained using an iterative procedure until we find smallest deviation between the synthetic and observed spectra, and the mean value for each star is shown in Table\,\ref{tab:atm}.

\subsubsection{Chemical abundance determination}

The list of chemical species analyzed was primarily compiled based on the transitions available within the \texttt{IGRINS} spectral coverage. When possible, the same species were also analyzed in the \texttt{FEROS} data, and in such cases, we adopted the average abundance value derived from both optical and NIR spectra. The only exceptions are lithium and lead: Li was included due to its importance as a sensitive tracer of internal mixing and stellar evolutionary processes, and, Pb was included to represent the third peak of heavy elements, which is important to our study of chemical enrichment in Ba-stars (see Sec.\,\ref{sec:ba_stars}).

\par We adopted two techniques to derive chemical abundances: equivalent width and spectral synthesis. Equivalent width measurements were performed in the optical for Na, Mg, Al, Si, Ca, Ti, Cr, Fe, Y, Ce, and Nd, while species such as Li, C, N, O, S, Sc, and Pb were analyzed using the spectral synthesis technique. On the other hand, all chemical abundances derived from the NIR region were obtained exclusively through spectral synthesis, since molecular bands are ubiquitous along NIR wavelengths. The line lists used in this work are similar to those adopted by previous works of our group \citep[see][and references therein]{holanda2024a}. We used the \texttt{LINEMAKE} program\footnote{Available at \href{https://github.com/vmplacco/linemake}{\rm https://github.com/vmplacco/linemake}} \citep[][and references therein]{placco2021} to generate the input line lists for all spectral regions used in the synthesis, with minor adaptations applied when necessary.

\par Concerning the analysis of light species in optical spectra, carbon abundances were derived from the C$_2$ (0,1) band head of the Swan system at 5635\,\AA\,\citep{lambert1978, huber1979}, while nitrogen abundances were obtained from CN lines in the 8002–8005\,\AA\,range \citep{sneden1982}. The wavelengths for $^{12}$CN and $^{13}$CN lines were adopted from \citet{davis1963} and \citet{wyller1966}, respectively. Due to telluric contamination affecting the [O\,{\sc i}] 6300\,\AA\, line, oxygen abundances were determined from NIR analysis and adopted as fixed values in the optical analysis, given the interdependence of CNO species. Additionally, lithium abundances were determined from the Li\,{\sc i} $\lambda$6708 resonance doublet, assuming \texttt{LTE} conditions and accounting for nearby CN and Fe lines. Hyperfine and isotopic components of the Li line were included, with oscillator strengths and wavelengths taken from \citet{smith1998} and \citet{hobbs1999}.

\par For the determination of light element abundances from NIR spectra, rotational-vibrational transitions of CO molecules were used to derive the abundances of $^{12}$C \citep{goorvitch1994}, as well as the isotopes $^{13}$C, $^{17}$O, and $^{18}$O, while OH molecular lines were adopted to determine the abundance of $^{16}$O \citep{brooke2016}. The CO and OH lines are abundant in the NIR spectra of evolved stars, allowing us to analyze different regions in the $H$ and $K$ bands to obtain mean abundance values. However, the OH molecule has a low dissociation potential, making the synthesis process challenging for the higher temperature stars in our sample. Additionally, nitrogen abundances were determined by analyzing the electronic transitions of $^{12}$C$^{14}$N \citep[$A^{2}\Pi-X^{2}\Sigma$;][]{sneden2014}.  

\par To address the interdependence of these light species, we followed a procedure similar to that described by \citet{smith2013}, summarized as follows: first, the $^{16}$O abundance was constrained using OH lines and fixed to estimate the $^{12}$C abundance from the synthesis of $^{12}$CO lines. The derived $^{12}$C abundance was then used to refine the $^{16}$O abundance from OH lines, and this updated $^{16}$O value was subsequently fixed to obtain the final $^{12}$C abundance from $^{12}$CO lines. Finally, nitrogen abundances were determined through CN synthesis, keeping the final $^{16}$O and $^{12}$C abundances fixed. Given that C/O\,$\leq$\,1.0, the impact of CO and OH lines on nitrogen abundance is minimal. For these molecular species, we adopted dissociation energy values of D$_0$(CO)\,=\,11.092\,eV, D$_0$(OH)\,=\,4.411\,eV, and D$_0$(CN)\,=\,7.724\,eV.  

\par The isotopic ratios $^{12}$C/$^{13}$C and $^{16}$O/$^{17, 18}$O were derived using $gf$-values from \citet{kurucz2011}. The $^{12}$C/$^{13}$C ratio was determined by analyzing two spectral regions containing the $^{13}$C$^{16}$O\,(2$-$0)\,R49-56 and (3$-$1)\,R46-R57 lines. The $^{16}$O/$^{17}$O ratio was obtained through spectral synthesis of two regions featuring the $^{12}$C$^{17}$O\,(2$-$0)\,R28 and R29 lines. Additionally, the $^{16}$O/$^{18}$O ratio was determined using the $^{12}$C$^{18}$O\,(2$-$0)\,R23 line. In Section\,\ref{sec:uncert} we propagate the uncertainties of atmospheric parameters and CNO variation in these ratios, since oxygen isotopic ratios are very sensitive to atmospheric parameters. 

\par The fluorine abundance was determined through analysis of the rotational-vibrational transitions of H$^{19}$F\,(1$-$0)\,R11 and R9, using the $gf$-values from \citet{jonsson2014}. These lines, located at 23134\,\AA\ and 23357\,\AA, respectively, are commonly used in the literature to determine fluorine abundances in cool stars. However, their intensities decrease significantly with increasing temperature, a characteristic behavior of certain molecular absorption features (D$_0$(F)\,=\,5.869\,eV). 

\par Additionally, the chemical abundances of Na\,{\sc i}, Mg\,{\sc i}, Al\,{\sc i}, Si\,{\sc i}, P\,{\sc i}, S\,{\sc i}, K\,{\sc i}, Ca\,{\sc i}, Sc\,{\sc i}, Ti\,{\sc i}, Cr\,{\sc i}, Fe\,{\sc i}, Ni\,{\sc i}, Ce\,{\sc ii}, Nd\,{\sc ii}, and Yb\,{\sc ii} were determined via NIR spectral synthesis of atomic lines. Much of the line information was sourced from \citet{afsar2018}, \citet{civis2013}, \citet{kramida2014}, \citet{pehlivan2015}, \citet{hasselquist2016}, and \citet{cunha2017}. In Figure\,\ref{fig:dev_synth}, we present examples of spectral synthesis performed in different regions of the optical and NIR spectra.

\par We have applied available corrections to the abundance values, considering the departure from \texttt{LTE} (\texttt{NLTE}) observed in late type stars. In this sense, we employed correction grids provided by \citet{lind2009} for Li\,{\sc i}, \citet{lind2011} for Na\,{\sc i}, \citet{osorio2015} for Mg\,{\sc i}, \citet{nordlander2017} for Al\,{\sc i}, \citet[private communication]{mashonkina2012} for Pb lines. 

\par
\begin{figure*}
    \centering
    \includegraphics[scale=0.4]{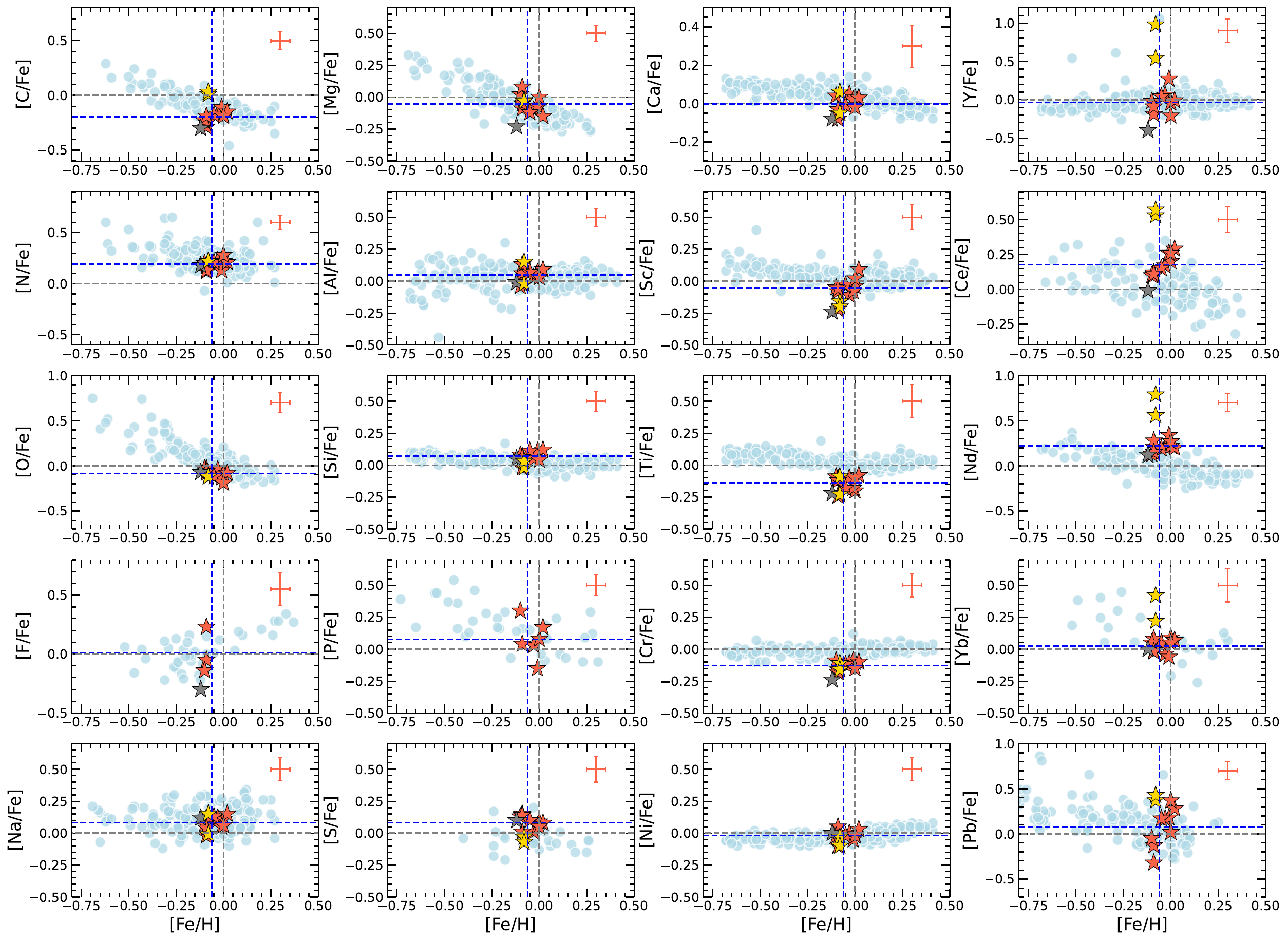}    
    \caption{Abundance ratios [X/Fe] versus [Fe/H] for the analysed sample: stars with high astrometric probability are represented in red and the stars with low probability are denoted by yellow (\#002 and \#201; Ba-stars) and gray (\#001) colour symbols. The light blue circles represent the samples of giant stars from \citet[][C, N, O, Na, Mg]{mishenina2006}, \citet[][F]{ryde2020}, \citet[][P]{nandakumar2022}, from \citet[][Yb]{montelius2022}, and from \citep[][Pb]{contursi2024}. We also include dwarf stars analyzed by \citet[][S]{lucertini2022}, \citet[][Al, Si, Ca, Ti, Cr, Ni]{bensby2014}, \citet[][Sc]{battistini2015}, and \citet[][Y, Ce, Nd]{battistini2016}, selected with Age\,$<$\,7.0\,Gyr and TD/D\,$<$\,0.5. The gray dashed lines indicate the solar values, while the blue dashed lines represent the mean values of the cluster members.}
    \label{fig:ab_literature}
\end{figure*}

 Figure\,\ref{fig:ab_literature} compares our individual abundance results with distributions from the literature, based on various sources that employed similar spectroscopic analysis methods. All abundance ratios were normalized to the solar abundances of \citet{asplund2009}. Finally, all abundance results and isotopic ratios are shown in Table\,\ref{tab:ab_values} and Table\,\ref{tab:isotopic}. These results will be discussed in detail in Section\,\ref{sec:discuss}.

\begin{figure*}
    \centering
    \includegraphics[width=1\linewidth]{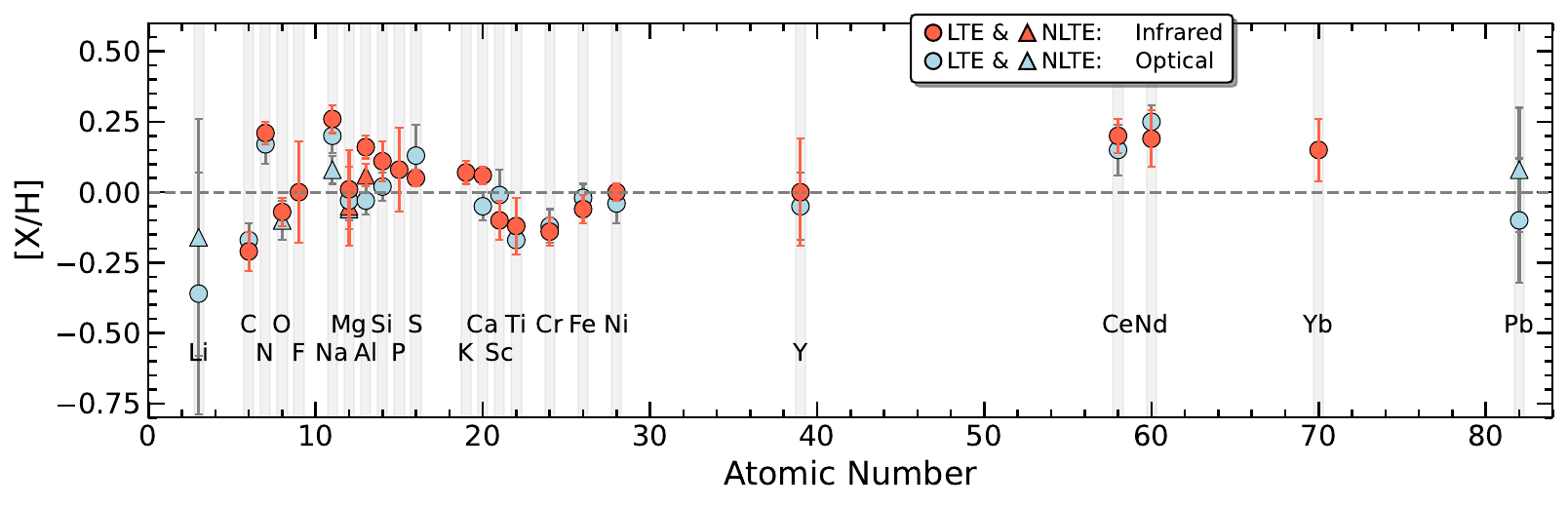}
    \caption{Mean elemental abundances for NGC\,5822 derived from the optical (light blue) and NIR (red) spectral regions. Triangles indicate mean abundances corrected for \texttt{NLTE} effects. Abundance ratios are normalized to the solar abundances of \citet{asplund2009}.} 
    \label{fig:ab_general}
\end{figure*}

\begin{table*}
\centering
\caption{Abundance derived in this study. OC Mean indicates the mean cluster value, excluding the stars \#001, \#002, and \#201.}
\label{tab:ab_values}
\begin{tabular}{>{\raggedright\arraybackslash}p{0.9cm} >{\centering\arraybackslash}p{0.5cm} >{\centering\arraybackslash}p{0.6cm} >{\centering\arraybackslash}p{0.6cm} >{\centering\arraybackslash}p{0.6cm} >{\centering\arraybackslash}p{0.6cm} >{\centering\arraybackslash}p{0.6cm} >{\centering\arraybackslash}p{0.6cm} >{\centering\arraybackslash}p{0.6cm} >{\centering\arraybackslash}p{0.6cm} >{\centering\arraybackslash}p{0.6cm} >{\centering\arraybackslash}p{0.6cm} >{\centering\arraybackslash}p{0.6cm} >{\centering\arraybackslash}p{0.6cm} >{\centering\arraybackslash}p{0.6cm} >{\centering\arraybackslash}p{1.2cm}}
\hline
$\log\,\varepsilon$(X) &  & \#001 & \#002 & \#006 & \#008 & \#102 & \#201 & \#224 & \#240 & \#316 & \#348 & \#375 & \#443 & TYC & OC Mean \\ \hline
Li\,{\sc i} & OPT & $-$0.10 & 0.15 & 1.31 & 0.50 & 1.24 & 0.69 & 0.28 & 1.13 & 0.81 & 0.24 & 0.27 & 0.26 & 0.89 & 0.69\,$\pm$\,0.43 \\ 
Li\,{\sc i}$_{\rm NLTE}$ & OPT & 0.24 & 0.27 & 1.45 & 0.67 & 1.39 & 0.83 & 0.44 & 1.43 & 0.98 & 0.44 & 0.54 & 0.50 & 1.04 & 0.89\,$\pm$\,0.42 \\ 
C (CO) & IR & 8.07 & 8.40 & 8.23 & 8.23 & 8.09 & 8.44 & 8.33 & 8.18 & 8.32 & 8.22 & 8.17 & 8.18 & 8.28 & 8.22\,$\pm$\,0.07 \\
C (C$_{2}$) & OPT & 8.18 & 8.48 & 8.28 & 8.26 & 8.18 & 8.48 & 8.30 & 8.17 & 8.23 & 8.27 & 8.23 & 8.30 & 8.20 & 8.26\,$\pm$\,0.06 \\ 
N (CN) & IR & 8.07 & 8.03 & 8.09 & 8.07 & 8.04 & 8.07 & 7.98 & 7.97 & 8.07 & 8.03 & 8.07 & 7.99 & 8.07 & 8.04\,$\pm$\,0.04 \\
N (CN) & OPT & 7.95 & 8.06 & 8.02 & 8.03 & 7.95 & 8.01 & 7.94 & 7.92 & 8.00 & 8.05 & 7.98 & 7.93 & 8.14 & 8.00\,$\pm$\,0.07 \\ 
O (OH) & IR & 8.57 & 8.55 & 8.74 & 8.61 & 8.58 & 8.61 & 8.56 & 8.63 & 8.65 & 8.57 & 8.64 & 8.64 & 8.57 & 8.62\,$\pm$\,0.05 \\
O\,{\sc i}$_{\rm NLTE}$ & OPT & 8.66 & 8.59 & 8.56 & 8.58 & 8.63 & 8.53 & 8.56 & 8.69 & 8.57 & 8.57 & 8.67 & 8.62 & 8.43 & 8.59\,$\pm$\,0.07 \\ 
F (HF) & IR & 4.26 & -- & -- & -- & -- & -- & -- & 4.51 & -- & -- & 4.42 & 4.79 & -- & 4.57\,$\pm$\,0.18 \\ 
Na\,{\sc i} & IR & 6.50 & 6.43 & 6.53 & 6.54 & 6.47 & 6.46 & 6.52 & 6.47 & 6.50 & 6.59 & 6.41 & 6.51 & 6.45 & 6.50\,$\pm$\,0.05 \\
Na\,{\sc i} & OPT & 6.48 & 6.50 & 6.48 & 6.47 & 6.32 & 6.32 & 6.40 & 6.49 & 6.51 & 6.49 & 6.41 & 6.40 & 6.43 & 6.44\,$\pm$\,0.06 \\
Na\,{\sc i}$_{\rm NLTE}$ & OPT & 6.36 & 6.39 & 6.36 & 6.35 & 6.22 & 6.23 & 6.29 & 6.36 & 6.39 & 6.37 & 6.31 & 6.28 & 6.30 & 6.32\,$\pm$\,0.05 \\ 
Mg\,{\sc i}$_{\rm NLTE}$ & IR & 7.33 & 7.59 & 7.58 & 7.53 & -- & 7.58 & 7.53 & 7.59 & 7.33 & 7.38 & 7.72 & 7.76 & 7.50 & 7.55\,$\pm$\,0.14 \\ 
Mg\,{\sc i} & IR & 7.43 & 7.67 & 7.64 & 7.60 & -- & 7.66 & 7.61 & 7.70 & 7.39 & 7.44 & 7.73 & 7.85 & 7.56 & 7.61\,$\pm$\,0.14 \\ 
Mg\,{\sc i}$_{\rm NLTE}$ & OPT & 7.40 & 7.48 & 7.42 & 7.52 & 7.51 & 7.56 & 7.52 & 7.52 & 7.57 & 7.57 & 7.51 & 7.59 & 7.69 & 7.54\,$\pm$\,0.07 \\
Mg\,{\sc i} & OPT & 7.48 & 7.50 & 7.44 & 7.55 & 7.53 & 7.59 & 7.54 & 7.58 & 7.59 & 7.59 & 7.56 & 7.63 & 7.71 & 7.57\,$\pm$\,0.07 \\
Al\,{\sc i}$_{\rm NLTE}$ & IR & 6.44 & 6.60 & 6.48 & 6.52 & 6.43 & 6.43 & 6.52 & 6.51 & 6.54 & 6.52 & 6.41 & 6.58 & 6.48 & 6.51\,$\pm$\,0.04 \\
Al\,{\sc i} & IR & 6.60 & 6.75 & 6.60 & 6.64 & 6.57 & 6.58 & 6.65 & 6.64 & 6.66 & 6.63 & 6.57 & 6.57 & 6.59 & 6.61\,$\pm$\,0.04 \\
Al\,{\sc i} & OPT & 6.38 & 6.52 & 6.45 & 6.45 & 6.34 & 6.49 & 6.42 & 6.46 & 6.40 & 6.46 & 6.45 & 6.44 & 6.33 & 6.42\,$\pm$\,0.05 \\
Si\,{\sc i} & IR & 7.56 & 7.44 & 7.59 & 7.66 & 7.54 & 7.57 & 7.68 & 7.56 & 7.69 & 7.70 & 7.65 & 7.60 & 7.62 & 7.62\,$\pm$\,0.07 \\
Si\,{\sc i} & OPT & 7.56 & 7.53 & 7.53 & 7.56 & 7.44 & 7.51 & 7.55 & 7.60 & 7.57 & 7.54 & 7.52 & 7.53 & 7.48 & 7.53\,$\pm$\,0.05 \\ 
P\,{\sc i} & IR & -- & -- & 5.44 & 5.49 & -- & -- & 5.26 & -- & 5.58 & -- & 5.71 & 5.45 & -- & 5.49\,$\pm$\,0.15 \\ 
S\,{\sc i} & IR & 7.11 & 7.06 & 7.11 & 7.15 & 7.15 & 7.05 & 7.14 & 7.17 & 7.20 & 7.21 & 7.18 & 7.19 & 7.16 & 7.17\,$\pm$\,0.03 \\
S\,{\sc i} & OPT & 7.33 & 7.12 & 7.15 & 7.27 & 7.10 & 7.10 & 7.22 & 7.37 & 7.19 & 7.22 & 7.32 & 7.47 & 7.17 & 7.25\,$\pm$\,0.11 \\ 
K\,{\sc i} & IR & 4.98 & 4.81 & 5.04 & 5.13 & 5.08 & 5.10 & 5.07 & 5.08 & 5.15 & 5.03 & 5.14 & 5.09 & 5.15 & 5.10\,$\pm$\,0.04 \\ 
Ca\,{\sc i} & IR & 6.29 & 6.42 & 6.41 & 6.41 & 6.36 & 6.32 & 6.39 & 6.38 & 6.44 & 6.41 & 6.44 & 6.36 & 6.39 & 6.40\,$\pm$\,0.03 \\
Ca\,{\sc i} & OPT & 6.22 & 6.38 & 6.37 & 6.32 & 6.22 & 6.25 & 6.32 & 6.24 & 6.30 & 6.24 & 6.32 & 6.15 & 6.25 & 6.29\,$\pm$\,0.05 \\ 
Sc\,{\sc i}  & IR & 2.80 & 2.95 & 2.95 & 3.01 & -- & 2.88 & 3.03 & 3.02 & 3.19 & 3.12 & 3.05 & 3.02 & 3.10 & 3.05\,$\pm$\,0.07 \\ 
Sc\,{\sc ii}  & OPT & 3.02 & 3.03 & 3.13 & 3.20 & 2.93 & 3.06 & 3.11 & 3.11 & 3.28 & 3.11 & 3.15 & 3.16 & 3.23 & 3.14\,$\pm$\,0.09 \\ 
Ti\,{\sc i}  & IR & 4.85 & 4.80 & 4.82 & 4.72 & 4.68 & 4.67 & 4.75 & 4.96 & 4.87 & 4.78 & 4.89 & 4.93 & 4.91 & 4.83\,$\pm$\,0.10 \\
Ti\,{\sc i}  & OPT & 4.60 & 4.91 & 4.87 & 4.77 & 4.74 & 4.76 & 4.78 & 4.73 & 4.86 & 4.76 & 4.79 & 4.73 & 4.79 & 4.78\,$\pm$\,0.05 \\ 
Cr\,{\sc i}  & IR & 5.39 & 5.49 & 5.49 & 5.53 & 5.41 & 5.43 & 5.49 & 5.44 & 5.52 & 5.51 & 5.55 & 5.50 & 5.55 & 5.50\,$\pm$\,0.05 \\ 
Cr\,{\sc i}  & OPT & 5.40 & 5.56 & 5.56 & 5.54 & 5.50 & 5.50 & 5.60 & 5.51 & 5.56 & 5.53 & 5.55 & 5.41 & 5.43 & 5.52\,$\pm$\,0.06 \\ 
Fe\,{\sc i} & IR & 7.38 & 7.39 & 7.44 & 7.51 & 7.41 & 7.45 & 7.50 & 7.41 & 7.52 & 7.47 & 7.41 & 7.47 & 7.51 & 7.44\,$\pm$\,0.05 \\
Fe\,{\sc i} & OPT & 7.38 & 7.44 & 7.50 & 7.49 & 7.41 & 7.38 & 7.47 & 7.40 & 7.50 & 7.43 & 7.41 & 7.35 & 7.48 & 7.44\,$\pm$\,0.05 \\ 
Fe\,{\sc ii} & OPT & 7.39 & 6.20 & 7.50 & 7.49 & 7.41 & 7.38 & 7.47 & 7.40 & 7.51 & 7.43 & 7.39 & 7.35 & 7.48 & 7.44\,$\pm$\,0.05 \\ 
Ni\,{\sc i}  & IR & 6.19 & 6.18 & 6.24 & 6.22 & 6.17 & 6.12 & 6.17 & 6.21 & 6.26 & 6.26 & 6.21 & 6.23 & 6.23 & 6.22\,$\pm$\,0.03 \\
Ni\,{\sc i}  & OPT & 6.24 & 6.20 & 6.20 & 6.20 & 6.07 & 6.11 & 6.16 & 6.32 & 6.24 & 6.14 & 6.19 & 6.12 & 6.16 & 6.18\,$\pm$\,0.07 \\ 
Y\,{\sc i}  & IR & 1.81 & 2.82 & 2.29 & 2.16 & -- & 3.17 & 2.56 & 1.96 & 2.24 & -- & 2.12 & 2.13 & -- & 2.21\,$\pm$\,0.19 \\
Y\,{\sc ii}  & OPT & 1.80 & 2.67 & 2.24 & 2.18 & 2.02 & 3.21 & 2.39 & 2.10 & 2.15 & 2.14 & 2.25 & 2.10 & 2.00 & 2.16\,$\pm$\,0.12 \\ 
Ce\,{\sc ii}  & IR & 1.54 & 2.06 & 1.80 & 1.79 & 1.72 & 2.15 & 1.83 & 1.72 & 1.88 & 1.77 & 1.69 & 1.75 & 1.84 & 1.78\,$\pm$\,0.06 \\
Ce\,{\sc ii}  & OPT & 1.59 & 2.15 & 1.69 & 1.77 & 1.63 & 2.14 & 1.84 & 1.68 & 1.85 & 1.69 & 1.66 & 1.64 & 1.83 & 1.73\,$\pm$\,0.09 \\ 
Nd\,{\sc ii}  & IR & 1.47 & 2.15 & 1.52 & 1.59 & 1.62 & 2.02 & 1.77 & 1.65 & 1.54 & 1.58 & 1.46 & 1.56 & 1.77 & 1.61\,$\pm$\,0.10 \\
Nd\,{\sc ii}  & OPT & 1.61 & 2.27 & 1.73 & 1.69 & 1.64 & 1.94 & 1.75 & 1.75 & 1.69 & 1.61 & 1.63 & 1.58 & 1.60 & 1.67\,$\pm$\,0.06 \\ 
Yb\,{\sc ii}  & IR & 0.84 & 1.06 & 0.97 & 0.93 & 0.92 & 1.26 & 0.92 & 0.83 & 0.91 & 0.89 & 0.89 & 0.82 & 0.88 & 0.90\,$\pm$\,0.05 \\
Pb\,{\sc i}  & OPT             & -- & 1.95 & 1.75 & 1.60 & 1.45 & 2.00 & 1.75 & 1.25 & 1.85 & 1.75 & 1.50 & -- & 1.95 & 1.65\,$\pm$\,0.22 \\
Pb\,{\sc i}$_{\rm NLTE}$ & OPT & -- & 2.12 & 1.92 & 1.77 & 1.62 & 2.18 & 1.93 & 1.43 & 2.03 & 1.92 & 1.70 & -- & 2.12 & 1.83\,$\pm$\,0.22 \\
\hline
\end{tabular}
\end{table*}

\begin{table*}
\centering
\caption{Isotopic ratios derived in this study. OC Mean indicates the mean cluster value, excluding the stars \#001, \#002, and \#201.}
\label{tab:isotopic}
\begin{tabular}{l c c c c c c c c c c c c c c}
\hline
Isotopic ratio & \#001 & \#002 & \#006 & \#008 & \#102 & \#201 & \#224 & \#240 & \#316 & \#348 & \#375 & \#443 & TYC & OC Mean \\ \hline
$^{12}$C/$^{13}$C\,{\tiny ($^{13}$CO)} & 24 & 12 & 20 & 14 & 12 & 24 & 24 & 25 & 22 & 20 & 21 & 16 & 14 & 19\,$\pm$\,5 \\  
$^{16}$O/$^{17}$O\,{\tiny (C$^{17}$O)} & 300 & -- & -- & -- & -- & -- & -- & 250 & -- & -- & 250 & 150 & -- & 216\,$\pm$\,58 \\ 
$^{16}$O/$^{18}$O\,{\tiny (C$^{18}$O)} & 400 & -- & -- & -- & -- & -- & -- & 600 & -- & 800 & -- & 400 & -- & 600\,$\pm$\,200 \\ \hline
\end{tabular}
\end{table*}

\subsubsection{Abundance uncertainties}
\label{sec:uncert}

We provide an analysis of the uncertainties in determining the chemical abundances for the star \#443, which serves as a representative example for our sample. This analysis follows the methodology described in \citet{holanda2020}; in summary, the uncertainty in $T_{\rm eff}$ is derived from the error in the slope of the [Fe\,{\sc i}/H] versus $\chi$ relation, while the uncertainty in $\xi_{\rm t}$ is estimated from the error in the slope of [Fe\,{\sc i}/H] versus $\log\,{\rm EW}/\lambda$. The uncertainty in surface gravity is determined iteratively by adjusting $\log\,g$ until the difference between the mean abundances of [Fe\,{\sc i}/H] and [Fe\,{\sc ii}/H] equals the standard deviation of [Fe\,{\sc i}/H]. 

\par These uncertainties in the atmospheric parameters are subsequently propagated to assess the overall uncertainties in the chemical abundances, which are listed in Table\,\ref{tab:ab_uncert}. The total abundance uncertainty, $\sigma_{\rm atm}$, is computed as the square root of the sum of the squared uncertainties associated with the atmospheric parameters, i.e.:

\begin{multline}
\sigma_{\rm atm}^{2} = \sigma_{T_{\text{eff}}, \log\,\varepsilon({\rm X})}^{2} + \sigma_{\log\,g,  \log\,\varepsilon({\rm X})}^{2} + \\
\sigma_{\Delta \log\,\varepsilon \text{(Fe)}, \log\,\varepsilon({\rm X})}^{2} + \sigma_{\xi_{t}, \log\,\varepsilon({\rm X})}^{2}.
\end{multline}

\par Table\,\ref{tab:ab_uncert_cno} presents the variations in the abundances of carbon, nitrogen, and oxygen. Additionally, the three columns in the table illustrate how an increase of 0.20\,dex in the CNO abundances affects the results, due to the interdependence of these species. The total uncertainty in the CNO abundances, $\sigma_{\rm CNO}$, is calculated as the square root of the sum of the squared uncertainties in the abundances:

\begin{equation}
\sigma_{\rm CNO}^{2}\,=\,\sigma_{\rm \Delta log\,\varepsilon(C)}^{2}\,+ \,\sigma_{\rm \Delta log\,\varepsilon(N)}^{2} + \sigma_{\rm \Delta log\,\varepsilon(O)}^{2}.
\end{equation}

For the uncertainties in the isotopic ratios, we considered the variations caused by a $+\,0.20$\,dex increase in the abundances of the CNO species. For instance, an increase in nitrogen abundance by $+\,0.20$\,dex led to a variation of $-\,0.10$\,dex in oxygen abundance. This variation in oxygen abundance was then propagated to estimate its impact on the isotopic ratios of $^{16}$O/$^{17, 18}$O.

\begin{table}
\centering
\caption{Abundance uncertainties for the star \#443. From the second to the fifth column, we show the variations in abundances caused by the uncertainties in the atmospheric parameters. In the last column, we present the total uncertainty.}
\label{tab:ab_uncert}
\begin{tabular}{l c c c c c}
\hline
$\log\,\varepsilon$(X) & $\Delta T_{\rm eff}$ & $\Delta \log\,g$ & $\Delta \xi_{t}$ & $\Delta \log\,\varepsilon$(Fe) & $\sigma_{\rm atm}$ \\ 
  & $+80$\,K & $+0.20$\,dex & $+0.08$\,km/s & $+0.10$\,dex & \\ \hline
Li\,{\sc i} & $+$0.13 & $-$0.08 & $-$0.10 & 0.00 & 0.18 \\
F\,{\tiny (HF)} & $+$0.12 & $+$0.02 & $+$0.01 & $+$0.04 & 0.13 \\ 
Na\,{\sc i} & $+$0.06 & $-$0.01 & $-$0.03 & $-$0.03 & 0.07 \\
Mg\,{\sc i} & $+$0.02 & $-$0.01 & $-$0.02 & $-$0.02 & 0.04 \\ 
Al\,{\sc i} & $+$0.04 & $-$0.02 & $-$0.02 & $-$0.02 & 0.05 \\
Si\,{\sc i} & $-$0.04 & $+$0.02 & $-$0.03 & $-$0.02 & 0.06 \\
P\,{\sc i} & $+$0.02 & $+$0.05 & $-$0.03 & 0.00 & 0.06 \\
S\,{\sc i} & $-$0.09 & $-$0.02 & $-$0.02 & $-$0.01 & 0.09 \\
K\,{\sc i} & $+$0.05 & $+$0.01 & $-$0.07 & $+$0.04 & 0.10 \\
Ca\,{\sc i} & $+$0.07 & $-$0.02 & $-$0.05 & $-$0.05 & 0.10 \\
Sc\,{\sc ii} & $-$0.01 & $+$0.09 & $-$0.01 & 0.00 & 0.09 \\
Ti\,{\sc i} & $+$0.11 & $+$0.01 & $-$0.03 & $-$0.03 & 0.12 \\
Cr\,{\sc i} & $+$0.07 & 0.00 & $-$0.02 & $-$0.02 & 0.08 \\
Fe\,{\sc i} & $+$0.03 & $+$0.01 & $-$0.04 & $+$0.01 & 0.05 \\
Ni\,{\sc i} & $+$0.03 & $+$0.03 & $-$0.05 & $-$0.04 & 0.08 \\
Y\,{\sc ii} & $-$0.01 & $+$0.06 & $-$0.09 & $-$0.09 & 0.14 \\ 
Ce\,{\sc ii} & 0.00 & $+$0.07 & $-$0.02 & $-$0.02 & 0.08 \\ 
Nd\,{\sc ii} & $+$0.02 & $+$0.07 & $-$0.04 & $-$0.04 & 0.09 \\ 
Yb\,{\sc ii} & $+$0.04 & $+$0.11 & 0.00 & 0.00 & 0.12 \\  
Pb\,{\sc i} & $+$0.05 & $-$0.05 & $-$0.05 & 0.00 & 0.09 \\ \hline
\end{tabular}
\end{table}

\begin{table*}
\centering
\caption{Influence of the errors in atmospheric parameters on the abundances of carbon, nitrogen, and oxygen and isotopic ratios for NGC\,5822\,-\,443. We also show the dependence of the CNO abundance uncertainty on each light element.}
\label{tab:ab_uncert_cno}
\begin{tabular}{l c c c c c c c c c}
\hline
$\log\,\varepsilon$(X) & $\Delta T_{\rm eff}$ & $\Delta \log\,g$ & $\Delta \xi_{t}$ & $\Delta \log\,\varepsilon$(Fe) & $\Delta \log\,\varepsilon$(C) & $\Delta \log\,\varepsilon$(N) & $\Delta \log\,\varepsilon$(O) & $\sigma_{\rm atm}$ & $\sigma_{\rm CNO}$ \\ 
  & $+80$\,K & $+0.20$\,dex & $+0.08$\,dex & $+0.10$\,dex & $+0.20$\,dex & $+0.20$\,dex & $+0.20$\,dex &  &  \\ \hline
C\,{\tiny ($^{12}$C$^{16}$O)} & $+$0.06 & $+$0.04 & $-$0.01 & $-$0.03 & --- & 0.00 & $-$0.06 & 0.08 & 0.06 \\ 
N\,{\tiny ($^{12}$CN)} & $+$0.04 & $+$0.06 & $+$0.00 & $+$0.02 & $-$0.27 & --- & $+$0.15 & 0.07 & 0.31 \\
O\,{\tiny ($^{16}$OH)} & $+$0.08 & $+$0.05 & $+$0.02 & $+$0.05 & $-$0.06 & $-$0.10 & --- & 0.11 & 0.12 \\ 
$^{12}$C/$^{13}$C & $+$1 & $-$2 & $+$2 & $+$0 & --- & 0 & 2 & 3 & 2 \\
$^{16}$O/$^{17}$O & $-$40 & $-$50 & $+$10 & $-$50 & $-$10 & $+$20 & --- & 82 & 22 \\
$^{16}$O/$^{18}$O & $-$100 & $+$100 & $-$100 & 0 & $-$100 & $-$100 & --- & 173 & 141 \\
\hline
\end{tabular}
\end{table*}

\begin{figure}
    \centering
    \includegraphics[width=\linewidth]{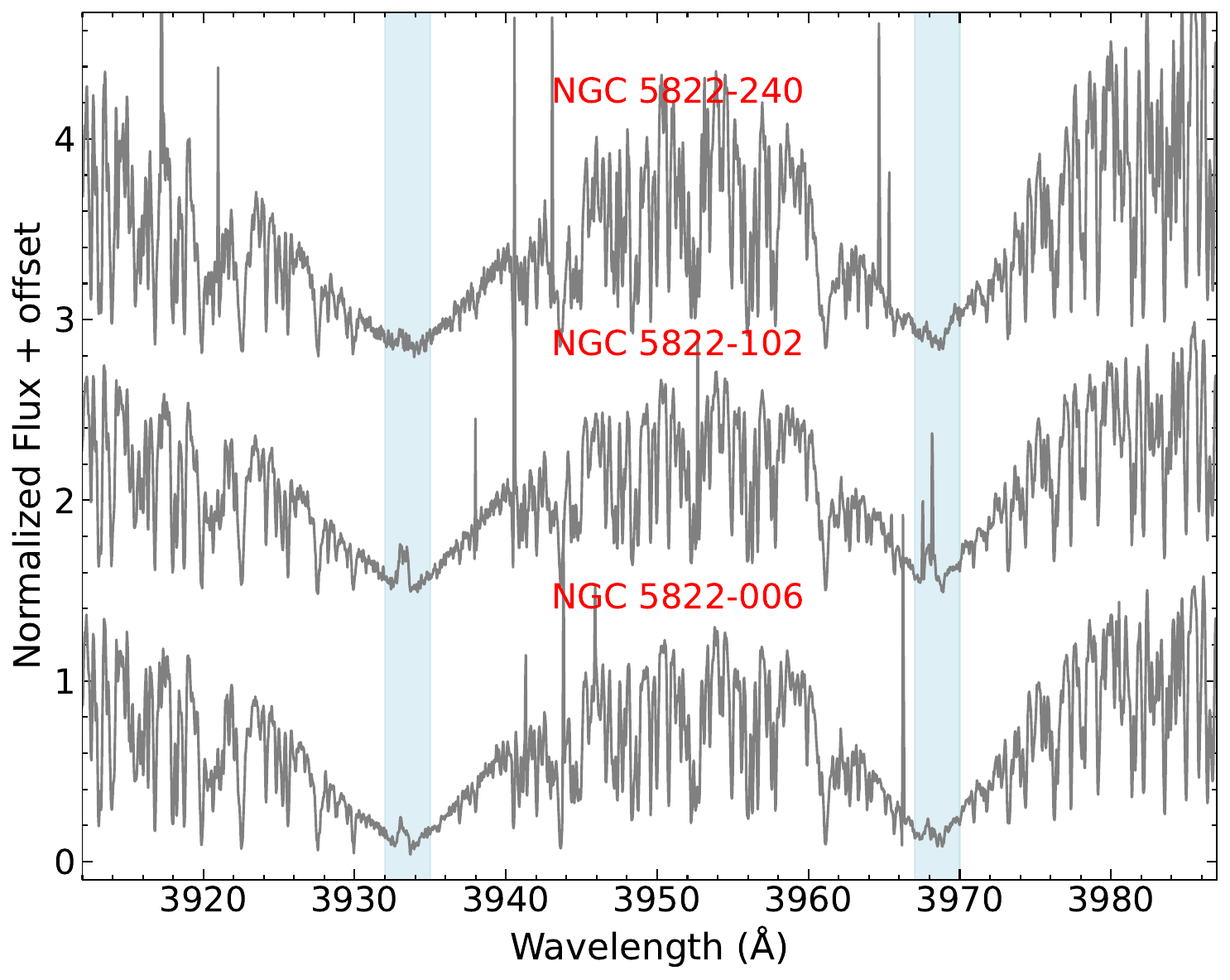}
    \caption{Normalized Ca\,{\sc ii} H\,\&\,K spectra of the three Li-enriched giants in NGC\,5822.}
    \label{fig:hk_activity}
\end{figure}

\begin{figure}
    \centering
    \includegraphics[width=1\linewidth]{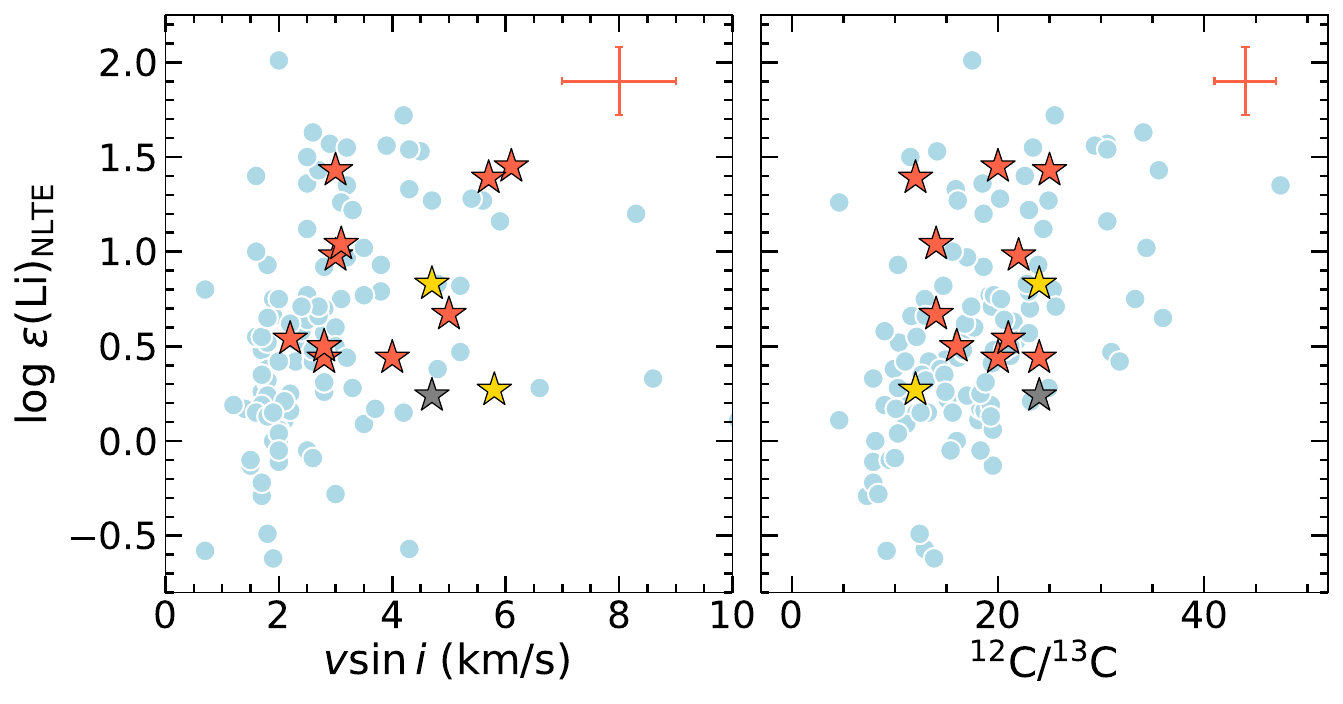}
    \caption{Lithium abundance as a function of projected rotational velocity and carbon isotopic ratio for giant stars. Open cluster members are shown in red, while stars with low membership probability are represented by yellow (\#002 and \#201; Ba stars) and gray (\#001) symbols. G-K type giant stars analyzed by \citet{takeda2019} and \citet{takeda2017} are shown in light blue.}
    \label{fig:lithium}
\end{figure}

\begin{figure*}
    \centering
    \includegraphics[width=1\linewidth]{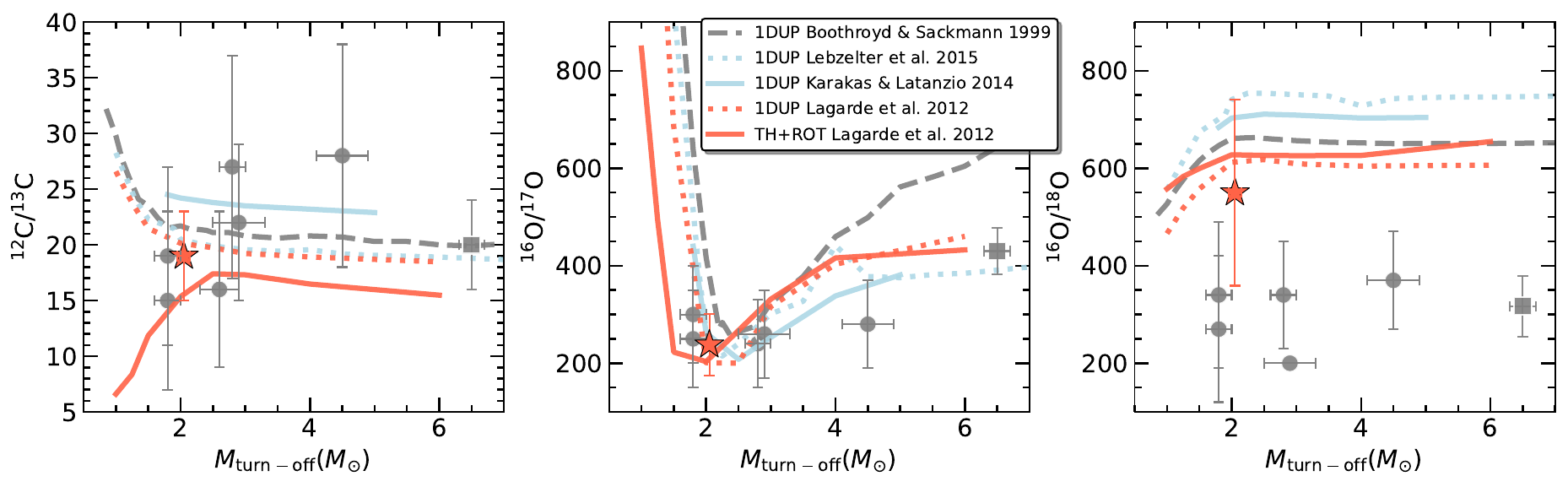}
    \caption{Carbon and oxygen isotopic ratios in the cluster NGC\,5822 compared with mixing models and observational data from the literature. Predictions after the first dredge-up (1DUP) are shown from: \citet[][blue dotted lines]{lebzelder2015}, \citet[][blue solid lines]{karakas2014}, \citet[][red dotted lines]{lagarde2012}, and \citet[][gray dashed lines]{boothroyd1999}. The thermohaline and rotation (TH+ROT) mixing model by \citet[][red solid lines]{lagarde2012} is also presented. Grey circles represent stars analyzed by \citet{lebzelder2015}, and the grey square represents the open cluster NGC\,2345 studied by \citet{holanda2024a}. The mean isotopic ratios and turn-off mass determined for NGC\,5822 are represented by a red symbol.}
    \label{fig:isotopic}
\end{figure*}

\section{Discussion}\label{sec:discuss}

\par In Figure\,\ref{fig:ab_literature}, we compare the stars analyzed in NGC\,5822's field with literature samples from different contexts. Stars with high astrometric membership probability are shown in red, while those with no probability are represented by yellow symbols. Light blue circles correspond to literature samples, including clump giant stars from \citet[][C, N, O, Na, Mg]{mishenina2006}, cool giant stars from \citet[][F]{ryde2020}, K giant stars in the solar neighborhood from \citet[][P]{nandakumar2022}, K giants from \citet[][Yb]{montelius2022}, and giants ($\log\,g\,\leq\,3.0$) from \citet[][Pb]{contursi2024}. We also include dwarf stars analyzed by \citet[][S]{lucertini2022}, \citet[][Al, Si, Ca, Ti, Cr, Ni]{bensby2014}, \citet[][Sc]{battistini2015}, and \citet[][Y, Ce, Nd]{battistini2016}, selected with Age\,$<$\,7.0\,Gyr and TD/D\,$<$\,0.5. The gray dashed lines indicate solar abundance values.

\par Figure\,\ref{fig:ab_general} compares the chemical abundances derived from optical (light blue) and NIR (red) spectra for the giant stars of NGC\,5822. For the elements for which \texttt{NLTE} corrections are available in the literature (Li, Na, Mg, Al, and Pb), \texttt{NLTE} values are indicated as triangles. Overall, optical and NIR measurements are consistent within the uncertainties for most species, with only a few elements showing differences between the two spectral regions. When determining the chemical pattern of the cluster, we adopt the \texttt{NLTE} abundances whenever available and use the mean values between the optical and NIR measurements to minimize possible wavelength-dependent biases and to increase the number of usable lines for each species.

\par Concerning differences within the same chemical species, the Al abundances derived from optical lines and from NIR lines in \texttt{LTE} show a significant difference. This occurs because lines in the NIR are more sensitive to \texttt{NLTE} effects and tend to overestimate the abundance in \texttt{LTE}. After applying the \texttt{NLTE} corrections to the NIR lines \citep{nordlander2017}, the resulting values are much closer to the optical results, indicating that the discrepancy was largely driven by \texttt{NLTE} effects. For Ca, the mean abundance from optical lines differs by 0.11 dex from the NIR values; the NIR Ca lines have higher excitation potentials and may probe different atmospheric layers, and they may also be more sensitive to \texttt{NLTE} departures, uncertainties in the $\log\,gf$ values, or continuum determination. Any combination of these effects could shift the NIR abundances relative to the optical ones.

\subsection{Light elements}

\par Light elements are excellent tracers of internal mixing events and their chemical abundances in red giant stars provide important insights into stellar evolutionary processes. Following the first dredge-up (1DUP) during the ascent of the red giant branch (RGB), a decrease in the $^{12}$C abundance is typically observed, accompanied by increases in $^{13}$C, N, and Na. Regarding oxygen isotopes, $^{16}$O remains almost unaffected, while $^{17}$O increases due to the sequence of reactions $^{16}$O(p, $\gamma$)$^{17}$F($\beta^{+}$)$^{17}$O, until equilibrium is reached via $^{17}$O(p, $\alpha$)$^{14}$N. In contrast, $^{18}$O decreases through the $^{14}$N($\alpha$, $\gamma$)$^{18}$F($\beta^{+}$)$^{18}$O reaction chain. Additionally, lithium abundance drops significantly, as $^{7}$Li becomes diluted in the deep convective envelope and is rapidly destroyed through proton capture, $^{7}$Li(p, $\alpha$)$^{4}$He, at temperatures exceeding $2.5\,\times\,10^{6}$\,K.

\par The stars in our sample show good agreement with N and O abundance values from the literature (Figure\,\ref{fig:ab_literature}). Regarding carbon abundance, a slight enrichment in $^{12}$C is noticeable in the two Ba-stars, \#002 and \#201. As mentioned before, this peculiar class of stars forms in binary systems and has undergone polluted from an asymptotic giant branch (AGB) companion, which can also be enriched in $^{12}$C \citep{karakas2014,lugaro2023}.

\subsubsection{The Li-enriched giants}

\par So-called ``Li-rich giant stars'', commonly defined as having lithium abundances of $\log\,\varepsilon(\mathrm{Li})\,\geq\,1.5$\,dex, comprise approximately 1\% of all G-K giant stars \citep[see, e.g.,][]{brown1989,kumar2011,holanda2020b,holanda2020,holanda2023,holanda2024b}. In our sample, three stars (\#006, \#102, and \#240) exhibit slight Li enrichment, with $\log\,\varepsilon(\mathrm{Li}) \gtrsim\,1.30$\,dex. Although these values fall just below the classical threshold, we classify them as Li-enriched given their marginally elevated abundances and typical measurement uncertainties ($\sigma_{\log\,\varepsilon(\mathrm{Li})} \sim 0.20$\,dex).

\par Beyond lithium abundance, we investigated other potential anomalies in stellar activity, isotopic ratios and projected rotational velocities. As stars evolve off the main sequence, their magnetic activity decays; thus, no significant activity signatures are generally expected \citep[e.g.][]{wilson1964,skumanich1972}. However, two Li-enriched giants in our sample (\#006 and \#102) display Ca\,{\sc ii} H\,\&\,K emission (Figure\,\ref{fig:hk_activity}), a classical indicator of chromospheric activity. These stars also show the highest $v\,\sin\,i$ values in the sample, 6.1 and 5.7\,km\,s$^{-1}$. Rapid rotation combined with chromospheric activity in evolved stars is often associated with binarity, where a companion can induce spin-up and internal mixing \citep{fekel2002,morel2004,fekel2005}. However, no known companion explains the observed activity.

\par Previous studies have explored possible correlations between lithium abundance and chromospheric activity \citep[e.g.,][]{goncalvez2020,xing2021,sneden2022}, as well as with rotation \citep[e.g.][]{rebolo1988,chaboyer1998,drake2002,magrini2021,flaulhabe2025}. Some of these works suggest that stronger chromospheric flux and faster rotation may favor higher Li abundances. In our case, however, the measured $v\,\sin i$ values remain below the $\sim$8-10\,km\,s$^{-1}$ threshold traditionally used to classify rapid rotators in this context \citep{drake2002}.

\par Recently, \citet{rolo2024} investigated the relation between lithium enrichment, stellar activity, and radial velocity (RV) variability in OC giants, including NGC\,5822. Their Table\,4 shows moderate correlations between RV and activity indicators (BIS, FWHM, and H$\alpha$) for the star \#102, together with a relatively high projected rotational velocity. However, these correlations are based on a limited number of observations and are not supported by common periodicities in the RV and activity time series. On the other hand, the star \#240 does not show significant RV-activity correlations. Taken together, these results do not provide sufficient evidence to establish a link between stellar activity and Li-enrichment in these stars.

\par Figure\,\ref{fig:lithium} shows lithium abundances, $v\,\sin\,i$, and $^{12}$C/$^{13}$C for a sample of G-K giants from \citet{takeda2017} and \citet{takeda2019}, compared with the thirteen giants analyzed here. The Li-enriched giants are broadly scattered in Figure\,\ref{fig:lithium} (left panel), showing no consistent trend linking higher Li abundance to higher $v\,\sin\,i$ values. Some Li-rich giants exhibit both low and moderate $v\,\sin\,i$, indicating that rotation alone cannot explain Li enrichment.

\par The surface carbon isotopic ratio $^{12}$C/$^{13}$C provides a key diagnostic of extra mixing. However, lithium abundance does not correlate systematically with $^{12}$C/$^{13}$C (Figure\,\ref{fig:lithium}, right panel). Stars with elevated Li abundances appear across the full isotopic range, implying that deep mixing, typically associated with lower ratios, is not the unique mechanism responsible for Li enhancement. A notable case is the star \#102, which exhibits a low $^{12}$C/$^{13}$C ratio of 12, making it a candidate for having experienced additional mixing. Altogether, these results reinforce the view that Li-rich giants constitute a heterogeneous class of chemically peculiar stars, likely shaped by multiple or combined enrichment mechanisms.

\begin{table}
\centering
\caption{Comparison of projected rotational velocities ($v\sin i$) and lithium abundances between this work, \citet{delgadomena2016} (DM16), and \citet{tsantaki2023} (TS23) for four stars in NGC\,5822.}
\label{tab:delgado}
\begin{tabular}{l c c c c c}
\hline
Star & \multicolumn{2}{c}{$v\sin\,i$\,(km\,s$^{-1}$)} & \multicolumn{3}{c}{$\log\varepsilon(\text{Li})_{\rm NLTE}$}\\
 & This Work & DM16 & This Work & DM16 & TS23 \\
\hline
NGC\,5822-001 & 4.70 & --- & 0.24 & --- & 0.13 \\ 
NGC\,5822-006 & 6.10 & --- & 1.45 & --- & 1.31 \\ 
NGC\,5822-008 & 5.00 & 3.79 & 0.67 & 0.80 & 0.43 \\ 
NGC\,5822-102 & 5.70 & 5.40 & 1.39 & 1.57 & 1.35 \\ 
NGC\,5822-201 & 4.70 & 4.17 & 0.83 & 1.18 & 1.01 \\ 
NGC\,5822-224 & 4.00 & 3.69 & 0.44 & 0.56 & $-$0.11 \\ 
NGC\,5822-240 & 3.00 & --- & 1.43 & --- & 1.40 \\ 
NGC\,5822-316 & 3.00 & --- & 0.98 & --- & 0.64 \\ 
NGC\,5822-348 & 2.80 & --- & 0.44 & --- & $-$0.22 \\ 
NGC\,5822-375 & 2.20 & --- & 0.54 & --- & $-$0.08 \\ 
NGC\,5822-443 & 2.80 & --- & 0.50 & --- & $-$0.14 \\ 
\hline
\end{tabular}
\end{table}

\par Table\,\ref{tab:delgado} presents a comparison of the $v\,\sin\,i$ and $\log\varepsilon(\text{Li})_{\rm NLTE}$ values obtained in this work with those reported by \citet{delgadomena2016} and \citet{tsantaki2023}. A clear and systematic difference is observed between our measurements and those of \citet{delgadomena2016} for both $v\,\sin\,i$ and Li abundances in the NGC\,5822 members. We highlight three key points in this first comparison: (1) Our $v\sin\,i$ values are consistently higher by $\sim$0.3--1.2\,km\,s$^{-1}$, which may be attributed to differences in the macroturbulence values assumed; (2) Li abundances show a uniform offset of $\sim$0.1--0.3\,dex, with the values from \citet{delgadomena2016} being systematically larger, probably due to discrepancies in the assumed atmospheric parameters by them; and (3) The relative star-to-star trends are preserved (for instance, \#102 maintains the highest rotation and Li abundance in both studies). \citet{delgadomena2016} report a $T_{\rm eff}\,=\,5252$\,K for star \#102, i.e., 221\,K higher than the value found in this work, which may explain the observed difference in Li abundance since this specie is easily affected by effective temperature (see Table\,\ref{tab:ab_uncert}). Nevertheless, the results obtained in our analysis are in excellent agreement with those of \citet{tsantaki2023}, especially for the Li-enriched stars. Larger differences arise for stars with very low Li abundances, which in most cases may be associated with uncertainties in the continuum placement or with the blending of CN and Fe lines near 6707.8\,\AA. Concerning the atmospheric parameters reported by \citet{tsantaki2023}, star \#001 shows the most significant discrepancies, with $\Delta T_{\rm eff}$\,=\,96\,K and $\Delta\log g$\,=\,0.40\,dex, while the remaining stars exhibit differences below 78\,K and 0.17\,dex, which are compatible with typical uncertainties in these parameters.

\par Isochrone fitting (Figure\,\ref{fig:membership}) indicates that the three Li-enriched stars are likely in the red clump phase (core He-burning), the evolutionary stage most frequently associated with this type of peculiar star \citep[][among others]{yan2021}. The most reliable method to determine whether Li-rich giants are on the RGB or in the red clump is asteroseismic data analysis, using the frequency of maximum oscillation power ($\nu_{\mathrm{max}}$) and the period spacing of gravity modes ($\Delta\Pi_1$) \citep[e.g.,][]{bedding2011, mosser2014}. In the absence of such data, particularly for stars in OCs, isochrone fitting provides a reasonable alternative for constraining the evolutionary stage. In this context, \citet{delgadomena2016} suggest that the giant stars analyzed (and with anomalous Li abundance) in NGC\,5822 have undergone less lithium dilution because they are in an earlier phase of 1DUP. However, the authors assumes an age of 0.68\,Gyr for the cluster (they plotted of isochrone 0.63-0.80\,Gyr), based on the literature available, which differs significantly from more recent determinations (Table\,\ref{tab:cluster_par}). On the other hand, \citet{tsantaki2023} adopted the OC parameters from \citet[][0.89\,Gyr]{bossini2019} and suggested that giants \#006 and \#102 could either be located at the base of the RGB or already lie in the red clump. For the star \#240, the authors reported clear lithium enrichment based on its Li absorption line and proposed that it is positioned near the RGB tip. We reinforces the importance of a comprehensive analysis combining membership, accurate determination of cluster parameters, and chemical mixing indicators to better constrain the evolutionary stages of the stars. As said before, all stars analyzed in this work present C and N abundances compatible with post-1DUP chemical pattern.

\par Several scenarios have been proposed to explain the formation of Li-rich/enriched giant stars, but for the red clump stage, two main mechanisms appear relevant to the stars analyzed in this work: mergers involving a helium white dwarf and an RGB star \citep[e.g.,][]{zhang2020} and helium-flash-induced mixing events \citep{schwab2020}. In the merger scenario, the final surface lithium abundance depends on the mass of the helium white dwarf. Models predict that mergers involving white dwarfs with masses between 0.35 and 0.40\,$M_{\odot}$ can produce Li-rich giants, while mergers outside this mass range are more likely to result in the formation of early-R type carbon stars. In the second scenario, the induced mixing is predicted to occur in low-mass stars after He flash and provides a single-star evolutionary pathway to explain Li enhancement in red clump giants, without requiring binary interaction.

\subsubsection{Isotopic ratios}

\par As already mentioned, isotopic ratios in atmospheres of red giant stars are known to be influenced by 1DUP, so they can serve as helpful probes to trace the induced envelope mixing in this kind of object \citep[e.g.,][]{takeda2019}. In Figure\,\ref{fig:isotopic} we compare observed $^{12}$C/$^{13}$C and $^{16}$O/$^{17, 18}$O ratios in the NGC\,5822 with mixing models, focusing on the effect of the 1DUP and extra mixing processes like thermohaline and rotation (TH+ROT) \citep{boothroyd1999,lebzelder2015,karakas2014,lagarde2012}. Additionally, isotopic ratio measurements from \citet{lebzelder2015} and \citet{holanda2024a} are included for comparison as gray circles and squares, respectively. The $^{12}$C/$^{13}$C and $^{16}$O/$^{17}$O ratios reveal a clear mass dependence: low-mass stars ($<$2\,$M_{\odot}$) show significant change in these ratios after the 1DUP, and NGC\,5822's results present a moderate agreement with predictions from both 1DUP and TH+ROT models. The two most evolved objects in the sample, stars \#375 and \#443, lie slightly above the red clump region and may already be evolving toward the early-AGB stage. Their $^{12}$C/$^{13}$C ratios of 21 and 16, respectively, are consistent with expectations for post-1DUP giants that may already exhibit extra mixing. On the other hand, \#240, which has just passed the RGB tip and is now approaching the red clump, is the least evolved star in the sample and shows the highest isotopic ratio, $^{12}$C/$^{13}$C of 25. The remaining stars, all located within the red clump region (see Figure\,\ref{fig:membership}), present a broader range of isotopic ratios, from 12 to 24. Red clump stars may present different mixing histories depending on mass, rotation, and internal angular-momentum transport. In general, the distribution of $^{12}$C/$^{13}$C values across the CMD is also in agreement with the expected evolutionary dependence of carbon isotopic ratios in low-mass giants.

\par A significant discrepancy is observed for the $^{18}$O isotope, where the measured $^{16}$O/$^{18}$O ratio significantly deviate from model predictions, mainly for the stars observed in the literature. As suggested by \citet{lebzelder2015}, this divergence may reflect uncertainties in the initial $^{18}$O abundance or limitations in current nucleosynthetic yields and mixing prescriptions involving this isotope.

\par The TH+ROT models by \citet{lagarde2012} predict more significant isotopic shifts compared to standard 1DUP models, particularly for carbon isotopic ratio, which is aligned with [Na/Fe] ratio (see Section\,\ref{sec:odd_z}), suggesting that additional mixing mechanisms play a critical role in evolution of post-main-sequence stars.

\subsection{Fluorine}

\begin{figure}
    \centering
    \includegraphics[width=\linewidth]{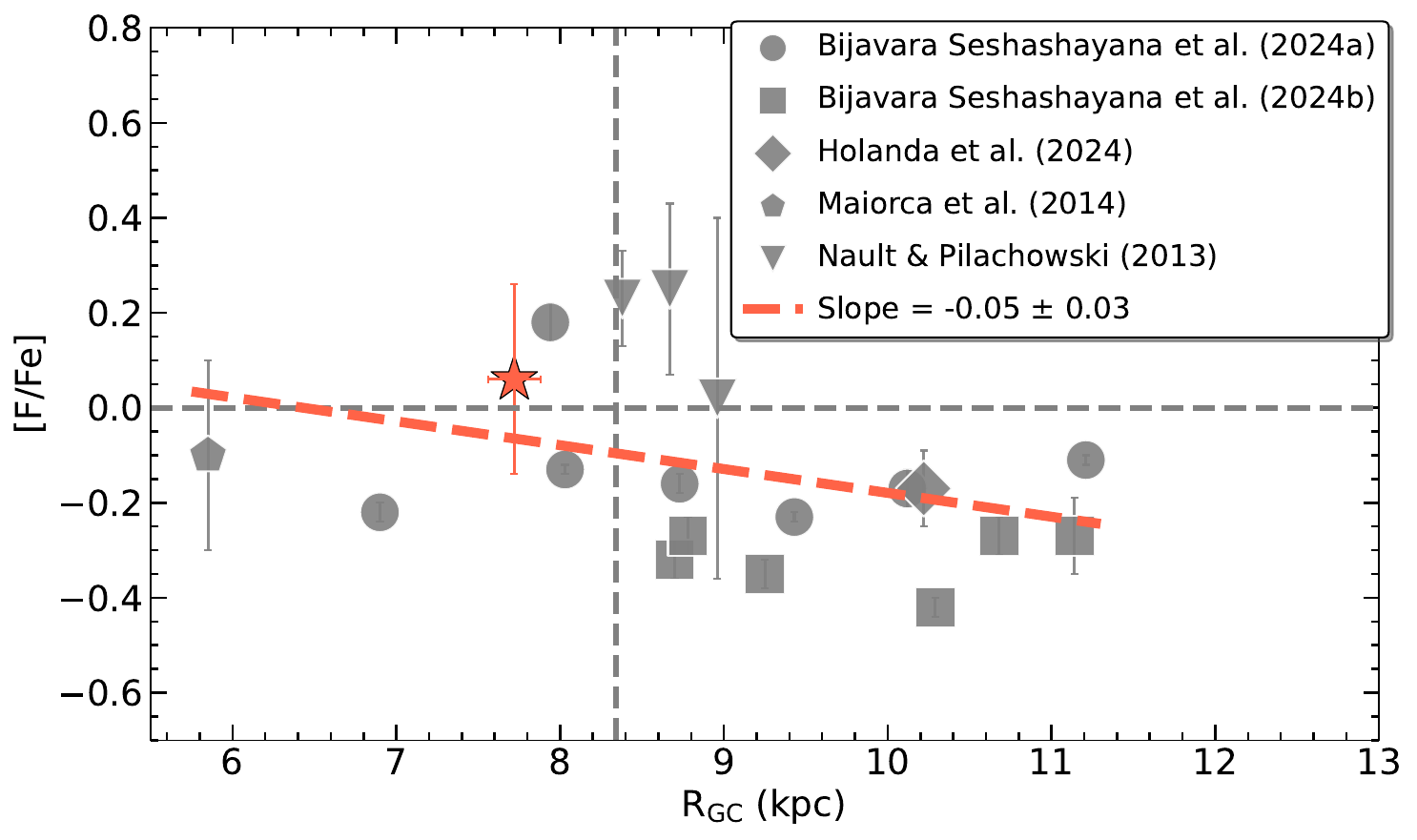}
    \caption{Galactocentric distance ($R_{\rm GC}$) versus [F/Fe] abundance for open clusters analyzed by \citet{bijavara2024a} (circles), \citet{bijavara2024b} (squares), \citet{holanda2024a} (diamond), \citet{maiorca2014} (pentagons), and \citet{nault2013} (triangles). The red dashed line shows the linear regression, while the vertical gray dashed line marks the solar Galactocentric distance \citep[8.34\,kpc;][]{reid2014}. All fluorine abundance values are scaled to the solar reference by \citet{asplund2009}: $\log\,\varepsilon$(F)\,=\,4.56.}
    \label{fig:fluorine}
\end{figure}

\begin{figure}
    \centering
    \includegraphics[width=\linewidth]{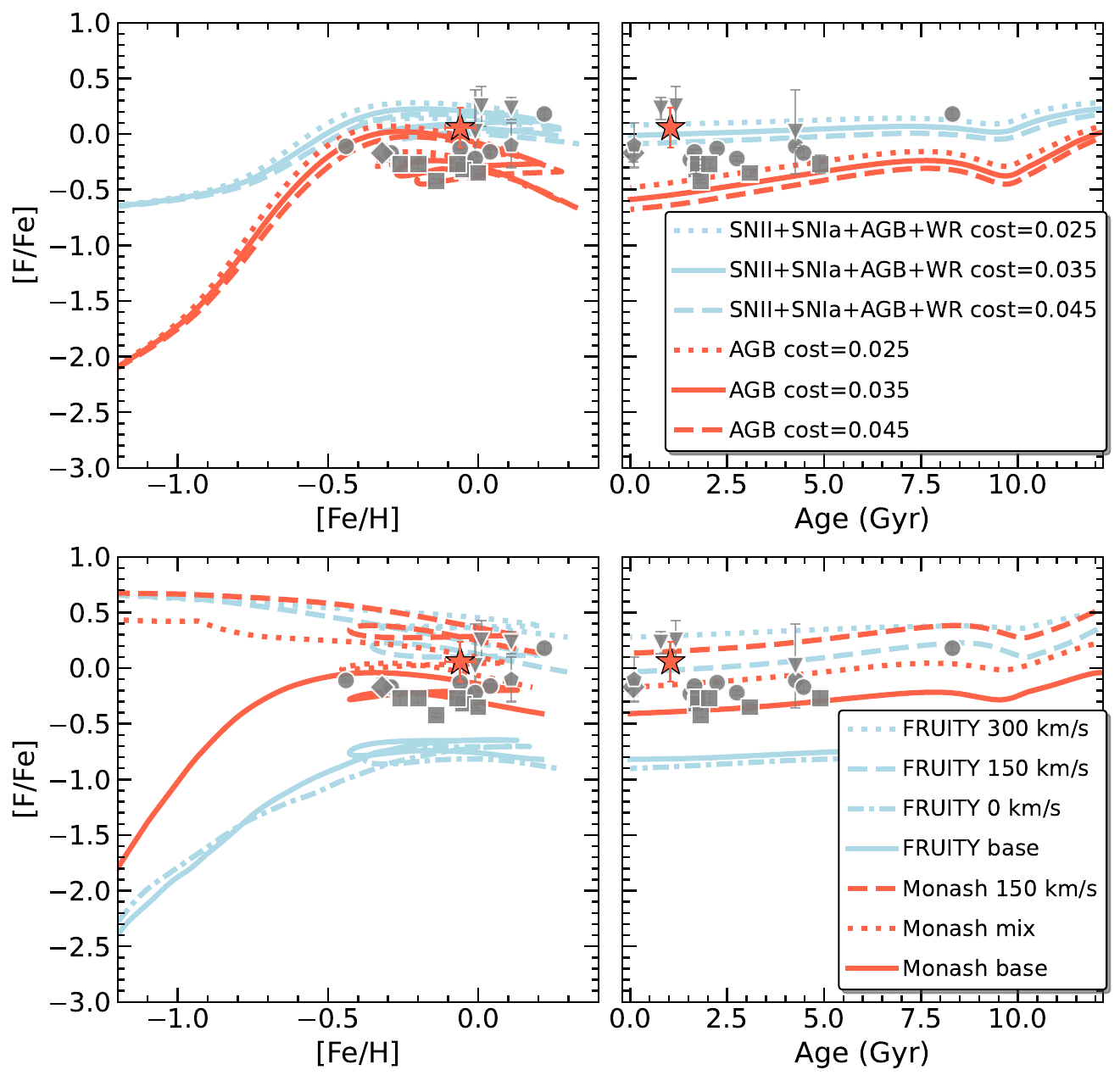}
    \caption{[Fe/H] vs. [F/Fe] and Age vs. [F/Fe] for open clusters. The upper panels compare the observational data with predictions from the Galactic chemical evolution two infall model \citep{spitoni2018}. The lower panels show comparisons with AGB nucleosynthesis models from the \texttt{Monash} and \texttt{FRUITY} grids. The symbols follow the same convention as in Figure\,\ref{fig:fluorine}.}
    \label{fig:nucleosynth}
\end{figure}

In recent years, the number of notable NIR studies of OCs and field stars has been increasing, contributing to important discussions on Galactic fluorine ($^{19}$F) abundances \citep{ryde2020,guerco2022,nandakumar2023,bijavara2024a,bijavara2024b,ryde2025}. \citet{clayton2003} suggested that approximately 15--30\% of the Galactic fluorine abundance originates from He-burning shells in massive stars through the nuclear reaction $^{18}$O(p, $\gamma$)$^{19}$F. Beyond this, multiple production channels for fluorine have been discussed in the literature, including low-mass AGB stars, Wolf-Rayet stars (WR), Type Ia supernovae (SN\,Ia), and core-collapse supernovae (SN\,II), highlighting the complexity of the chemical evolution of fluorine in the Galaxy.

\par In Figure\,\ref{fig:fluorine}, we present the fluorine abundance gradient based on the NGC\,5822 mean result along with OCs from the literature. The comparison includes results for NGC\,6939, NGC\,2420, NGC\,7762, NGC\,7142, Collinder\,110, and Berkeley\,32 from \citet{bijavara2024b}; NGC\,7789, NGC\,7044, NGC\,6819, NGC\,6791, Ruprecht\,171, Trumpler\,5, and King\,11 from \citet{bijavara2024a}; NGC\,2345 from \citet{holanda2024a}; NGC\,6404 from \citet{maiorca2014}; and NGC\,752, M\,67, and Hyades from \citet{nault2013}. For this combined dataset, \citet{bijavara2024b} reported a slope of $-0.09\,\pm\,0.02$\,dex\,kpc$^{-1}$ for the linear fit, which is consistent with our result of $-0.05\,\pm\,0.03$\,dex\,kpc$^{-1}$, within the uncertainties. These findings suggest a gradual decline in fluorine enrichment efficiency with increasing $R_{\rm GC}$, likely reflecting radial variations in stellar populations and star formation history. NGC\,5822 occupies an important position in the inner disk, where few clusters have available fluorine abundance measurements. Its well-determined distance and fluorine abundance help to partially fill the gap for OCs at $R_{\rm GC}\,<\,8$\,kpc, providing constraints for chemical evolution models. Moreover, the significant scatter among clusters at similar $R_{\rm GC}$ (8--9\,kpc) underscores the importance of local astrophysical conditions in influencing fluorine production. An interesting case is presented by \citet{bijavara2024a}: the cluster NGC\,6791, a metal-rich, fluorine-enhanced, and old OC, which is thought to have originated in the inner disk regions and later migrated outward, making it a notable outlier in the fluorine distribution.

\par In Figure\,\ref{fig:nucleosynth}, we compare the observational trends of [Fe/H] and Age versus [F/Fe], taking into account the same OCs from the literature, with predictions from three nucleosynthesis models: the two-infall Galactic chemical evolution model by \citet{spitoni2019}, and stellar yields from the \texttt{Monash} and the FUll-Network Repository of Updated Isotopic Tables \& Yields (\texttt{FRUITY}) AGB nucleosynthesis models by \citet{lugaro2012,fishlock2014,karakas2016,karakas2018} and \citet{cristallo2009,cristallo2011,cristallo2015}, respectively. A detailed description and discussion of these models are provided in \citet{bijavara2024b}. In the top panels, the two infall models incorporating all sources (SN\,II\,+\,SN\,Ia\,+\,AGB\,+\,WR; blue) predict higher [F/Fe] ratios at low [Fe/H] compared to AGB-only models (red), emphasizing the significant role of massive stars and supernovae in early fluorine enrichment. Additionally, given the uncertainties associated with the observed rates of SN\,Ia, different values for the model parameter were tested that governs the fraction of stellar systems capable of producing such supernovae. Following the same approach of \citet{bijavara2024b}, we considered three representative values: 0.025, 0.035, and 0.045. In terms of metallicity content, the NGC\,5822 is in agreement with both models. On the other hand, only the models with all sources can fit NGC\,5822 and other clusters with age less than 1.5\,-\,2.0\,Gyr.

\par In the bottom panels of Figure\,\ref{fig:nucleosynth}, we compare predictions of [F/Fe] as functions of [Fe/H] and Age under different nucleosynthesis scenarios. For WR stars, we adopt the yield set of \citet{limongi2018}, considering three rotational velocities: 0, 150, and 300\,km\,s$^{-1}$. For low- and intermediate-mass stars, we test sets of yields from the \texttt{FRUITY} database and the \texttt{Monash} group. Our results indicate that fluorine yields from the \texttt{Monash} models are excellent to reproduce the observational data. Previous studies \citep{prantzos2018,romano2019} have suggested that in order to reproduce the observed abundance trends of several chemical elements in the Galaxy, massive stars must rotate more rapidly at low metallicities and more slowly at high metallicities. Motivated by this, we tested a mixed-rotation model (``\texttt{Monash} mix''), in which 80\% of low-metallicity massive stars are assumed to be fast rotators, while the remaining 20\% are non-rotating. This model provides a better prediction to NGC\,5822's values and overall fit to the fluorine observations in [Fe/H] vs. [F/Fe] plot. We also note that average stellar rotational velocities are likely lower than 150\,km\,s$^{-1}$, which may help explain the overproduction seen in models with 300\,km\,s$^{-1}$ rotation rates.

\par The panel Age vs. [F/Fe] exhibits a clear dichotomy. The younger open clusters ($<$\,2\,Gyr) have the tendency to present higher values of [F/Fe] in relation to the older ones. This finding is consistent with the growing influence of AGB stars, which have been shown to enrich the interstellar medium in fluorine on timescales of up to several Gyr, depending on mass and metallicity. The fluorine produced in AGB stars is the result of helium burning and subsequent mixing processes, and their contribution becomes more prominent in later Galactic epochs. In contrast, older OCs, formed earlier in the Galaxy's evolution, show systematically lower [F/Fe] values. These clusters are likely to reflect a chemical composition that has been imprinted by earlier generations of stars when fluorine production was primarily governed by massive stars and SN\,II. The negligible contribution of AGB stars to the fluorine budget in these populations lends further support to the time-delay scenario, wherein the accumulation of fluorine from AGB nucleosynthesis is a gradual process.

\subsection{Odd-Z elements}\label{sec:odd_z}

\begin{figure}
    \centering
    \includegraphics[width=\linewidth]{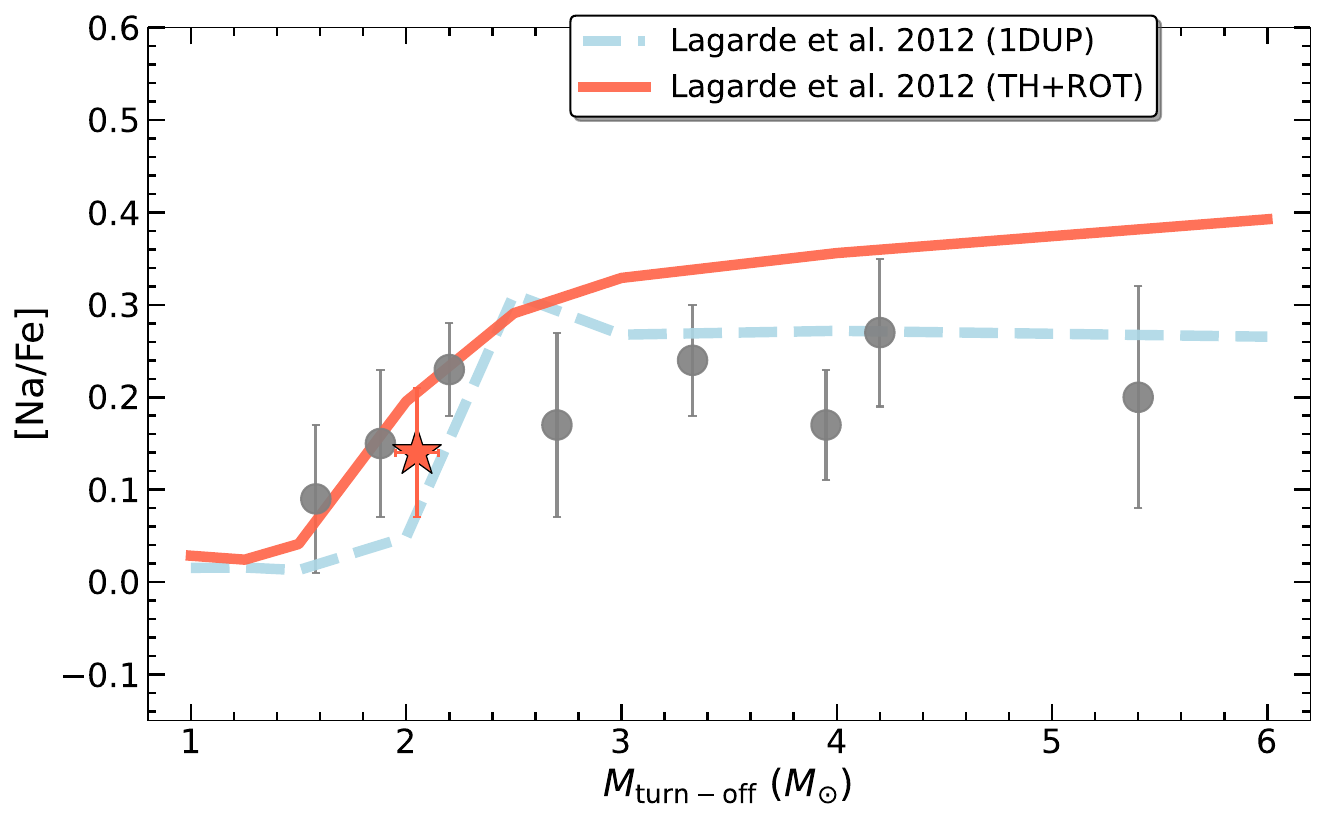}
    \caption{Turn-off mass versus [Na/Fe] ratio in giant stars of open clusters. We compare the mean result for the cluster NGC\,5822 with open clusters previously studied by our team  \citep{santrich2013,penasuarez2018,dasilveira2018,martinez2020,holanda2019,holanda2021,holanda2022}. Blue dashed line represents the predicted abundances of these elements for giants at the first dredge-up, using standard solar metallicity evolutionary models, and red solid line indicates prediction models for thermohaline and rotation-induced mixing from \citet{lagarde2012}.}
    \label{fig:na_mixing}
\end{figure}

\par Sodium ($^{23}$Na) is primarily produced during the carbon-burning phase in massive stars via the reaction $^{12}$C($^{12}$C, p)$^{23}$Na. Although part of the synthesized $^{23}$Na is subsequently destroyed, a significant amount survives and is later ejected during SN\,II, which are the main contributors to Galactic sodium enrichment. Additional contributions arise from hydrogen-burning shells in evolved stars and, to a lesser extent, from the s-process via neutron capture on $^{22}$Ne in AGB stars \citep{clayton2003}. The [Na/Fe] abundances obtained for the stars in NGC\,5822 are in excellent agreement with the values reported by \citet{mishenina2006} for clump giants. Moreover, when compared with theoretical models, the average Na abundance found in our sample aligns well with the predictions from 1DUP and thermohaline plus rotational mixing models (TH+ROT) by \citet{lagarde2012}, as shown in Figure\,\ref{fig:na_mixing}.

\par We did not detect any Na abundance enrichment in the Ba-stars, whose abundance values are consistent with those of typical red clump stars ([Na/Fe]\,$\sim$\,0.05–0.15\,dex). An enhancement in this chemical specie in Ba-stars may be associated with pollution by AGB companions with mass $M\,\geq\,4.0\,M_{\odot}$. This result helps to better constrain the mass of the AGB companions of these stars, as discussed in Section\,\ref{sec:ba_stars}.

\par The nucleosynthesis of aluminum, $^{27}$Al, is also primarily associated with the evolution of massive stars. As a result, aluminum abundances in stellar populations can serve as effective tracers of SN\,II. In Figure\,\ref{fig:ab_literature}, the [Al/Fe] ratios exhibit excellent agreement with those stars of similar metallicity analyzed by \citet{bensby2014}, showing no signs of enrichment and showing only a small dispersion for the stars in NGC\,5822.

\par Phosphorus nucleosynthesis derives almost entirely from neon-burning, which occurs at the end of carbon-burning in massive stars \citep{clayton2003}. In Figure\,\ref{fig:ab_literature}, P abundances are in good agreement with the literature results \citep{nandakumar2022}, though it exhibits a notable dispersion among the cluster stars. This internal dispersion in OC's results may be attributed to the influence of spectral line blending with CO lines in the abundance determination in 16483\,\AA\,(see Figure\,\ref{fig:dev_synth}).

\subsection{Alpha elements}

The $\alpha$-to-iron abundance ratio ($\alpha$/Fe) can serve as an indirect method for estimating stellar ages. This is because $\alpha$-elements such as Mg, Si, S, Ca, and Ti are rapidly produced by SN\,II, whereas iron is synthesized over longer timescales through SN\,Ia \citep{kobayashi2020}. 

\par The Mg abundances derived by \citet{mishenina2006} show good agreement with the values obtained for the analyzed stars in NGC\,5822. Again, the star \#001, a probable non-member of the cluster, displays the largest deviation from the OC's mean Mg abundance, which is $\langle$[Mg/Fe]$\rangle$\,=\,$+0.01$\,dex. When comparing our abundance results with the dwarf stars analyzed by \citet{bensby2014}, we find excellent agreement for Si and Ca. Ti abundances are subsolar and relatively low at a metallicity of [Fe/H]\,$=\,-0.06$\,dex, a trend also observed in the comparison with field stars. Although the differences are small and consistent with the uncertainties, a similar behavior is reported by \citet{ramos2024} and \citet{alonso2020} for the cluster NGC\,6664.

\par Sulfur abundances are compared with those of 30 dwarf stars analyzed by \citet{lucertini2022}. While their results include \texttt{NLTE} corrections, the S\,{\sc i} multiplet at 6757\,\AA\, used in our analysis of NGC\,5822 exhibits negligible departures from \texttt{LTE} \citep[e.g.,][]{takeda2005}, making our \texttt{LTE} results reliable for this comparison. This reliability is further supported by the good agreement observed between the two samples at similar metallicities, as shown in Figure\,\ref{fig:ab_literature}.

\subsection{Iron-peak elements}

\par Approximately half of the iron-peak elements are produced in the evolution of low- and intermediate mass stars in binary systems, such as in SN\,Ia, reflecting their prolonged enrichment timescale. However, these elements can also be synthesized through incomplete or complete silicon-burning during SN\,II \citep[][and references therein]{kobayashi2020}. 

\par All neutral species analyzed in this work exhibit measurable lines in both optical and NIR spectral domains. The abundance results show excellent agreement between the optical and NIR analyses, with the largest discrepancy observed between Sc\,{\sc ii} (optical) and Sc\,{\sc i} (NIR); however, within the uncertainties, there is agreement. When comparing the abundance results for NGC\,5822 with those of dwarf stars from \citet{bensby2014} and \citet{battistini2015}, we find good agreement across all analyzed species. Concerning Cr abundances, all individual values are consistent with those reported for dwarf stars in \citet{bensby2014}, when considering the associated uncertainty for this element.

\subsection{s- and r-process elements}

\par Elements heavier than iron-peak elements (Z\,>\,30) are primarily synthesized through neutron-capture processes, with only a minor contribution from proton-rich nuclei produced via photodisintegration. These processes are classified as either slow (s-process) or rapid (r-process), depending on whether the neutron-capture timescale is longer or shorter than the $\beta$-decay timescale of unstable isotopes \citep[see][]{lugaro2023}. The s-process occurs mainly in low- and intermediate mass AGB stars \citep{burbidge1957,busso1999} and accounts for 71.9\,\%, 83.5\,\%, 57.5\,\%, 40.5\,\%, and 87.2\,\% of the solar system abundances of Y, Ce, Nd, Yb, and Pb, respectively \citep{bisterzo2014}. With the exception of Yb, all heavy elements analyzed in this work are classified as s-process elements. 

\par The normal stars (in terms of s-process species) in our sample exhibit s-process element abundances in close agreement with the Y and Yb abundance values reported in the literature \citep{battistini2016,montelius2022}. The ytterbium results are exclusively obtained from the NIR analysis, while yttrium results are a mean of optical and NIR results. However, Ce and Nd abundances appear slightly enriched, a trend observed in both optical and NIR results, with a small mean dispersion for both species (Table\,\ref{tab:ab_values}). In \citet{bijavara2024a}, the authors investigate a possible connection between fluorine F and Ce abundances, considering factors such as metallicity, age, and Galactocentric distance, given that both elements are produced in AGB stars. In the stellar context, as previously noted, HF lines are not detectable in either of the Ba-stars due to their high effective temperatures. Also, any potential fluorine enrichment in their atmospheres could not be confirmed.

\par Pb abundances present significant dispersion for the OC mean value, likely due to the uncertainty of continuum determination -- Pb\,$\lambda$\,4057\,\AA\, is in a region of many blending for the interval of metallicity and temperature of the sample analyzed here.  This spread is also observed in the literature results by \citet{contursi2024}. Additionaly, \texttt{NLTE} corrections in Pb results are in excellent agreement with the values reported by \citet{contursi2024}, who also provide \texttt{NLTE} Pb results.

\par Star \#001 stands out by exhibiting the lowest abundances of Y, Ce, and Nd in the sample, reinforcing the same trend of deficiency observed in other elements previously discussed.

\begin{figure}
    \centering
    \includegraphics[width=\linewidth]{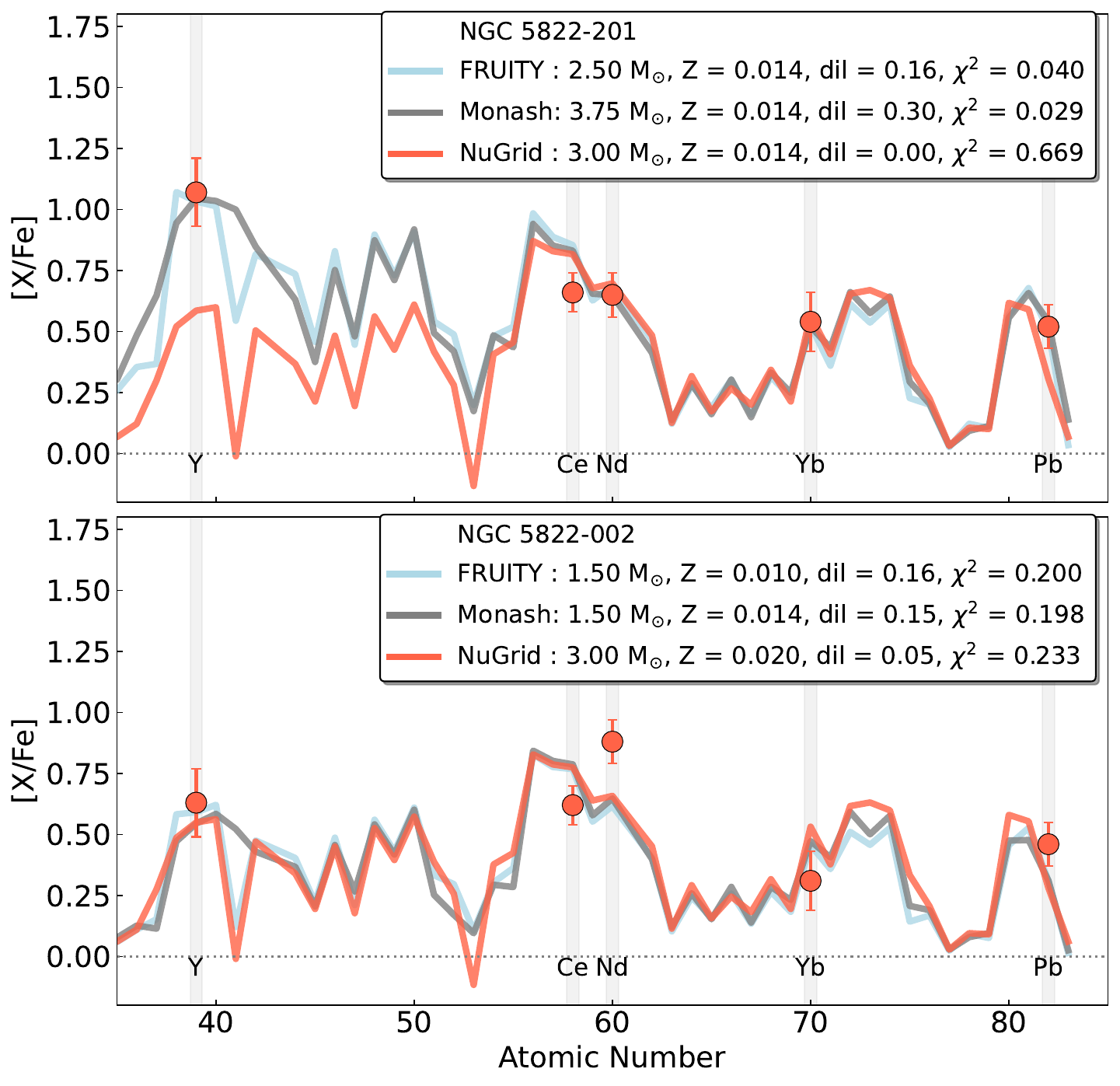}
    \caption{Comparison between the observed chemical abundances (red circles) and predictions from the \texttt{FRUITY} (blue), \texttt{Monash} (gray), and \texttt{NuGrid} (red) nucleosynthesis models, each assuming different stellar masses. The metallicity (Z), dilution factor (dil), and $\chi^{2}$ value for each model are also indicated.}
    \label{fig:ba_models}
\end{figure}

\subsubsection{The barium stars}\label{sec:ba_stars}

The two Ba-stars in our sample, \#002 and \#201, exhibit all the typical characteristics of this class: moderate to high enrichment in heavy elements such as Y, Ce, Nd and Yb, and slight enrichment in $^{12}$C. As previously mentioned, both stars also show elevated \texttt{RUWE} values, suggesting binarity and complicating astrometric membership assessment. Nevertheless, for elements not affected by binary mass transfer, their abundances are consistent with those of confirmed cluster members.

\par Keeping this in mind, we explore here the abundances of five heavy elements observed in the Ba-stars \#002 and \#201, in light of the \texttt{Monash}, \texttt{FRUITY}, and \texttt{NuGrid} \citep{battino2019} $s$-process nucleosynthesis models. Together, the \texttt{Monash} and \texttt{FRUITY} models cover a wide range of stellar masses (1.0\,-\,8.0\,M$_{\odot}$) and metallicities ($-1.20\,\lesssim\,\rm{[Fe/H]}\,\lesssim\,+0.30$). On the other hand, \texttt{NuGrid} models are available for 2 and 3\,M$_{\odot}$, close to solar metallicity ($\rm{[Fe/H]}\,\simeq\,-0.15, +0.16$, and $+0.30$), and include convective boundary mixing (CBM) at the bottom of the He intershell during the thermal pulse episodes.

\par The abundance patterns of the neutron-capture elements observed in \#002 and \#201 were compared with predictions from these three sets of nucleosynthesis models. To carry out this task, we followed the same approach as \citet{roriz2024}, by applying dilution to the models and searching for the best fits between observation and predictions, allowing us to find constraints on mass of the polluter TP-AGB stars. The elemental abundances, which probe the three peaks of the $s$-process, as well as the region between the second and third peaks, together with the theoretical predictions are shown in Figure\,\ref{fig:ba_models}. In general, low-mass models are able to reproduce the observations, in agreement with other studies \citep[e.g.][]{cseh2022}. For AGB stars in this mass range, the initial–final mass relation of \citet{badry2018} predicts white dwarf progeny masses $\lesssim$\,0.7\,M$_{\odot}$. This is further supported by the low values of the mass function derived for Ba stars; for the star \#002, \citet{swaelmen2017} reported $f(m)\,=\,0.009$\,M$\odot$. Such a low value points out for a white dwarf as the unseen companion of \#002, which is expected as the final remnant of a former low-mass polluter TP-AGB stars. For the star \#201, in particular, a \texttt{Monash} model of $3.75$\,M$_{\odot}$ accompanied by the largest dilution factor (dil\,=\,0.30) is need to fit the observations. This mass value is aligned with no sodium enrichment which is related to scenario of intermediate-low-mass polluting AGB. Interestingly, the \texttt{NuGrid} models that best fit the profiles observed in \#002 and \#201 are those labeled ``h-CBM", which produce larger $^{13}$C-pockets; these, in turn, needed of lower dilutions ($\rm{dil}\simeq0.00$). However, the \texttt{NuGrid} model does not reach the level of the [Y/Fe] ratio observed for \#201, unlike the \texttt{Monash} and \texttt{FRUITY} models which closely reproduce it. As far as the second peak is concerned, all the models overproduce Ce in both stars. Finally, in light of Yb abundances in \#002 and \#201, there does not appear to be additional contribution from another nucleosynthetic source, in addition to $s$-process.

\par The presence of the Ba-star \#002, apparently polluted by a $\sim\,1.5\,M_{\odot}$ AGB companion according to the \texttt{Monash} and \texttt{FRUITY} models, appears at first glance to be inconsistent with the OC's turn-off mass ($2.05\,M_{\odot}$). The AGBS's mass of $1.5\,M_{\odot}$ estimate from nucleosynthesis models is subject to significant uncertainties, particularly due to the complex dependence of $s$-process yields on stellar parameters such as metallicity, mixing efficiency, and the formation of the $^{13}$C pocket. In this context, although it results in a higher $\chi^2$ value, the NuGrid model yielding a 3.0\,$M_\odot$ TP-AGB polluter suggests a more plausible mass for the AGB companion.

\section{Conclusion}\label{sec:conc}

In this work, we present a comprehensive chemical analysis of thirteen red giant stars in the open cluster NGC\,5822. The study is based on high-resolution spectra obtained with two different spectrographs, \texttt{FEROS} and \texttt{IGRINS}, covering optical and near-infrared wavelengths ($H$ and $K$ bands). For the first time, we provide abundances of F, P, K, and Yb, as well as $^{17}$O/$^{16}$O and $^{18}$O/$^{16}$O isotopic ratios for stars in this well-studied cluster. We also performed an independent membership analysis using astrometric data from the Gaia mission. Three stars were removed from this astrometric analysis due to high \texttt{RUWE} values (\#001, \#002, and \#201), but two of them are in binary systems and have been classified as Ba-stars (\#002 and \#201). The Ba-stars present chemical pattern compatible with the members of the cluster, while the star \#001 differ significantly. 

\par The membership results and physical parameters derived for the cluster, such as age, distance, reddening, and mean astrometric values, are consistent with those reported in the literature \citep{Cantat2020, dias2021, hunt2023}, validating the robustness of our membership determination. Based on these primary parameters, additional quantities were calculated, including the turn-off mass and the Galactocentric distance. These values are fundamental for interpreting the chemical abundance results within the broader context of stellar and galactic chemical evolution. First, the turn-off mass was used to test mixing models for Na, $^{12}$C/$^{13}$C, $^{16}$O/$^{17}$O, and $^{16}$O/$^{18}$O isotopic ratios, as well as to constrain AGB nucleosynthesis models. All evolutionary models are very sensitive to stellar mass, particularly near the critical threshold of 2.0\,$M_{\odot}$, which is very close to the turn-off mass of NGC\,5822 (2.05\,$M_{\odot}$). We find agreement with the predictions of 1DUP and thermohaline mixing from \citet{lagarde2012} in the $^{12}$C/$^{13}$C ratios. The observed spread, from 12 to 25, reflects differences in evolutionary stage, as well as additional factors such as mass, rotation, and internal angular-momentum transport. Regarding the AGB nucleosynthesis models used to estimate the mass of the polluting companion in Ba-stars, the turn-off mass allowed us to rule out 1.5\,$M_{\odot}$ models, even in the case of star \#002, where such models yielded a lower $\chi^{2}$. The absence of Na enrichment in the Ba-stars led us to associate 4.0\,$M_{\odot}$ as an upper limit for the mass of the AGB polluting companions. This is consistent with the derived values of 3.00 and 3.75\,$M_{\odot}$ for stars \#002 and \#201, respectively.

\par The Li-enriched stars \#006, \#102, and \#240 do not exhibit any anomalous chemical abundances or isotopic ratios that could be linked to a specific Li enrichment scenario. However, isochrone fitting indicates that their most probable evolutionary stage is the red clump, which is consistent with the phase most commonly associated with Li-rich/enriched giant stars \citep[][among others]{yan2021}. In the absence of asteroseismic data the isochrone fitting serves as a reliable alternative and represents an advantage when analyzing peculiar stars in open clusters. Based on their evolutinary stage, the candidate mechanisms to explain the formation of these stars include a merger event involving a helium white dwarf (0.35\,$\,\leq\,M/M_{\odot}\,\leq\,$\,0.40) and a RGB star \citep{zhang2020}, as well as the helium-flash-induced mixing scenario \citep{schwab2020}. The analysis of the Ca\,{\sc ii} H \& K lines indicates the presence of chromospheric activity in stars \#006 and \#102. Although stellar activity is often discussed in connection with enhanced lithium abundances, rapid rotation, or binarity, these stars do not exhibit high projected rotational velocities ($v\,\sin\,i \gtrsim 8$\,km\,s$^{-1}$) nor any evidence of binarity, suggesting that activity alone is not sufficient to explain their lithium enrichment.

\par A Galactic gradient of fluorine was derived by combining results from the literature. We obtained a slope of $-0.05\,\pm\,0.03$\,dex\,kpc$^{-1}$ for the inner disk. Our result is consistent with the slope of $-0.09\,\pm\,0.02$\,dex\,kpc$^{-1}$ reported by \citet{bijavara2024b}, although we emphasize the scarcity of open clusters with Galactocentric distances $\leq$\,8.0\,kpc in the current sample. The results of fluorine nucleosynthesis models indicate a composite origin for fluorine, with early rapid enrichment by massive stars followed by a more extended and metallicity-dependent contribution from AGB stars. Fluorine in Ba-stars may be enriched since AGB stars are a important astrophysical site of this specie. However, even with probable enrichment, the effective temperature values observed in the stars \#002 and \#201 make this analysis impossible. In this context, a near-infrared study of cool field Ba-stars is currently being prepared by our group.

\section*{Acknowledgments}
NH thanks Chessy for all the technical support. The authors acknowledge Lyudmila Mashonkina for providing the \texttt{NLTE} corrections for the Pb abundances. We also thank Donatella Romano and Emanuele Spitoni for kindly sharing their model prediction files. The authors are grateful to Jane Gregorio-Hetem for reading the manuscript and providing helpful suggestions. This work has been developed under a fellowship of the Fundação de Amparo à Pesquisa do Estado do Rio de Janeiro -- FAPERJ, Rio de Janeiro, Brazil, grant E-26/200.097/2025. V.L.T. acknowledges a fellowship 302195/2024-6 of the PCI Program -- MCTI and a fellowship 152242/2024-4 of the PDJ -- MCTI and CNPq. S.B.S. acknowledges funding from the Crafoord Foundation. M.P.R. acknowledges financial support from the Coordenação de Aperfeiçoamento de Pessoal de Nível Superior -- Brasil (CAPES) -- Finance Code 001. M.B.F. acknowledges financial support from the National Council for Scientific and Technological Development (CNPq), Brazil (grant number 307711/2022-6). C.M. acknowledges financial support from the Consejo Nacional de Investigaciones Científicas y Técnicas (CONICET). S.D. acknowledges CNPq/MCTI for grant 306859/2022-0 and FAPERJ for grant 210.688/2024.

This work used the Immersion Grating Infrared Spectrometer (IGRINS) that was developed under a collaboration between the University of Texas at Austin and the Korea Astronomy and Space Science Institute (KASI) with the financial support of the Mt. Cuba Astronomical Foundation, of the US National Science Foundation under grants AST-1229522 and AST-1702267, McDonald Observatory of the University of Texas at Austin, of the Korean GMT Project of KASI, and Gemini Observatory. 
\section*{Data Availability}

Data are available on request. The data underlying this article will be shared on reasonable request to the corresponding author.

\bibliographystyle{mnras}
\bibliography{example} 

@ARTICLE{katime2013,
       author = {{Katime Santrich}, O.~J. and {Pereira}, C.~B. and {de Castro}, D.~B.},
        title = "{Two Barium Stars in the Open Cluster NGC 5822}",
      journal = {\aj},
         year = 2013,
        month = aug,
       volume = {146},
       number = {2},
          eid = {39},
        pages = {39},
          doi = {10.1088/0004-6256/146/2/39},
       adsurl = {https://ui.adsabs.harvard.edu/abs/2013AJ....146...39K}
}

@ARTICLE{contursi2024,
       author = {{Contursi}, G. and {de Laverny}, P. and {Recio-Blanco}, A. and {Molero}, M. and {Spitoni}, E. and {Matteucci}, F. and {Cristallo}, S.},
        title = "{The AMBRE Project: Lead abundance in Galactic stars}",
      journal = {\aap},
         year = 2024,
        month = oct,
       volume = {690},
          eid = {A97},
        pages = {A97},
          doi = {10.1051/0004-6361/202450782},
archivePrefix = {arXiv},
       eprint = {2408.16292},
 primaryClass = {astro-ph.SR},
       adsurl = {https://ui.adsabs.harvard.edu/abs/2024A&A...690A..97C}
}

@ARTICLE{majewski2017,
       author = {{Majewski}, Steven R. and {Schiavon}, Ricardo P. and {Frinchaboy}, Peter M. and {Allende Prieto}, Carlos and {Barkhouser}, Robert and {Bizyaev}, Dmitry and {Blank}, Basil and {Brunner}, Sophia and {Burton}, Adam and {Carrera}, Ricardo and {Chojnowski}, S. Drew and {Cunha}, K{\'a}tia and {Epstein}, Courtney and {Fitzgerald}, Greg and {Garc{\'\i}a P{\'e}rez}, Ana E. and {Hearty}, Fred R. and {Henderson}, Chuck and {Holtzman}, Jon A. and {Johnson}, Jennifer A. and {Lam}, Charles R. and {Lawler}, James E. and {Maseman}, Paul and {M{\'e}sz{\'a}ros}, Szabolcs and {Nelson}, Matthew and {Nguyen}, Duy Coung and {Nidever}, David L. and {Pinsonneault}, Marc and {Shetrone}, Matthew and {Smee}, Stephen and {Smith}, Verne V. and {Stolberg}, Todd and {Skrutskie}, Michael F. and {Walker}, Eric and {Wilson}, John C. and {Zasowski}, Gail and {Anders}, Friedrich and {Basu}, Sarbani and {Beland}, Stephane and {Blanton}, Michael R. and {Bovy}, Jo and {Brownstein}, Joel R. and {Carlberg}, Joleen and {Chaplin}, William and {Chiappini}, Cristina and {Eisenstein}, Daniel J. and {Elsworth}, Yvonne and {Feuillet}, Diane and {Fleming}, Scott W. and {Galbraith-Frew}, Jessica and {Garc{\'\i}a}, Rafael A. and {Garc{\'\i}a-Hern{\'a}ndez}, D. An{\'\i}bal and {Gillespie}, Bruce A. and {Girardi}, L{\'e}o and {Gunn}, James E. and {Hasselquist}, Sten and {Hayden}, Michael R. and {Hekker}, Saskia and {Ivans}, Inese and {Kinemuchi}, Karen and {Klaene}, Mark and {Mahadevan}, Suvrath and {Mathur}, Savita and {Mosser}, Beno{\^\i}t and {Muna}, Demitri and {Munn}, Jeffrey A. and {Nichol}, Robert C. and {O'Connell}, Robert W. and {Parejko}, John K. and {Robin}, A.~C. and {Rocha-Pinto}, Helio and {Schultheis}, Matthias and {Serenelli}, Aldo M. and {Shane}, Neville and {Silva Aguirre}, Victor and {Sobeck}, Jennifer S. and {Thompson}, Benjamin and {Troup}, Nicholas W. and {Weinberg}, David H. and {Zamora}, Olga},
        title = "{The Apache Point Observatory Galactic Evolution Experiment (APOGEE)}",
      journal = {\aj},
         year = 2017,
        month = sep,
       volume = {154},
       number = {3},
          eid = {94},
        pages = {94},
          doi = {10.3847/1538-3881/aa784d},
archivePrefix = {arXiv},
       eprint = {1509.05420},
 primaryClass = {astro-ph.IM},
       adsurl = {https://ui.adsabs.harvard.edu/abs/2017AJ....154...94M}
}

@ARTICLE{takeda2005,
       author = {{Takeda}, Yoichi and {Hashimoto}, Osamu and {Taguchi}, Hikaru and {Yoshioka}, Kazuo and {Takada-Hidai}, Masahide and {Saito}, Yuji and {Honda}, Satoshi},
        title = "{Non-LTE Line-Formation and Abundances of Sulfur and Zinc in F, G, and K Stars}",
      journal = {\pasj},
         year = 2005,
        month = oct,
       volume = {57},
        pages = {751-768},
          doi = {10.1093/pasj/57.5.751},
archivePrefix = {arXiv},
       eprint = {astro-ph/0509239},
 primaryClass = {astro-ph},
       adsurl = {https://ui.adsabs.harvard.edu/abs/2005PASJ...57..751T}
}

@ARTICLE{civis2013,
       author = {{Civi{\v{s}}}, S. and {Ferus}, M. and {Chernov}, V.~E. and {Zanozina}, E.~M.},
        title = "{Infrared transitions and oscillator strengths of Ca and Mg}",
      journal = {\aap},
         year = 2013,
        month = jun,
       volume = {554},
          eid = {A24},
        pages = {A24},
          doi = {10.1051/0004-6361/201321052},
       adsurl = {https://ui.adsabs.harvard.edu/abs/2013A&A...554A..24C}
}

@BOOK{clayton2003,
       author = {{Clayton}, Donald},
        title = "{Handbook of Isotopes in the Cosmos}",
         year = 2003,
     publisher= "{Cambridge, UK: Cambridge University Press}",
       adsurl = {https://ui.adsabs.harvard.edu/abs/2003hic..book.....C}
}

@ARTICLE{nordlander2017,
       author = {{Nordlander}, T. and {Lind}, K.},
        title = "{Non-LTE aluminium abundances in late-type stars}",
      journal = {\aap},
         year = 2017,
        month = nov,
       volume = {607},
          eid = {A75},
        pages = {A75},
          doi = {10.1051/0004-6361/201730427},
archivePrefix = {arXiv},
       eprint = {1708.01949},
 primaryClass = {astro-ph.SR},
       adsurl = {https://ui.adsabs.harvard.edu/abs/2017A&A...607A..75N}
}

@BOOK{cox2000,
       author = {{Cox}, Arthur N.},
        title = "{Allen's astrophysical quantities}",
         year = 2000,
     publisher={Springer Science \& Business Media},
       adsurl = {https://ui.adsabs.harvard.edu/abs/2000asqu.book.....C}
}

@ARTICLE{holanda2022,
       author = {{Holanda}, N. and {Ramos}, Andr{\'e} A. and {Pe{\~n}a Su{\'a}rez}, V.~J. and {Martinez}, Cintia F. and {Pereira}, C.~B.},
        title = "{A chemical analysis of seven red giants of the Galactic cluster NGC 4349}",
      journal = {\mnras},
         year = 2022,
        month = nov,
       volume = {516},
       number = {3},
        pages = {4484-4496},
          doi = {10.1093/mnras/stac2496},
       adsurl = {https://ui.adsabs.harvard.edu/abs/2022MNRAS.516.4484H}
}

@ARTICLE{cunha2017,
       author = {{Cunha}, Katia and {Smith}, Verne V. and {Hasselquist}, Sten and {Souto}, Diogo and {Shetrone}, Matthew D. and {Allende Prieto}, Carlos and {Bizyaev}, Dmitry and {Frinchaboy}, Peter and {Garc{\'\i}a-Hern{\'a}ndez}, D. Anibal and {Holtzman}, Jon and {Johnson}, Jennifer A. and {J{\H{o}}nsson}, Henrik and {Majewski}, Steven R. and {M{\'e}sz{\'a}ros}, Szabolcs and {Nidever}, David and {Pinsonneault}, Mark and {Schiavon}, Ricardo P. and {Sobeck}, Jennifer and {Skrutskie}, Michael F. and {Zamora}, Olga and {Zasowski}, Gail and {Fern{\'a}ndez-Trincado}, J.~G.},
        title = "{Adding the s-Process Element Cerium to the APOGEE Survey: Identification and Characterization of Ce II Lines in the H-band Spectral Window}",
      journal = {\apj},
         year = 2017,
        month = aug,
       volume = {844},
       number = {2},
          eid = {145},
        pages = {145},
          doi = {10.3847/1538-4357/aa7beb},
       adsurl = {https://ui.adsabs.harvard.edu/abs/2017ApJ...844..145C}
}

@ARTICLE{montelius2022,
       author = {{Montelius}, M. and {Forsberg}, R. and {Ryde}, N. and {J{\"o}nsson}, H. and {Af{\c{s}}ar}, M. and {Johansen}, A. and {Kaplan}, K.~F. and {Kim}, H. and {Mace}, G. and {Sneden}, C. and {Thorsbro}, B.},
        title = "{Chemical evolution of ytterbium in the Galactic disk}",
      journal = {\aap},
         year = 2022,
        month = sep,
       volume = {665},
          eid = {A135},
        pages = {A135},
          doi = {10.1051/0004-6361/202243140},
archivePrefix = {arXiv},
       eprint = {2202.00691},
 primaryClass = {astro-ph.GA},
       adsurl = {https://ui.adsabs.harvard.edu/abs/2022A&A...665A.135M}
}

@ARTICLE{goorvitch1994,
       author = {{Goorvitch}, D.},
        title = "{Infrared CO Line List for the X 1 Sigma + State}",
      journal = {\apjs},
         year = 1994,
        month = dec,
       volume = {95},
        pages = {535},
          doi = {10.1086/192110},
       adsurl = {https://ui.adsabs.harvard.edu/abs/1994ApJS...95..535G}
}

@ARTICLE{topcu2020,
       author = {{B{\"o}cek Topcu}, G. and {Af{\c{s}}ar}, M. and {Sneden}, C. and {Pilachowski}, C.~A. and {Denissenkov}, P.~A. and {VandenBerg}, D.~A. and {Wright}, D. and {Mace}, G.~N. and {Jaffe}, D.~T. and {Strickland}, E. and {Kim}, H. and {Sokal}, K.~R.},
        title = "{Chemical abundances of open clusters from high-resolution infrared spectra - II. NGC 752}",
      journal = {\mnras},
         year = 2020,
        month = jan,
       volume = {491},
       number = {1},
        pages = {544-559},
          doi = {10.1093/mnras/stz3008},
archivePrefix = {arXiv},
       eprint = {1910.10554},
 primaryClass = {astro-ph.SR},
       adsurl = {https://ui.adsabs.harvard.edu/abs/2020MNRAS.491..544B}
}

@ARTICLE{ryde2020,
       author = {{Ryde}, Nils and {J{\"o}nsson}, Henrik and {Mace}, Gregory and {Cunha}, Katia and {Spitoni}, Emanuele and {Af{\c{s}}ar}, Melike and {Jaffe}, Daniel and {Forsberg}, Rebecca and {Kaplan}, Kyle F. and {Kidder}, Benjamin T. and {Lee}, Jae-Joon and {Oh}, Heeyoung and {Smith}, Verne V. and {Sneden}, Christopher and {Sokal}, Kimberly R. and {Strickland}, Emily and {Thorsbro}, Brian},
        title = "{Fluorine in the Solar Neighborhood: The Need for Several Cosmic Sources}",
      journal = {\apj},
         year = 2020,
        month = apr,
       volume = {893},
       number = {1},
          eid = {37},
        pages = {37},
          doi = {10.3847/1538-4357/ab7eb1},
archivePrefix = {arXiv},
       eprint = {2003.04656},
 primaryClass = {astro-ph.GA},
       adsurl = {https://ui.adsabs.harvard.edu/abs/2020ApJ...893...37R}
}

@ARTICLE{reid2014,
       author = {{Reid}, M.~J. and {Menten}, K.~M. and {Brunthaler}, A. and {Zheng}, X.~W. and {Dame}, T.~M. and {Xu}, Y. and {Wu}, Y. and {Zhang}, B. and {Sanna}, A. and {Sato}, M. and {Hachisuka}, K. and {Choi}, Y.~K. and {Immer}, K. and {Moscadelli}, L. and {Rygl}, K.~L.~J. and {Bartkiewicz}, A.},
        title = "{Trigonometric Parallaxes of High Mass Star Forming Regions: The Structure and Kinematics of the Milky Way}",
      journal = {\apj},
         year = 2014,
        month = mar,
       volume = {783},
       number = {2},
          eid = {130},
        pages = {130},
          doi = {10.1088/0004-637X/783/2/130},
archivePrefix = {arXiv},
       eprint = {1401.5377},
 primaryClass = {astro-ph.GA},
       adsurl = {https://ui.adsabs.harvard.edu/abs/2014ApJ...783..130R}
}

@ARTICLE{spitoni2018,
       author = {{Spitoni}, E. and {Matteucci}, F. and {J{\"o}nsson}, H. and {Ryde}, N. and {Romano}, D.},
        title = "{Fluorine in the solar neighborhood: Chemical evolution models}",
      journal = {\aap},
         year = 2018,
        month = apr,
       volume = {612},
          eid = {A16},
        pages = {A16},
          doi = {10.1051/0004-6361/201732092},
archivePrefix = {arXiv},
       eprint = {1711.09671},
 primaryClass = {astro-ph.GA},
       adsurl = {https://ui.adsabs.harvard.edu/abs/2018A&A...612A..16S}
}

@ARTICLE{cristallo2015,
       author = {{Cristallo}, S. and {Straniero}, O. and {Piersanti}, L. and {Gobrecht}, D.},
        title = "{Evolution, Nucleosynthesis, and Yields of AGB Stars at Different Metallicities. III. Intermediate-mass Models, Revised Low-mass Models, and the ph-FRUITY Interface}",
      journal = {\apjs},
         year = 2015,
        month = aug,
       volume = {219},
       number = {2},
          eid = {40},
        pages = {40},
          doi = {10.1088/0067-0049/219/2/40},
archivePrefix = {arXiv},
       eprint = {1507.07338},
 primaryClass = {astro-ph.SR},
       adsurl = {https://ui.adsabs.harvard.edu/abs/2015ApJS..219...40C}
}

@ARTICLE{karakas2018,
       author = {{Karakas}, Amanda I. and {Lugaro}, Maria and {Carlos}, Mar{\'\i}lia and {Cseh}, Borb{\'a}la and {Kamath}, Devika and {Garc{\'\i}a-Hern{\'a}ndez}, D.~A.},
        title = "{Heavy-element yields and abundances of asymptotic giant branch models with a Small Magellanic Cloud metallicity}",
      journal = {\mnras},
         year = 2018,
        month = jun,
       volume = {477},
       number = {1},
        pages = {421-437},
          doi = {10.1093/mnras/sty625},
archivePrefix = {arXiv},
       eprint = {1803.02028},
 primaryClass = {astro-ph.SR},
       adsurl = {https://ui.adsabs.harvard.edu/abs/2018MNRAS.477..421K}
}

@ARTICLE{maiorca2014,
       author = {{Maiorca}, E. and {Uitenbroek}, H. and {Uttenthaler}, S. and {Randich}, S. and {Busso}, M. and {Magrini}, L.},
        title = "{A New Solar Fluorine Abundance and a Fluorine Determination in the Two Open Clusters M67 and NGC 6404}",
      journal = {\apj},
         year = 2014,
        month = jun,
       volume = {788},
       number = {2},
          eid = {149},
        pages = {149},
          doi = {10.1088/0004-637X/788/2/149},
archivePrefix = {arXiv},
       eprint = {1404.5755},
 primaryClass = {astro-ph.SR},
       adsurl = {https://ui.adsabs.harvard.edu/abs/2014ApJ...788..149M}
}

@ARTICLE{nault2013,
       author = {{Nault}, K.~A. and {Pilachowski}, C.~A.},
        title = "{The Abundance of Fluorine in the Hyades, NGC 752, and M67}",
      journal = {\aj},
         year = 2013,
        month = dec,
       volume = {146},
       number = {6},
          eid = {153},
        pages = {153},
          doi = {10.1088/0004-6256/146/6/153},
       adsurl = {https://ui.adsabs.harvard.edu/abs/2013AJ....146..153N}
}

@ARTICLE{ryde2025,
       author = {{Ryde}, Nils and {Nandakumar}, Govind and {Schultheis}, Mathias and {Kordopatis}, Georges and {di Matteo}, Paola and {Haywood}, Misha and {Sch{\"o}del}, Rainer and {Nogueras-Lara}, Francisco and {Rich}, R. Michael and {Thorsbro}, Brian and {Mace}, Gregory N. and {Agertz}, Oscar and {Amarsi}, Anish M. and {Kocher}, Jessica and {Molero}, Marta and {Orglia}, Livia and {Pagnini}, Giulia and {Spitoni}, Emanuele},
        title = "{Chemical Abundances in the Nuclear Star Cluster of the Milky Way: Alpha-element Trends and Their Similarities with the Inner Bulge}",
      journal = {\apj},
         year = 2025,
        month = feb,
       volume = {979},
       number = {2},
          eid = {174},
        pages = {174},
          doi = {10.3847/1538-4357/ad9b2b},
archivePrefix = {arXiv},
       eprint = {2412.04528},
 primaryClass = {astro-ph.GA},
       adsurl = {https://ui.adsabs.harvard.edu/abs/2025ApJ...979..174R}
}

@ARTICLE{bijavara2024a,
       author = {{Bijavara Seshashayana}, S. and {J{\"o}nsson}, H. and {D'Orazi}, V. and {Nandakumar}, G. and {Oliva}, E. and {Bragaglia}, A. and {Sanna}, N. and {Romano}, D. and {Spitoni}, E. and {Karakas}, A. and {Lugaro}, M. and {Origlia}, L.},
        title = "{Stellar Population Astrophysics (SPA) with TNG. Fluorine abundances in seven open clusters}",
      journal = {\aap},
         year = 2024,
        month = mar,
       volume = {683},
          eid = {A218},
        pages = {A218},
          doi = {10.1051/0004-6361/202349068},
archivePrefix = {arXiv},
       eprint = {2402.09058},
 primaryClass = {astro-ph.GA},
       adsurl = {https://ui.adsabs.harvard.edu/abs/2024A&A...683A.218B}
}

@ARTICLE{bijavara2024b,
       author = {{Bijavara Seshashayana}, S. and {J{\"o}nsson}, H. and {D'Orazi}, V. and {Sanna}, N. and {Andreuzzi}, G. and {Nandakumar}, G. and {Bragaglia}, A. and {Romano}, D. and {Spitoni}, E.},
        title = "{Exploring fluorine chemical evolution in the Galactic disk: The open cluster perspective}",
      journal = {\aap},
         year = 2024,
        month = sep,
       volume = {689},
          eid = {A120},
        pages = {A120},
          doi = {10.1051/0004-6361/202451056},
archivePrefix = {arXiv},
       eprint = {2407.10229},
 primaryClass = {astro-ph.GA},
       adsurl = {https://ui.adsabs.harvard.edu/abs/2024A&A...689A.120B}
}

@ARTICLE{myers2022,
       author = {{Myers}, Natalie and {Donor}, John and {Spoo}, Taylor and {Frinchaboy}, Peter M. and {Cunha}, Katia and {Price-Whelan}, Adrian M. and {Majewski}, Steven R. and {Beaton}, Rachael L. and {Zasowski}, Gail and {O'Connell}, Julia and {Ray}, Amy E. and {Bizyaev}, Dmitry and {Chiappini}, Cristina and {Garc{\'\i}a-Hern{\'a}ndez}, D.~A. and {Geisler}, Doug and {J{\"o}nsson}, Henrik and {Lane}, Richard R. and {Longa-Pe{\~n}a}, Pen{\'e}lope and {Minchev}, Ivan and {Minniti}, Dante and {Nitschelm}, Christian and {Roman-Lopes}, A.},
        title = "{The Open Cluster Chemical Abundances and Mapping Survey. VI. Galactic Chemical Gradient Analysis from APOGEE DR17}",
      journal = {\aj},
         year = 2022,
        month = sep,
       volume = {164},
       number = {3},
          eid = {85},
        pages = {85},
          doi = {10.3847/1538-3881/ac7ce5},
archivePrefix = {arXiv},
       eprint = {2206.13650},
 primaryClass = {astro-ph.GA},
       adsurl = {https://ui.adsabs.harvard.edu/abs/2022AJ....164...85M}
}

@ARTICLE{donor2020,
       author = {{Donor}, John and {Frinchaboy}, Peter M. and {Cunha}, Katia and {O'Connell}, Julia E. and {Allende Prieto}, Carlos and {Almeida}, Andr{\'e}s and {Anders}, Friedrich and {Beaton}, Rachael and {Bizyaev}, Dmitry and {Brownstein}, Joel R. and {Carrera}, Ricardo and {Chiappini}, Cristina and {Cohen}, Roger and {Garc{\'\i}a-Hern{\'a}ndez}, D.~A. and {Geisler}, Doug and {Hasselquist}, Sten and {J{\"o}nsson}, Henrik and {Lane}, Richard R. and {Majewski}, Steven R. and {Minniti}, Dante and {Bidin}, Christian Moni and {Pan}, Kaike and {Roman-Lopes}, Alexandre and {Sobeck}, Jennifer S. and {Zasowski}, Gail},
        title = "{The Open Cluster Chemical Abundances and Mapping Survey. IV. Abundances for 128 Open Clusters Using SDSS/APOGEE DR16}",
      journal = {\aj},
         year = 2020,
        month = may,
       volume = {159},
       number = {5},
          eid = {199},
        pages = {199},
          doi = {10.3847/1538-3881/ab77bc},
archivePrefix = {arXiv},
       eprint = {2002.08980},
 primaryClass = {astro-ph.GA},
       adsurl = {https://ui.adsabs.harvard.edu/abs/2020AJ....159..199D}
}

@ARTICLE{lugaro2023,
       author = {{Lugaro}, Maria and {Pignatari}, Marco and {Reifarth}, Ren{\'e} and {Wiescher}, Michael},
        title = "{The s Process and Beyond}",
      journal = {Annual Review of Nuclear and Particle Science},
         year = 2023,
        month = sep,
       volume = {73},
        pages = {315-340},
          doi = {10.1146/annurev-nucl-102422-080857},
       adsurl = {https://ui.adsabs.harvard.edu/abs/2023ARNPS..73..315L}
}

@ARTICLE{cristallo2009,
       author = {{Cristallo}, S. and {Straniero}, O. and {Gallino}, R. and {Piersanti}, L. and {Dom{\'\i}nguez}, I. and {Lederer}, M.~T.},
        title = "{Evolution, Nucleosynthesis, and Yields of Low-Mass Asymptotic Giant Branch Stars at Different Metallicities}",
      journal = {\apj},
         year = 2009,
        month = may,
       volume = {696},
       number = {1},
        pages = {797-820},
          doi = {10.1088/0004-637X/696/1/797},
archivePrefix = {arXiv},
       eprint = {0902.0243},
 primaryClass = {astro-ph.SR},
       adsurl = {https://ui.adsabs.harvard.edu/abs/2009ApJ...696..797C}
}

@ARTICLE{cristallo2011,
       author = {{Cristallo}, S. and {Piersanti}, L. and {Straniero}, O. and {Gallino}, R. and {Dom{\'\i}nguez}, I. and {Abia}, C. and {Di Rico}, G. and {Quintini}, M. and {Bisterzo}, S.},
        title = "{Evolution, Nucleosynthesis, and Yields of Low-mass Asymptotic Giant Branch Stars at Different Metallicities. II. The FRUITY Database}",
      journal = {\apjs},
         year = 2011,
        month = dec,
       volume = {197},
       number = {2},
          eid = {17},
        pages = {17},
          doi = {10.1088/0067-0049/197/2/17},
archivePrefix = {arXiv},
       eprint = {1109.1176},
 primaryClass = {astro-ph.SR},
       adsurl = {https://ui.adsabs.harvard.edu/abs/2011ApJS..197...17C}
}

@ARTICLE{karakas2016,
       author = {{Karakas}, Amanda I. and {Lugaro}, Maria},
        title = "{Stellar Yields from Metal-rich Asymptotic Giant Branch Models}",
      journal = {\apj},
         year = 2016,
        month = jul,
       volume = {825},
       number = {1},
          eid = {26},
        pages = {26},
          doi = {10.3847/0004-637X/825/1/26},
archivePrefix = {arXiv},
       eprint = {1604.02178},
 primaryClass = {astro-ph.SR},
       adsurl = {https://ui.adsabs.harvard.edu/abs/2016ApJ...825...26K}
}

@ARTICLE{lugaro2012,
       author = {{Lugaro}, Maria and {Karakas}, Amanda I. and {Stancliffe}, Richard J. and {Rijs}, Carlos},
        title = "{The s-process in Asymptotic Giant Branch Stars of Low Metallicity and the Composition of Carbon-enhanced Metal-poor Stars}",
      journal = {\apj},
         year = 2012,
        month = mar,
       volume = {747},
       number = {1},
          eid = {2},
        pages = {2},
          doi = {10.1088/0004-637X/747/1/2},
archivePrefix = {arXiv},
       eprint = {1112.2757},
 primaryClass = {astro-ph.SR},
       adsurl = {https://ui.adsabs.harvard.edu/abs/2012ApJ...747....2L}
}

@ARTICLE{fishlock2014,
       author = {{Fishlock}, Cherie K. and {Karakas}, Amanda I. and {Lugaro}, Maria and {Yong}, David},
        title = "{Evolution and Nucleosynthesis of Asymptotic Giant Branch Stellar Models of Low Metallicity}",
      journal = {\apj},
         year = 2014,
        month = dec,
       volume = {797},
       number = {1},
          eid = {44},
        pages = {44},
          doi = {10.1088/0004-637X/797/1/44},
archivePrefix = {arXiv},
       eprint = {1410.7457},
 primaryClass = {astro-ph.SR},
       adsurl = {https://ui.adsabs.harvard.edu/abs/2014ApJ...797...44F}
}

@ARTICLE{magrini2021,
       author = {{Magrini}, L. and {Lagarde}, N. and {Charbonnel}, C. and {Franciosini}, E. and {Randich}, S. and {Smiljanic}, R. and {Casali}, G. and {Viscasillas V{\'a}zquez}, C. and {Spina}, L. and {Biazzo}, K. and {Pasquini}, L. and {Bragaglia}, A. and {Van der Swaelmen}, M. and {Tautvai{\v{s}}ien{\.{e}}}, G. and {Inno}, L. and {Sanna}, N. and {Prisinzano}, L. and {Degl'Innocenti}, S. and {Prada Moroni}, P. and {Roccatagliata}, V. and {Tognelli}, E. and {Monaco}, L. and {de Laverny}, P. and {Delgado-Mena}, E. and {Baratella}, M. and {D'Orazi}, V. and {Vallenari}, A. and {Gonneau}, A. and {Worley}, C. and {Jim{\'e}nez-Esteban}, F. and {Jofre}, P. and {Bensby}, T. and {Fran{\c{c}}ois}, P. and {Guiglion}, G. and {Bayo}, A. and {Jeffries}, R.~D. and {Binks}, A.~S. and {Gilmore}, G. and {Damiani}, F. and {Korn}, A. and {Pancino}, E. and {Sacco}, G.~G. and {Hourihane}, A. and {Morbidelli}, L. and {Zaggia}, S.},
        title = "{The Gaia-ESO survey: Mixing processes in low-mass stars traced by lithium abundance in cluster and field stars}",
      journal = {\aap},
         year = 2021,
        month = jul,
       volume = {651},
          eid = {A84},
        pages = {A84},
          doi = {10.1051/0004-6361/202140935},
archivePrefix = {arXiv},
       eprint = {2105.04866},
 primaryClass = {astro-ph.SR},
       adsurl = {https://ui.adsabs.harvard.edu/abs/2021A&A...651A..84M}
}

@ARTICLE{topcu2019,
       author = {{B{\"o}cek Topcu}, G. and {Af{\c{s}}ar}, M. and {Sneden}, C. and {Pilachowski}, C.~A. and {Denissenkov}, P.~A. and {VandenBerg}, D.~A. and {Strickland}, E. and {{\"O}zdemir}, S. and {Mace}, G.~N. and {Kim}, H. and {Jaffe}, D.~T.},
        title = "{Chemical abundances of open clusters from high-resolution infrared spectra - I. NGC 6940}",
      journal = {\mnras},
         year = 2019,
        month = jun,
       volume = {485},
       number = {4},
        pages = {4625-4640},
          doi = {10.1093/mnras/stz727},
archivePrefix = {arXiv},
       eprint = {1903.03786},
 primaryClass = {astro-ph.SR},
       adsurl = {https://ui.adsabs.harvard.edu/abs/2019MNRAS.485.4625B}
}

@ARTICLE{afsar2018,
       author = {{Af{\c{s}}ar}, Melike and {Sneden}, Christopher and {Wood}, Michael P. and {Lawler}, James E. and {Bozkurt}, Zeynep and {B{\"o}cek Topcu}, Gamze and {Mace}, Gregory N. and {Kim}, Hwihyun and {Jaffe}, Daniel T.},
        title = "{Chemical Compositions of Evolved Stars from Near-infrared IGRINS High-resolution Spectra. I. Abundances in Three Red Horizontal Branch Stars}",
      journal = {\apj},
         year = 2018,
        month = sep,
       volume = {865},
       number = {1},
          eid = {44},
        pages = {44},
          doi = {10.3847/1538-4357/aada0c},
archivePrefix = {arXiv},
       eprint = {1808.03855},
 primaryClass = {astro-ph.SR},
       adsurl = {https://ui.adsabs.harvard.edu/abs/2018ApJ...865...44A}
}

@ARTICLE{holanda2021,
       author = {{Holanda}, N. and {Drake}, N.~A. and {Corradi}, W.~J.~B. and {Ferreira}, F.~A. and {Maia}, F. and {Magrini}, L. and {da Silva}, J.~R.~P. and {Pereira}, C.~B.},
        title = "{NGC 6124: a young open cluster with anomalous- and fast-rotating giant stars}",
      journal = {\mnras},
         year = 2021,
        month = dec,
       volume = {508},
       number = {4},
        pages = {5786-5801},
          doi = {10.1093/mnras/stab2836},
       adsurl = {https://ui.adsabs.harvard.edu/abs/2021MNRAS.508.5786H}
}

@ARTICLE{holanda2023,
       author = {{Holanda}, N. and {Drake}, N.~A. and {Pereira}, C.~B.},
        title = "{HD 16424: A new weak G-band star with high Li abundance}",
      journal = {\mnras},
         year = 2023,
        month = jan,
       volume = {518},
       number = {3},
        pages = {4038-4044},
          doi = {10.1093/mnras/stac3343},
       adsurl = {https://ui.adsabs.harvard.edu/abs/2023MNRAS.518.4038H}
}

@article{kramida2014,
  title={NIST Atomic Spectra Database (ver. 5.2), National Institute of Standards and Technology, Gaithersburg, MD, 2014},
  author={Kramida, A and Ralchenko, Yu and Reader, J and others},
  journal={URL https://physics. nist. gov/asd},
  year={2014}
}

@ARTICLE{guerco2022,
       author = {{Guer{\c{c}}o}, Rafael and {Smith}, Verne V. and {Cunha}, Katia and {Ekstr{\"o}m}, Sylvia and {Abia}, Carlos and {Plez}, Bertrand and {Meynet}, Georges and {Ramirez}, Solange V. and {Prantzos}, Nikos and {Sellgren}, Kris and {Hayes}, Cristian R. and {Majewski}, Steven R.},
        title = "{Evidence of deep mixing in IRS 7, a cool massive supergiant member of the Galactic nuclear star cluster}",
      journal = {\mnras},
         year = 2022,
        month = oct,
       volume = {516},
       number = {2},
        pages = {2801-2811},
          doi = {10.1093/mnras/stac2393},
archivePrefix = {arXiv},
       eprint = {2208.10529},
 primaryClass = {astro-ph.SR},
       adsurl = {https://ui.adsabs.harvard.edu/abs/2022MNRAS.516.2801G}
}

@ARTICLE{sneden2014,
       author = {{Sneden}, Christopher and {Lucatello}, Sara and {Ram}, Ram S. and {Brooke}, James S.~A. and {Bernath}, Peter},
        title = "{Line Lists for the A $^{2}${\ensuremath{\Pi}}-X $^{2}${\ensuremath{\Sigma}}$^{+}$ (Red) and B $^{2}${\ensuremath{\Sigma}}$^{+}$-X $^{2}${\ensuremath{\Sigma}}$^{+}$ (Violet) Systems of CN, $^{13}$C$^{14}$N, and $^{12}$C$^{15}$N, and Application to Astronomical Spectra}",
      journal = {\apjs},
         year = 2014,
        month = oct,
       volume = {214},
       number = {2},
          eid = {26},
        pages = {26},
          doi = {10.1088/0067-0049/214/2/26},
archivePrefix = {arXiv},
       eprint = {1408.3828},
 primaryClass = {astro-ph.SR},
       adsurl = {https://ui.adsabs.harvard.edu/abs/2014ApJS..214...26S}
}

@ARTICLE{vazquez2022,
       author = {{Viscasillas V{\'a}zquez}, C. and {Magrini}, L. and {Casali}, G. and {Tautvai{\v{s}}ien{\.{e}}}, G. and {Spina}, L. and {Van der Swaelmen}, M. and {Randich}, S. and {Bensby}, T. and {Bragaglia}, A. and {Friel}, E. and {Feltzing}, S. and {Sacco}, G.~G. and {Turchi}, A. and {Jim{\'e}nez-Esteban}, F. and {D'Orazi}, V. and {Delgado-Mena}, E. and {Mikolaitis}, {\v{S}}. and {Drazdauskas}, A. and {Minkevi{\v{c}}i{\={u}}t{\.{e}}}, R. and {Stonkut{\.{e}}}, E. and {Bagdonas}, V. and {Montes}, D. and {Guiglion}, G. and {Baratella}, M. and {Tabernero}, H.~M. and {Gilmore}, G. and {Alfaro}, E. and {Francois}, P. and {Korn}, A. and {Smiljanic}, R. and {Bergemann}, M. and {Franciosini}, E. and {Gonneau}, A. and {Hourihane}, A. and {Worley}, C.~C. and {Zaggia}, S.},
        title = "{The Gaia-ESO survey: Age-chemical-clock relations spatially resolved in the Galactic disc}",
      journal = {\aap},
         year = 2022,
        month = apr,
       volume = {660},
          eid = {A135},
        pages = {A135},
          doi = {10.1051/0004-6361/202142937},
archivePrefix = {arXiv},
       eprint = {2202.04863},
 primaryClass = {astro-ph.GA},
       adsurl = {https://ui.adsabs.harvard.edu/abs/2022A&A...660A.135V}
}

@ARTICLE{jonsson2014,
       author = {{J{\"o}nsson}, H. and {Ryde}, N. and {Harper}, G.~M. and {Cunha}, K. and {Schultheis}, M. and {Eriksson}, K. and {Kobayashi}, C. and {Smith}, V.~V. and {Zoccali}, M.},
        title = "{Chemical evolution of fluorine in the bulge. High-resolution K-band spectra of giants in three fields}",
      journal = {\aap},
         year = 2014,
        month = apr,
       volume = {564},
          eid = {A122},
        pages = {A122},
          doi = {10.1051/0004-6361/201423597},
archivePrefix = {arXiv},
       eprint = {1403.2594},
 primaryClass = {astro-ph.GA},
       adsurl = {https://ui.adsabs.harvard.edu/abs/2014A&A...564A.122J}
}

@ARTICLE{smith2013,
       author = {{Smith}, Verne V. and {Cunha}, Katia and {Shetrone}, Matthew D. and {Meszaros}, Szabolcs and {Allende Prieto}, Carlos and {Bizyaev}, Dmitry and {Garc{\'\i}a P{\'e}rez}, Ana and {Majewski}, Steven R. and {Schiavon}, Ricardo and {Holtzman}, Jon and {Johnson}, Jennifer A.},
        title = "{Chemical Abundances in Field Red Giants from High-resolution H-band Spectra Using the APOGEE Spectral Linelist}",
      journal = {\apj},
         year = 2013,
        month = mar,
       volume = {765},
       number = {1},
          eid = {16},
        pages = {16},
          doi = {10.1088/0004-637X/765/1/16},
archivePrefix = {arXiv},
       eprint = {1212.4091},
 primaryClass = {astro-ph.SR},
       adsurl = {https://ui.adsabs.harvard.edu/abs/2013ApJ...765...16S}
}

@ARTICLE{pehlivan2015,
       author = {{Pehlivan}, A. and {Nilsson}, H. and {Hartman}, H.},
        title = "{Laboratory oscillator strengths of Sc i in the near-infrared region for astrophysical applications}",
      journal = {\aap},
         year = 2015,
        month = oct,
       volume = {582},
          eid = {A98},
        pages = {A98},
          doi = {10.1051/0004-6361/201526813},
archivePrefix = {arXiv},
       eprint = {1509.06341},
 primaryClass = {astro-ph.IM},
       adsurl = {https://ui.adsabs.harvard.edu/abs/2015A&A...582A..98P}
}

@article{brooke2016,
  title={Line strengths of rovibrational and rotational transitions in the X2$\Pi$ ground state of OH},
  author={Brooke, James SA and Bernath, Peter F and Western, Colin M and Sneden, Christopher and Af{\c{s}}ar, Melike and Li, Gang and Gordon, Iouli E},
  journal={JQSRT},
  volume={168},
  pages={142--157},
  year={2016},
  publisher={Elsevier}
}

@ARTICLE{lebzelder2015,
       author = {{Lebzelter}, T. and {Straniero}, O. and {Hinkle}, K.~H. and {Nowotny}, W. and {Aringer}, B.},
        title = "{Oxygen isotopic ratios in intermediate-mass red giants}",
      journal = {\aap},
         year = 2015,
        month = jun,
       volume = {578},
          eid = {A33},
        pages = {A33},
          doi = {10.1051/0004-6361/201525832},
archivePrefix = {arXiv},
       eprint = {1504.05377},
 primaryClass = {astro-ph.SR},
       adsurl = {https://ui.adsabs.harvard.edu/abs/2015A&A...578A..33L}
}

@ARTICLE{kobayashi2020,
       author = {{Kobayashi}, Chiaki and {Karakas}, Amanda I. and {Lugaro}, Maria},
        title = "{The Origin of Elements from Carbon to Uranium}",
      journal = {\apj},
         year = 2020,
        month = sep,
       volume = {900},
       number = {2},
          eid = {179},
        pages = {179},
          doi = {10.3847/1538-4357/abae65},
archivePrefix = {arXiv},
       eprint = {2008.04660},
 primaryClass = {astro-ph.GA},
       adsurl = {https://ui.adsabs.harvard.edu/abs/2020ApJ...900..179K}
}

@ARTICLE{hasselquist2016,
       author = {{Hasselquist}, Sten and {Shetrone}, Matthew and {Cunha}, Katia and {Smith}, Verne V. and {Holtzman}, Jon and {Lawler}, J.~E. and {Allende Prieto}, Carlos and {Beers}, Timothy C. and {Chojnowski}, Drew and {Fern{\'a}ndez-Trincado}, J.~G. and {Garc{\'\i}a-Hern{\'a}ndez}, D.~A. and {Hearty}, Fred R. and {Majewski}, Steven R. and {Pereira}, C.~B. and {Placco}, Vinicius M. and {Villanova}, Sandro and {Zamora}, Olga},
        title = "{Identification of Neodymium in the Apogee H-Band Spectra}",
      journal = {\apj},
         year = 2016,
        month = dec,
       volume = {833},
       number = {1},
          eid = {81},
        pages = {81},
          doi = {10.3847/1538-4357/833/1/81},
       adsurl = {https://ui.adsabs.harvard.edu/abs/2016ApJ...833...81H}
}

@ARTICLE{yan2021,
       author = {{Yan}, Hong-Liang and {Zhou}, Yu-Tao and {Zhang}, Xianfei and {Li}, Yaguang and {Gao}, Qi and {Shi}, Jian-Rong and {Zhao}, Gang and {Aoki}, Wako and {Matsuno}, Tadafumi and {Li}, Yan and {Xu}, Xiao-Dong and {Li}, Haining and {Wu}, Ya-Qian and {Jin}, Meng-Qi and {Mosser}, Benoit and {Bi}, Shao-Lan and {Fu}, Jian-Ning and {Pan}, Kaike and {Suda}, Takuma and {Liu}, Yu-Juan and {Zhao}, Jing-Kun and {Liang}, Xi-Long},
        title = "{Most lithium-rich low-mass evolved stars revealed as red clump stars by asteroseismology and spectroscopy}",
      journal = {Nature Astronomy},
         year = 2021,
        month = jan,
       volume = {5},
        pages = {86-93},
          doi = {10.1038/s41550-020-01217-8},
archivePrefix = {arXiv},
       eprint = {2010.02106},
 primaryClass = {astro-ph.SR},
       adsurl = {https://ui.adsabs.harvard.edu/abs/2021NatAs...5...86Y}
}

@ARTICLE{nandakumar2022,
       author = {{Nandakumar}, G. and {Ryde}, N. and {Montelius}, M. and {Thorsbro}, B. and {J{\"o}nsson}, H. and {Mace}, G.},
        title = "{The Galactic chemical evolution of phosphorus observed with IGRINS}",
      journal = {\aap},
         year = 2022,
        month = dec,
       volume = {668},
          eid = {A88},
        pages = {A88},
          doi = {10.1051/0004-6361/202244724},
archivePrefix = {arXiv},
       eprint = {2210.04940},
 primaryClass = {astro-ph.SR},
       adsurl = {https://ui.adsabs.harvard.edu/abs/2022A&A...668A..88N}
}

@ARTICLE{placco2021,
       author = {{Placco}, Vinicius M. and {Sneden}, Christopher and {Roederer}, Ian U. and {Lawler}, James E. and {Den Hartog}, Elizabeth A. and {Hejazi}, Neda and {Maas}, Zachary and {Bernath}, Peter},
        title = "{Linemake: An Atomic and Molecular Line List Generator}",
      journal = {Research Notes of the American Astronomical Society},
         year = 2021,
        month = apr,
       volume = {5},
       number = {4},
          eid = {92},
        pages = {92},
          doi = {10.3847/2515-5172/abf651},
archivePrefix = {arXiv},
       eprint = {2104.08286},
 primaryClass = {astro-ph.IM},
       adsurl = {https://ui.adsabs.harvard.edu/abs/2021RNAAS...5...92P}
      }

@ARTICLE{kurucz2011,
       author = {{Kurucz}, Robert L.},
        title = "{Including all the lines}",
      journal = {Canadian Journal of Physics},
         year = 2011,
        month = apr,
       volume = {89},
        pages = {417-428},
          doi = {10.1139/p10-104},
       adsurl = {https://ui.adsabs.harvard.edu/abs/2011CaJPh..89..417K}
}

@ARTICLE{martinez2020,
       author = {{Martinez}, Cintia F. and {Holanda}, N. and {Pereira}, C.~B. and
         {Drake}, N.~A.},
        title = "{High-resolution spectroscopy of red giants and 'yellow stragglers' in the southern open cluster NGC 2539}",
      journal = {\mnras},
         year = 2020,
        month = mar,
       volume = {494},
       number = {1},
        pages = {1470-1489},
          doi = {10.1093/mnras/staa647},
       adsurl = {https://ui.adsabs.harvard.edu/abs/2020MNRAS.494.1470M}
}

@ARTICLE{hunt2023,
       author = {{Hunt}, Emily L. and {Reffert}, Sabine},
        title = "{Improving the open cluster census. II. An all-sky cluster catalogue with Gaia DR3}",
      journal = {\aap},
         year = 2023,
        month = may,
       volume = {673},
          eid = {A114},
        pages = {A114},
          doi = {10.1051/0004-6361/202346285},
archivePrefix = {arXiv},
       eprint = {2303.13424},
 primaryClass = {astro-ph.GA},
       adsurl = {https://ui.adsabs.harvard.edu/abs/2023A&A...673A.114H}
}

@ARTICLE{casali2019,
       author = {{Casali}, G. and {Magrini}, L. and {Tognelli}, E. and {Jackson}, R. and {Jeffries}, R.~D. and {Lagarde}, N. and {Tautvai{\v{s}}ien{\.{e}}}, G. and {Masseron}, T. and {Degl'Innocenti}, S. and {Prada Moroni}, P.~G. and {Kordopatis}, G. and {Pancino}, E. and {Randich}, S. and {Feltzing}, S. and {Sahlholdt}, C. and {Spina}, L. and {Friel}, E. and {Roccatagliata}, V. and {Sanna}, N. and {Bragaglia}, A. and {Drazdauskas}, A. and {Mikolaitis}, {\v{S}}. and {Minkevi{\v{c}}i{\={u}}t{\.{e}}}, R. and {Stonkut{\.{e}}}, E. and {Chorniy}, Y. and {Bagdonas}, V. and {Jimenez-Esteban}, F. and {Martell}, S. and {Van der Swaelmen}, M. and {Gilmore}, G. and {Vallenari}, A. and {Bensby}, T. and {Koposov}, S.~E. and {Korn}, A. and {Worley}, C. and {Smiljanic}, R. and {Bergemann}, M. and {Carraro}, G. and {Damiani}, F. and {Prisinzano}, L. and {Bonito}, R. and {Franciosini}, E. and {Gonneau}, A. and {Hourihane}, A. and {Jofre}, P. and {Lewis}, J. and {Morbidelli}, L. and {Sacco}, G. and {Sousa}, S.~G. and {Zaggia}, S. and {Lanzafame}, A.~C. and {Heiter}, U. and {Frasca}, A. and {Bayo}, A.},
        title = "{The Gaia-ESO survey: Calibrating a relationship between age and the [C/N] abundance ratio with open clusters}",
      journal = {\aap},
         year = 2019,
        month = sep,
       volume = {629},
          eid = {A62},
        pages = {A62},
          doi = {10.1051/0004-6361/201935282},
archivePrefix = {arXiv},
       eprint = {1907.07350},
 primaryClass = {astro-ph.GA},
       adsurl = {https://ui.adsabs.harvard.edu/abs/2019A&A...629A..62C}
}

@ARTICLE{sousa2015,
       author = {{Sousa}, S.~G. and {Santos}, N.~C. and {Adibekyan}, V. and {Delgado-Mena}, E. and {Israelian}, G.},
        title = "{ARES v2: new features and improved performance}",
      journal = {\aap},
         year = 2015,
        month = may,
       volume = {577},
          eid = {A67},
        pages = {A67},
          doi = {10.1051/0004-6361/201425463},
archivePrefix = {arXiv},
       eprint = {1504.02725},
 primaryClass = {astro-ph.IM},
       adsurl = {https://ui.adsabs.harvard.edu/abs/2015A&A...577A..67S}
}

@ARTICLE{pace2010,
       author = {{Pace}, G. and {Danziger}, J. and {Carraro}, G. and {Melendez}, J. and {Fran{\c{c}}ois}, P. and {Matteucci}, F. and {Santos}, N.~C.},
        title = "{Abundances and physical parameters for stars in the open clusters NGC 5822 and IC 4756}",
      journal = {\aap},
         year = 2010,
        month = jun,
       volume = {515},
          eid = {A28},
        pages = {A28},
          doi = {10.1051/0004-6361/200913029},
archivePrefix = {arXiv},
       eprint = {1002.2547},
 primaryClass = {astro-ph.SR},
       adsurl = {https://ui.adsabs.harvard.edu/abs/2010A&A...515A..28P}
}

@ARTICLE{sales2014,
       author = {{Sales Silva}, J.~V. and {Pe{\~n}a Su{\'a}rez}, V.~J. and {Katime Santrich}, O.~J. and {Pereira}, C.~B. and {Drake}, N.~A. and {Roig}, F.},
        title = "{High-resolution Spectroscopic Observations of Binary Stars and Yellow Stragglers in Three Open Clusters : NGC 2360, NGC 3680, and NGC 5822}",
      journal = {\aj},
         year = 2014,
        month = nov,
       volume = {148},
       number = {5},
          eid = {83},
        pages = {83},
          doi = {10.1088/0004-6256/148/5/83},
       adsurl = {https://ui.adsabs.harvard.edu/abs/2014AJ....148...83S}
}

@ARTICLE{pera2021,
       author = {{Pera}, M.~S. and {Perren}, G.~I. and {Moitinho}, A. and {Navone}, H.~D. and {Vazquez}, R.~A.},
        title = "{pyUPMASK: an improved unsupervised clustering algorithm}",
      journal = {\aap},
         year = 2021,
        month = jun,
       volume = {650},
          eid = {A109},
        pages = {A109},
          doi = {10.1051/0004-6361/202040252},
archivePrefix = {arXiv},
       eprint = {2101.01660},
 primaryClass = {astro-ph.GA},
       adsurl = {https://ui.adsabs.harvard.edu/abs/2021A&A...650A.109P}
}

@ARTICLE{sales2022,
       author = {{Sales-Silva}, J.~V. and {Daflon}, S. and {Cunha}, K. and {Souto}, D. and {Smith}, V.~V. and {Chiappini}, C. and {Donor}, J. and {Frinchaboy}, P.~M. and {Garc{\'\i}a-Hern{\'a}ndez}, D.~A. and {Hayes}, C. and {Majewski}, S.~R. and {Masseron}, T. and {Schiavon}, R.~P. and {Weinberg}, D.~H. and {Beaton}, R.~L. and {Fern{\'a}ndez-Trincado}, J.~G. and {J{\"o}nsson}, H. and {Lane}, R.~R. and {Minniti}, D. and {Manchado}, A. and {Moni Bidin}, C. and {Nitschelm}, C. and {O'Connell}, J. and {Villanova}, S.},
        title = "{Exploring the S-process History in the Galactic Disk: Cerium Abundances and Gradients in Open Clusters from the OCCAM/APOGEE Sample}",
      journal = {\apj},
         year = 2022,
        month = feb,
       volume = {926},
       number = {2},
          eid = {154},
        pages = {154},
          doi = {10.3847/1538-4357/ac4254},
archivePrefix = {arXiv},
       eprint = {2112.02196},
 primaryClass = {astro-ph.GA},
       adsurl = {https://ui.adsabs.harvard.edu/abs/2022ApJ...926..154S}
}

@ARTICLE{randich2022,
       author = {{Randich}, S. and {Gilmore}, G. and {Magrini}, L. and {Sacco}, G.~G. and {Jackson}, R.~J. and {Jeffries}, R.~D. and {Worley}, C.~C. and {Hourihane}, A. and {Gonneau}, A. and {Viscasillas Vazquez}, C. and {Franciosini}, E. and {Lewis}, J.~R. and {Alfaro}, E.~J. and {Allende Prieto}, C. and {Bensby}, T. and {Blomme}, R. and {Bragaglia}, A. and {Flaccomio}, E. and {Fran{\c{c}}ois}, P. and {Irwin}, M.~J. and {Koposov}, S.~E. and {Korn}, A.~J. and {Lanzafame}, A.~C. and {Pancino}, E. and {Recio-Blanco}, A. and {Smiljanic}, R. and {Van Eck}, S. and {Zwitter}, T. and {Asplund}, M. and {Bonifacio}, P. and {Feltzing}, S. and {Binney}, J. and {Drew}, J. and {Ferguson}, A.~M.~N. and {Micela}, G. and {Negueruela}, I. and {Prusti}, T. and {Rix}, H. -W. and {Vallenari}, A. and {Bayo}, A. and {Bergemann}, M. and {Biazzo}, K. and {Carraro}, G. and {Casey}, A.~R. and {Damiani}, F. and {Frasca}, A. and {Heiter}, U. and {Hill}, V. and {Jofr{\'e}}, P. and {de Laverny}, P. and {Lind}, K. and {Marconi}, G. and {Martayan}, C. and {Masseron}, T. and {Monaco}, L. and {Morbidelli}, L. and {Prisinzano}, L. and {Sbordone}, L. and {Sousa}, S.~G. and {Zaggia}, S. and {Adibekyan}, V. and {Bonito}, R. and {Caffau}, E. and {Daflon}, S. and {Feuillet}, D.~K. and {Gebran}, M. and {Gonzalez Hernandez}, J.~I. and {Guiglion}, G. and {Herrero}, A. and {Lobel}, A. and {Maiz Apellaniz}, J. and {Merle}, T. and {Mikolaitis}, {\v{S}}. and {Montes}, D. and {Morel}, T. and {Soubiran}, C. and {Spina}, L. and {Tabernero}, H.~M. and {Tautvai{\v{s}}iene}, G. and {Traven}, G. and {Valentini}, M. and {Van der Swaelmen}, M. and {Villanova}, S. and {Wright}, N.~J. and {Abbas}, U. and {Aguirre B{\o}rsen-Koch}, V. and {Alves}, J. and {Balaguer-Nunez}, L. and {Barklem}, P.~S. and {Barrado}, D. and {Berlanas}, S.~R. and {Binks}, A.~S. and {Bressan}, A. and {Capuzzo-Dolcetta}, R. and {Casagrande}, L. and {Casamiquela}, L. and {Collins}, R.~S. and {D'Orazi}, V. and {Dantas}, M.~L.~L. and {Debattista}, V.~P. and {Delgado-Mena}, E. and {Di Marcantonio}, P. and {Drazdauskas}, A. and {Evans}, N.~W. and {Famaey}, B. and {Franchini}, M. and {Fr{\'e}mat}, Y. and {Friel}, E.~D. and {Fu}, X. and {Geisler}, D. and {Gerhard}, O. and {Gonzalez Solares}, E.~A. and {Grebel}, E.~K. and {Gutierrez Albarran}, M.~L. and {Hatzidimitriou}, D. and {Held}, E.~V. and {Jim{\'e}nez-Esteban}, F. and {J{\"o}nsson}, H. and {Jordi}, C. and {Khachaturyants}, T. and {Kordopatis}, G. and {Kos}, J. and {Lagarde}, N. and {Mahy}, L. and {Mapelli}, M. and {Marfil}, E. and {Martell}, S.~L. and {Messina}, S. and {Miglio}, A. and {Minchev}, I. and {Moitinho}, A. and {Montalban}, J. and {Monteiro}, M.~J.~P.~F.~G. and {Morossi}, C. and {Mowlavi}, N. and {Mucciarelli}, A. and {Murphy}, D.~N.~A. and {Nardetto}, N. and {Ortolani}, S. and {Paletou}, F. and {Palou{\v{s}}}, J. and {Paunzen}, E. and {Pickering}, J.~C. and {Quirrenbach}, A. and {Re Fiorentin}, P. and {Read}, J.~I. and {Romano}, D. and {Ryde}, N. and {Sanna}, N. and {Santos}, W. and {Seabroke}, G.~M. and {Spagna}, A. and {Steinmetz}, M. and {Stonkut{\'e}}, E. and {Sutorius}, E. and {Th{\'e}venin}, F. and {Tosi}, M. and {Tsantaki}, M. and {Vink}, J.~S. and {Wright}, N. and {Wyse}, R.~F.~G. and {Zoccali}, M. and {Zorec}, J. and {Zucker}, D.~B. and {Walton}, N.~A.},
        title = "{The Gaia-ESO Public Spectroscopic Survey: Implementation, data products, open cluster survey, science, and legacy}",
      journal = {\aap},
         year = 2022,
        month = oct,
       volume = {666},
          eid = {A121},
        pages = {A121},
          doi = {10.1051/0004-6361/202243141},
archivePrefix = {arXiv},
       eprint = {2206.02901},
 primaryClass = {astro-ph.GA},
       adsurl = {https://ui.adsabs.harvard.edu/abs/2022A&A...666A.121R}
}

@ARTICLE{dias2021,
       author = {{Dias}, W.~S. and {Monteiro}, H. and {Moitinho}, A. and {L{\'e}pine}, J.~R.~D. and {Carraro}, G. and {Paunzen}, E. and {Alessi}, B. and {Villela}, L.},
        title = "{Updated parameters of 1743 open clusters based on Gaia DR2}",
      journal = {\mnras},
         year = 2021,
        month = jun,
       volume = {504},
       number = {1},
        pages = {356-371},
          doi = {10.1093/mnras/stab770},
archivePrefix = {arXiv},
       eprint = {2103.12829},
 primaryClass = {astro-ph.SR},
       adsurl = {https://ui.adsabs.harvard.edu/abs/2021MNRAS.504..356D}
}

@ARTICLE{nandakumar2023,
       author = {{Nandakumar}, G. and {Ryde}, N. and {Mace}, G.},
        title = "{M Giants with IGRINS II. Chemical Evolution of Fluorine at High Metallicities}",
      journal = {arXiv e-prints},
         year = 2023,
        month = jun,
          eid = {arXiv:2306.08446},
        pages = {arXiv:2306.08446},
          doi = {10.48550/arXiv.2306.08446},
archivePrefix = {arXiv},
       eprint = {2306.08446},
 primaryClass = {astro-ph.GA},
       adsurl = {https://ui.adsabs.harvard.edu/abs/2023arXiv230608446N}
}

@ARTICLE{bensby2014,
       author = {{Bensby}, T. and {Feltzing}, S. and {Oey}, M.~S.},
        title = "{Exploring the Milky Way stellar disk. A detailed elemental abundance study of 714 F and G dwarf stars in the solar neighbourhood}",
      journal = {\aap},
         year = 2014,
        month = feb,
       volume = {562},
          eid = {A71},
        pages = {A71},
          doi = {10.1051/0004-6361/201322631},
archivePrefix = {arXiv},
       eprint = {1309.2631},
 primaryClass = {astro-ph.GA},
       adsurl = {https://ui.adsabs.harvard.edu/abs/2014A&A...562A..71B}
}

@ARTICLE{battistini2015,
       author = {{Battistini}, Chiara and {Bensby}, Thomas},
        title = "{The origin and evolution of the odd-Z iron-peak elements Sc, V, Mn, and Co in the Milky Way stellar disk}",
      journal = {\aap},
         year = 2015,
        month = may,
       volume = {577},
          eid = {A9},
        pages = {A9},
          doi = {10.1051/0004-6361/201425327},
archivePrefix = {arXiv},
       eprint = {1502.01152},
 primaryClass = {astro-ph.GA},
       adsurl = {https://ui.adsabs.harvard.edu/abs/2015A&A...577A...9B}
}

@ARTICLE{burbidge1957,
       author = {{Burbidge}, E. Margaret and {Burbidge}, G.~R. and {Fowler}, William A. and {Hoyle}, F.},
        title = "{Synthesis of the Elements in Stars}",
      journal = {Reviews of Modern Physics},
         year = 1957,
        month = jan,
       volume = {29},
       number = {4},
        pages = {547-650},
          doi = {10.1103/RevModPhys.29.547},
       adsurl = {https://ui.adsabs.harvard.edu/abs/1957RvMP...29..547B}
}

@ARTICLE{mermilliod2008,
       author = {{Mermilliod}, J.~C. and {Mayor}, M. and {Udry}, S.},
        title = "{Red giants in open clusters. XIV. Mean radial velocities for 1309 stars and 166 open clusters}",
      journal = {\aap},
         year = 2008,
        month = jul,
       volume = {485},
       number = {1},
        pages = {303-314},
          doi = {10.1051/0004-6361:200809664},
       adsurl = {https://ui.adsabs.harvard.edu/abs/2008A&A...485..303M}
}

@ARTICLE{penasuarez2018,
       author = {{Pe{\~n}a Su{\'a}rez}, V.~J. and {Sales Silva}, J.~V. and
         {Katime Santrich}, O.~J. and {Drake}, N.~A. and {Pereira}, C.~B.},
        title = "{High-resolution Spectroscopic Observations of Single Red Giants in Three Open Clusters: NGC 2360, NGC 3680, and NGC 5822}",
      journal = {\apj},
         year = 2018,
        month = feb,
       volume = {854},
       number = {2},
          eid = {184},
        pages = {184},
          doi = {10.3847/1538-4357/aaa017},
       adsurl = {https://ui.adsabs.harvard.edu/abs/2018ApJ...854..184P}
}

@ARTICLE{dasilveira2018,
       author = {{da Silveira}, M.~D. and {Pereira}, C.~B. and {Drake}, N.~A.},
        title = "{Red giants and yellow stragglers in the young open cluster NGC 2447}",
      journal = {\mnras},
         year = 2018,
        month = jun,
       volume = {476},
       number = {4},
        pages = {4907-4931},
          doi = {10.1093/mnras/sty265},
       adsurl = {https://ui.adsabs.harvard.edu/abs/2018MNRAS.476.4907D}
}

@ARTICLE{battistini2016,
       author = {{Battistini}, Chiara and {Bensby}, Thomas},
        title = "{The origin and evolution of r- and s-process elements in the Milky Way stellar disk}",
      journal = {\aap},
         year = 2016,
        month = feb,
       volume = {586},
          eid = {A49},
        pages = {A49},
          doi = {10.1051/0004-6361/201527385},
archivePrefix = {arXiv},
       eprint = {1511.00966},
 primaryClass = {astro-ph.SR},
       adsurl = {https://ui.adsabs.harvard.edu/abs/2016A&A...586A..49B}
}

@ARTICLE{perren2015,
       author = {{Perren}, G.~I. and {V{\'a}zquez}, R.~A. and {Piatti}, A.~E.},
        title = "{ASteCA: Automated Stellar Cluster Analysis}",
      journal = {\aap},
         year = 2015,
        month = apr,
       volume = {576},
          eid = {A6},
        pages = {A6},
          doi = {10.1051/0004-6361/201424946},
archivePrefix = {arXiv},
       eprint = {1412.2366},
 primaryClass = {astro-ph.GA},
       adsurl = {https://ui.adsabs.harvard.edu/abs/2015A&A...576A...6P}
}

@ARTICLE{mishenina2006,
       author = {{Mishenina}, T.~V. and {Bienaym{\'e}}, O. and {Gorbaneva}, T.~I. and
         {Charbonnel}, C. and {Soubiran}, C. and {Korotin}, S.~A. and
         {Kovtyukh}, V.~V.},
        title = "{Elemental abundances in the atmosphere of clump giants}",
      journal = {\aap},
         year = 2006,
        month = sep,
       volume = {456},
       number = {3},
        pages = {1109-1120},
          doi = {10.1051/0004-6361:20065141},
archivePrefix = {arXiv},
       eprint = {astro-ph/0605615},
 primaryClass = {astro-ph},
       adsurl = {https://ui.adsabs.harvard.edu/abs/2006A&A...456.1109M}
}

@ARTICLE{wyller1966,
       author = {{Wyller}, Arne A.},
        title = "{New C\^\{13\} Indicators in Stellar Spectra}",
      journal = {\apj},
         year = 1966,
        month = mar,
       volume = {143},
        pages = {828},
          doi = {10.1086/148560},
       adsurl = {https://ui.adsabs.harvard.edu/abs/1966ApJ...143..828W}
}

@BOOK{davis1963,
       author = {{Davis}, Sumner P. and {Phillips}, John G.},
        title = "{The red system (A2[pi]-X2[Sigma]) of the CN molecule}",
         year = 1963,
    publisher = "Univ. California Press",
       adsurl = {https://ui.adsabs.harvard.edu/abs/1963rspx.book.....D}
}

@ARTICLE{sneden1982,
       author = {{Sneden}, C. and {Lambert}, D.~L.},
        title = "{The CN red system in the solar spectrum}",
      journal = {\apj},
         year = 1982,
        month = aug,
       volume = {259},
        pages = {381-391},
          doi = {10.1086/160175},
       adsurl = {https://ui.adsabs.harvard.edu/abs/1982ApJ...259..381S}
}

@article{huber1979,
  title={IV. Constants of diatomic molecules},
  author={Huber, KP and Herzberg, G},
  journal={Molecular Spectra and Molecular Structure. New York, USA: Van Nostrand Reinhold Company},
  year={1979}
}

@ARTICLE{santrich2013,
       author = {{Katime Santrich}, O.~J. and {Pereira}, C.~B. and {Drake}, N.~A.},
        title = "{Chemical analysis of giant stars in the young open cluster NGC 3114}",
      journal = {\aap},
         year = 2013,
        month = jun,
       volume = {554},
          eid = {A2},
        pages = {A2},
          doi = {10.1051/0004-6361/201220252},
archivePrefix = {arXiv},
       eprint = {1304.1004},
 primaryClass = {astro-ph.SR},
       adsurl = {https://ui.adsabs.harvard.edu/abs/2013A&A...554A...2S}
}

@ARTICLE{delgadomena2016,
       author = {{Delgado Mena}, E. and {Tsantaki}, M. and {Sousa}, S.~G. and
         {Kunitomo}, M. and {Adibekyan}, V. and {Zaworska}, P. and
         {Santos}, N.~C. and {Israelian}, G. and {Lovis}, C.},
        title = "{Searching for Li-rich giants in a sample of 12 open clusters. Li enhancement in two stars with substellar companions}",
      journal = {\aap},
         year = 2016,
        month = mar,
       volume = {587},
          eid = {A66},
        pages = {A66},
          doi = {10.1051/0004-6361/201527196},
archivePrefix = {arXiv},
       eprint = {1512.05296},
 primaryClass = {astro-ph.SR},
       adsurl = {https://ui.adsabs.harvard.edu/abs/2016A&A...587A..66D}
}

@ARTICLE{karakas2014,
       author = {{Karakas}, Amanda I. and {Lattanzio}, John C.},
        title = "{The Dawes Review 2: Nucleosynthesis and Stellar Yields of Low- and Intermediate-Mass Single Stars}",
      journal = {\pasa},
         year = 2014,
        month = jul,
       volume = {31},
          eid = {e030},
        pages = {e030},
          doi = {10.1017/pasa.2014.21},
archivePrefix = {arXiv},
       eprint = {1405.0062},
 primaryClass = {astro-ph.SR},
       adsurl = {https://ui.adsabs.harvard.edu/abs/2014PASA...31...30K}
}

@ARTICLE{busso1999,
       author = {{Busso}, M. and {Gallino}, R. and {Wasserburg}, G.~J.},
        title = "{Nucleosynthesis in Asymptotic Giant Branch Stars: Relevance for Galactic Enrichment and Solar System Formation}",
      journal = {\araa},
         year = 1999,
        month = jan,
       volume = {37},
        pages = {239-309},
          doi = {10.1146/annurev.astro.37.1.239},
       adsurl = {https://ui.adsabs.harvard.edu/abs/1999ARA&A..37..239B}
}

@ARTICLE{lambert1978,
       author = {{Lambert}, D.~L.},
        title = "{The abundances of the elements in the solar photosphere - VIII. Revised abundances of carbon, nitrogen and oxygen.}",
      journal = {\mnras},
         year = "1978",
        month = "Jan",
       volume = {182},
        pages = {249-272},
          doi = {10.1093/mnras/182.2.249},
       adsurl = {https://ui.adsabs.harvard.edu/abs/1978MNRAS.182..249L}
}

@ARTICLE{holanda2019,
       author = {{Holanda}, N. and {Pereira}, C.~B. and {Drake}, N.~A.},
        title = "{Chemical analysis of K giants in the young open cluster NGC 2345}",
      journal = {\mnras},
         year = "2019",
        month = "Feb",
       volume = {482},
       number = {4},
        pages = {5275-5289},
          doi = {10.1093/mnras/sty2991},
       adsurl = {https://ui.adsabs.harvard.edu/abs/2019MNRAS.482.5275H}
}

@ARTICLE{spitoni2019,
       author = {{Spitoni}, E. and {Silva Aguirre}, V. and {Matteucci}, F. and {Calura}, F. and {Grisoni}, V.},
        title = "{Galactic Archaeology with asteroseismic ages: Evidence for delayed gas infall in the formation of the Milky Way disc}",
      journal = {\aap},
         year = 2019,
        month = mar,
       volume = {623},
          eid = {A60},
        pages = {A60},
          doi = {10.1051/0004-6361/201834188},
archivePrefix = {arXiv},
       eprint = {1809.00914},
 primaryClass = {astro-ph.GA},
       adsurl = {https://ui.adsabs.harvard.edu/abs/2019A&A...623A..60S}
}

@ARTICLE{holanda2020b,
       author = {{Holanda}, N. and {Drake}, N.~A. and {Pereira}, C.~B.},
        title = "{TYC 8327 - 1678 - 1: A new super lithium-rich K giant}",
      journal = {\mnras},
         year = 2020,
        month = aug,
          doi = {10.1093/mnras/staa2271},
       adsurl = {https://ui.adsabs.harvard.edu/abs/2020MNRAS.tmp.2307H}
}

@ARTICLE{alonso1999,
       author = {{Alonso}, A. and {Arribas}, S. and {Mart{\'\i}nez-Roger}, C.},
        title = "{The effective temperature scale of giant stars (F0-K5). II. Empirical calibration of T$_{eff}$ versus colours and [Fe/H]}",
      journal = {\aaps},
         year = "1999",
        month = "Dec",
       volume = {140},
        pages = {261-277},
          doi = {10.1051/aas:1999521},
       adsurl = {https://ui.adsabs.harvard.edu/abs/1999A&AS..140..261A}
}

@ARTICLE{romano2019,
       author = {{Romano}, Donatella and {Matteucci}, Francesca and {Zhang}, Zhi-Yu and {Ivison}, Rob J. and {Ventura}, Paolo},
        title = "{The evolution of CNO isotopes: the impact of massive stellar rotators}",
      journal = {\mnras},
         year = 2019,
        month = dec,
       volume = {490},
       number = {2},
        pages = {2838-2854},
          doi = {10.1093/mnras/stz2741},
archivePrefix = {arXiv},
       eprint = {1907.09476},
 primaryClass = {astro-ph.GA},
       adsurl = {https://ui.adsabs.harvard.edu/abs/2019MNRAS.490.2838R}
}

@ARTICLE{prantzos2018,
       author = {{Prantzos}, N. and {Abia}, C. and {Limongi}, M. and {Chieffi}, A. and {Cristallo}, S.},
        title = "{Chemical evolution with rotating massive star yields - I. The solar neighbourhood and the s-process elements}",
      journal = {\mnras},
         year = 2018,
        month = may,
       volume = {476},
       number = {3},
        pages = {3432-3459},
          doi = {10.1093/mnras/sty316},
archivePrefix = {arXiv},
       eprint = {1802.02824},
 primaryClass = {astro-ph.GA},
       adsurl = {https://ui.adsabs.harvard.edu/abs/2018MNRAS.476.3432P}
}

@ARTICLE{limongi2018,
       author = {{Limongi}, Marco and {Chieffi}, Alessandro},
        title = "{Presupernova Evolution and Explosive Nucleosynthesis of Rotating Massive Stars in the Metallicity Range -3 {\ensuremath{\leq}} [Fe/H] {\ensuremath{\leq}} 0}",
      journal = {\apjs},
         year = 2018,
        month = jul,
       volume = {237},
       number = {1},
          eid = {13},
        pages = {13},
          doi = {10.3847/1538-4365/aacb24},
archivePrefix = {arXiv},
       eprint = {1805.09640},
 primaryClass = {astro-ph.SR},
       adsurl = {https://ui.adsabs.harvard.edu/abs/2018ApJS..237...13L}
}

@ARTICLE{katime2022,
       author = {{Katime Santrich}, Orlando J. and {Kerber}, Leandro and {Abuchaim}, Yuri and {Gon{\c{c}}alves}, Geraldo},
        title = "{On the validity of the spectroscopic age indicators [Y/Mg], [Y/Al], [Y/Si], [Y/Ca], and [Y/Ti] for giant stars}",
      journal = {\mnras},
         year = 2022,
        month = aug,
       volume = {514},
       number = {4},
        pages = {4816-4827},
          doi = {10.1093/mnras/stac1183},
archivePrefix = {arXiv},
       eprint = {2208.12891},
 primaryClass = {astro-ph.SR},
       adsurl = {https://ui.adsabs.harvard.edu/abs/2022MNRAS.514.4816K}
}

@ARTICLE{alonso2020,
       author = {{Alonso-Santiago}, J. and {Negueruela}, I. and {Marco}, A. and {Tabernero}, H.~M. and {Castro}, N.},
        title = "{Three open clusters containing Cepheids: NGC 6649, NGC 6664, and Berkeley 55}",
      journal = {\aap},
         year = 2020,
        month = dec,
       volume = {644},
          eid = {A136},
        pages = {A136},
          doi = {10.1051/0004-6361/202038495},
archivePrefix = {arXiv},
       eprint = {2009.12418},
 primaryClass = {astro-ph.GA},
       adsurl = {https://ui.adsabs.harvard.edu/abs/2020A&A...644A.136A}
}

@ARTICLE{tsantaki2023,
       author = {{Tsantaki}, M. and {Delgado-Mena}, E. and {Bossini}, D. and {Sousa}, S.~G. and {Pancino}, E. and {Martins}, J.~H.~C.},
        title = "{Search for lithium-rich giants in 32 open clusters with high-resolution spectroscopy}",
      journal = {arXiv e-prints},
         year = 2023,
        month = mar,
          eid = {arXiv:2303.16124},
        pages = {arXiv:2303.16124},
          doi = {10.48550/arXiv.2303.16124},
archivePrefix = {arXiv},
       eprint = {2303.16124},
 primaryClass = {astro-ph.SR},
       adsurl = {https://ui.adsabs.harvard.edu/abs/2023arXiv230316124T}
}

@ARTICLE{cantat2020,
       author = {{Cantat-Gaudin}, T. and {Anders}, F. and {Castro-Ginard}, A. and {Jordi}, C. and {Romero-G{\'o}mez}, M. and {Soubiran}, C. and {Casamiquela}, L. and {Tarricq}, Y. and {Moitinho}, A. and {Vallenari}, A. and {Bragaglia}, A. and {Krone-Martins}, A. and {Kounkel}, M.},
        title = "{Painting a portrait of the Galactic disc with its stellar clusters}",
      journal = {\aap},
         year = 2020,
        month = aug,
       volume = {640},
          eid = {A1},
        pages = {A1},
          doi = {10.1051/0004-6361/202038192},
archivePrefix = {arXiv},
       eprint = {2004.07274},
 primaryClass = {astro-ph.GA},
       adsurl = {https://ui.adsabs.harvard.edu/abs/2020A&A...640A...1C}
}

@INPROCEEDINGS{tody1986,
       author = {{Tody}, Doug},
        title = "{The IRAF Data Reduction and Analysis System}",
    booktitle = {Instrumentation in astronomy VI},
         year = 1986,
       editor = {{Crawford}, David L.},
       series = {Society of Photo-Optical Instrumentation Engineers (SPIE) Conference Series},
       volume = {627},
        month = jan,
        pages = {733},
          doi = {10.1117/12.968154},
       adsurl = {https://ui.adsabs.harvard.edu/abs/1986SPIE..627..733T}
}

@ARTICLE{krone2014,
       author = {{Krone-Martins}, A. and {Moitinho}, A.},
        title = "{UPMASK: unsupervised photometric membership assignment in stellar clusters}",
      journal = {\aap},
         year = 2014,
        month = jan,
       volume = {561},
          eid = {A57},
        pages = {A57},
          doi = {10.1051/0004-6361/201321143},
archivePrefix = {arXiv},
       eprint = {1309.4471},
 primaryClass = {astro-ph.IM},
       adsurl = {https://ui.adsabs.harvard.edu/abs/2014A&A...561A..57K}
}

@INPROCEEDINGS{yuk2010,
       author = {{Yuk}, In-Soo and {Jaffe}, Daniel T. and {Barnes}, Stuart and {Chun}, Moo-Young and {Park}, Chan and {Lee}, Sungho and {Lee}, Hanshin and {Wang}, Weisong and {Park}, Kwi-Jong and {Pak}, Soojong and {Strubhar}, Joseph and {Deen}, Casey and {Oh}, Heeyoung and {Seo}, Haingja and {Pyo}, Tae-Soo and {Park}, Won-Kee and {Lacy}, John and {Goertz}, John and {Rand}, Jared and {Gully-Santiago}, Michael},
        title = "{Preliminary design of IGRINS (Immersion GRating INfrared Spectrograph)}",
    booktitle = {Ground-based and Airborne Instrumentation for Astronomy III},
         year = 2010,
       editor = {{McLean}, Ian S. and {Ramsay}, Suzanne K. and {Takami}, Hideki},
       series = {Society of Photo-Optical Instrumentation Engineers (SPIE) Conference Series},
       volume = {7735},
        month = jul,
          eid = {77351M},
        pages = {77351M},
          doi = {10.1117/12.856864},
       adsurl = {https://ui.adsabs.harvard.edu/abs/2010SPIE.7735E..1MY}
}

@ARTICLE{mashonkina2012,
       author = {{Mashonkina}, L. and {Ryabtsev}, A. and {Frebel}, A.},
        title = "{Non-LTE effects on the lead and thorium abundance determinations for cool stars}",
      journal = {\aap},
         year = 2012,
        month = apr,
       volume = {540},
          eid = {A98},
        pages = {A98},
          doi = {10.1051/0004-6361/201218790},
archivePrefix = {arXiv},
       eprint = {1202.2630},
 primaryClass = {astro-ph.GA},
       adsurl = {https://ui.adsabs.harvard.edu/abs/2012A&A...540A..98M}
}

@ARTICLE{zhang2020,
       author = {{Zhang}, Xianfei and {Jeffery}, C. Simon and {Li}, Yaguang and
         {Bi}, Shaolan},
        title = "{Population Synthesis of Helium White Dwarf─Red Giant Star Mergers and the Formation of Lithium-rich Giants and Carbon Stars}",
      journal = {\apj},
         year = "2020",
        month = "Jan",
       volume = {889},
       number = {1},
          eid = {33},
        pages = {33},
          doi = {10.3847/1538-4357/ab5e89},
archivePrefix = {arXiv},
       eprint = {2001.05600},
 primaryClass = {astro-ph.SR},
       adsurl = {https://ui.adsabs.harvard.edu/abs/2020ApJ...889...33Z}
}

@ARTICLE{bisterzo2014,
       author = {{Bisterzo}, S. and {Travaglio}, C. and {Gallino}, R. and {Wiescher}, M. and {K{\"a}ppeler}, F.},
        title = "{Galactic Chemical Evolution and Solar s-process Abundances: Dependence on the $^{13}$C-pocket Structure}",
      journal = {\apj},
         year = 2014,
        month = may,
       volume = {787},
       number = {1},
          eid = {10},
        pages = {10},
          doi = {10.1088/0004-637X/787/1/10},
archivePrefix = {arXiv},
       eprint = {1403.1764},
 primaryClass = {astro-ph.SR},
       adsurl = {https://ui.adsabs.harvard.edu/abs/2014ApJ...787...10B}
}

@ARTICLE{lindegren2021,
       author = {{Lindegren}, L. and {Klioner}, S.~A. and {Hern{\'a}ndez}, J. and {Bombrun}, A. and {Ramos-Lerate}, M. and {Steidelm{\"u}ller}, H. and {Bastian}, U. and {Biermann}, M. and {de Torres}, A. and {Gerlach}, E. and {Geyer}, R. and {Hilger}, T. and {Hobbs}, D. and {Lammers}, U. and {McMillan}, P.~J. and {Stephenson}, C.~A. and {Casta{\~n}eda}, J. and {Davidson}, M. and {Fabricius}, C. and {Gracia-Abril}, G. and {Portell}, J. and {Rowell}, N. and {Teyssier}, D. and {Torra}, F. and {Bartolom{\'e}}, S. and {Clotet}, M. and {Garralda}, N. and {Gonz{\'a}lez-Vidal}, J.~J. and {Torra}, J. and {Abbas}, U. and {Altmann}, M. and {Anglada Varela}, E. and {Balaguer-N{\'u}{\~n}ez}, L. and {Balog}, Z. and {Barache}, C. and {Becciani}, U. and {Bernet}, M. and {Bertone}, S. and {Bianchi}, L. and {Bouquillon}, S. and {Brown}, A.~G.~A. and {Bucciarelli}, B. and {Busonero}, D. and {Butkevich}, A.~G. and {Buzzi}, R. and {Cancelliere}, R. and {Carlucci}, T. and {Charlot}, P. and {Cioni}, M. -R.~L. and {Crosta}, M. and {Crowley}, C. and {del Peloso}, E.~F. and {del Pozo}, E. and {Drimmel}, R. and {Esquej}, P. and {Fienga}, A. and {Fraile}, E. and {Gai}, M. and {Garcia-Reinaldos}, M. and {Guerra}, R. and {Hambly}, N.~C. and {Hauser}, M. and {Jan{\ss}en}, K. and {Jordan}, S. and {Kostrzewa-Rutkowska}, Z. and {Lattanzi}, M.~G. and {Liao}, S. and {Licata}, E. and {Lister}, T.~A. and {L{\"o}ffler}, W. and {Marchant}, J.~M. and {Masip}, A. and {Mignard}, F. and {Mints}, A. and {Molina}, D. and {Mora}, A. and {Morbidelli}, R. and {Murphy}, C.~P. and {Pagani}, C. and {Panuzzo}, P. and {Pe{\~n}alosa Esteller}, X. and {Poggio}, E. and {Re Fiorentin}, P. and {Riva}, A. and {Sagrist{\`a} Sell{\'e}s}, A. and {Sanchez Gimenez}, V. and {Sarasso}, M. and {Sciacca}, E. and {Siddiqui}, H.~I. and {Smart}, R.~L. and {Souami}, D. and {Spagna}, A. and {Steele}, I.~A. and {Taris}, F. and {Utrilla}, E. and {van Reeven}, W. and {Vecchiato}, A.},
        title = "{Gaia Early Data Release 3. The astrometric solution}",
      journal = {\aap},
         year = 2021,
        month = may,
       volume = {649},
          eid = {A2},
        pages = {A2},
          doi = {10.1051/0004-6361/202039709},
archivePrefix = {arXiv},
       eprint = {2012.03380},
 primaryClass = {astro-ph.IM},
       adsurl = {https://ui.adsabs.harvard.edu/abs/2021A&A...649A...2L}
}

@ARTICLE{gaia2023,
       author = {{Gaia Collaboration} and {Vallenari}, A. and {Brown}, A.~G.~A. and {Prusti}, T. and {de Bruijne}, J.~H.~J. and {Arenou}, F. and {Babusiaux}, C. and {Biermann}, M. and {Creevey}, O.~L. and {Ducourant}, C. and {Evans}, D.~W. and {Eyer}, L. and {Guerra}, R. and {Hutton}, A. and {Jordi}, C. and {Klioner}, S.~A. and {Lammers}, U.~L. and {Lindegren}, L. and {Luri}, X. and {Mignard}, F. and {Panem}, C. and {Pourbaix}, D. and {Randich}, S. and {Sartoretti}, P. and {Soubiran}, C. and {Tanga}, P. and {Walton}, N.~A. and {Bailer-Jones}, C.~A.~L. and {Bastian}, U. and {Drimmel}, R. and {Jansen}, F. and {Katz}, D. and {Lattanzi}, M.~G. and {van Leeuwen}, F. and {Bakker}, J. and {Cacciari}, C. and {Casta{\~n}eda}, J. and {De Angeli}, F. and {Fabricius}, C. and {Fouesneau}, M. and {Fr{\'e}mat}, Y. and {Galluccio}, L. and {Guerrier}, A. and {Heiter}, U. and {Masana}, E. and {Messineo}, R. and {Mowlavi}, N. and {Nicolas}, C. and {Nienartowicz}, K. and {Pailler}, F. and {Panuzzo}, P. and {Riclet}, F. and {Roux}, W. and {Seabroke}, G.~M. and {Sordo}, R. and {Th{\'e}venin}, F. and {Gracia-Abril}, G. and {Portell}, J. and {Teyssier}, D. and {Altmann}, M. and {Andrae}, R. and {Audard}, M. and {Bellas-Velidis}, I. and {Benson}, K. and {Berthier}, J. and {Blomme}, R. and {Burgess}, P.~W. and {Busonero}, D. and {Busso}, G. and {C{\'a}novas}, H. and {Carry}, B. and {Cellino}, A. and {Cheek}, N. and {Clementini}, G. and {Damerdji}, Y. and {Davidson}, M. and {de Teodoro}, P. and {Nu{\~n}ez Campos}, M. and {Delchambre}, L. and {Dell'Oro}, A. and {Esquej}, P. and {Fern{\'a}ndez-Hern{\'a}ndez}, J. and {Fraile}, E. and {Garabato}, D. and {Garc{\'\i}a-Lario}, P. and {Gosset}, E. and {Haigron}, R. and {Halbwachs}, J. -L. and {Hambly}, N.~C. and {Harrison}, D.~L. and {Hern{\'a}ndez}, J. and {Hestroffer}, D. and {Hodgkin}, S.~T. and {Holl}, B. and {Jan{\ss}en}, K. and {Jevardat de Fombelle}, G. and {Jordan}, S. and {Krone-Martins}, A. and {Lanzafame}, A.~C. and {L{\"o}ffler}, W. and {Marchal}, O. and {Marrese}, P.~M. and {Moitinho}, A. and {Muinonen}, K. and {Osborne}, P. and {Pancino}, E. and {Pauwels}, T. and {Recio-Blanco}, A. and {Reyl{\'e}}, C. and {Riello}, M. and {Rimoldini}, L. and {Roegiers}, T. and {Rybizki}, J. and {Sarro}, L.~M. and {Siopis}, C. and {Smith}, M. and {Sozzetti}, A. and {Utrilla}, E. and {van Leeuwen}, M. and {Abbas}, U. and {{\'A}brah{\'a}m}, P. and {Abreu Aramburu}, A. and {Aerts}, C. and {Aguado}, J.~J. and {Ajaj}, M. and {Aldea-Montero}, F. and {Altavilla}, G. and {{\'A}lvarez}, M.~A. and {Alves}, J. and {Anders}, F. and {Anderson}, R.~I. and {Anglada Varela}, E. and {Antoja}, T. and {Baines}, D. and {Baker}, S.~G. and {Balaguer-N{\'u}{\~n}ez}, L. and {Balbinot}, E. and {Balog}, Z. and {Barache}, C. and {Barbato}, D. and {Barros}, M. and {Barstow}, M.~A. and {Bartolom{\'e}}, S. and {Bassilana}, J. -L. and {Bauchet}, N. and {Becciani}, U. and {Bellazzini}, M. and {Berihuete}, A. and {Bernet}, M. and {Bertone}, S. and {Bianchi}, L. and {Binnenfeld}, A. and {Blanco-Cuaresma}, S. and {Blazere}, A. and {Boch}, T. and {Bombrun}, A. and {Bossini}, D. and {Bouquillon}, S. and {Bragaglia}, A. and {Bramante}, L. and {Breedt}, E. and {Bressan}, A. and {Brouillet}, N. and {Brugaletta}, E. and {Bucciarelli}, B. and {Burlacu}, A. and {Butkevich}, A.~G. and {Buzzi}, R. and {Caffau}, E. and {Cancelliere}, R. and {Cantat-Gaudin}, T. and {Carballo}, R. and {Carlucci}, T. and {Carnerero}, M.~I. and {Carrasco}, J.~M. and {Casamiquela}, L. and {Castellani}, M. and {Castro-Ginard}, A. and {Chaoul}, L. and {Charlot}, P. and {Chemin}, L. and {Chiaramida}, V. and {Chiavassa}, A. and {Chornay}, N. and {Comoretto}, G. and {Contursi}, G. and {Cooper}, W.~J. and {Cornez}, T. and {Cowell}, S. and {Crifo}, F. and {Cropper}, M. and {Crosta}, M. and {Crowley}, C. and {Dafonte}, C. and {Dapergolas}, A. and {David}, M. and {David}, P. and {de Laverny}, P. and {De Luise}, F. and {De March}, R. and {De Ridder}, J. and {de Souza}, R. and {de Torres}, A. and {del Peloso}, E.~F. and {del Pozo}, E. and {Delbo}, M. and {Delgado}, A. and {Delisle}, J. -B. and {Demouchy}, C. and {Dharmawardena}, T.~E. and {Di Matteo}, P. and {Diakite}, S. and {Diener}, C. and {Distefano}, E. and {Dolding}, C. and {Edvardsson}, B. and {Enke}, H. and {Fabre}, C. and {Fabrizio}, M. and {Faigler}, S. and {Fedorets}, G. and {Fernique}, P. and {Fienga}, A. and {Figueras}, F. and {Fournier}, Y. and {Fouron}, C. and {Fragkoudi}, F. and {Gai}, M. and {Garcia-Gutierrez}, A. and {Garcia-Reinaldos}, M. and {Garc{\'\i}a-Torres}, M. and {Garofalo}, A. and {Gavel}, A. and {Gavras}, P. and {Gerlach}, E. and {Geyer}, R. and {Giacobbe}, P. and {Gilmore}, G. and {Girona}, S. and {Giuffrida}, G. and {Gomel}, R. and {Gomez}, A. and {Gonz{\'a}lez-N{\'u}{\~n}ez}, J. and {Gonz{\'a}lez-Santamar{\'\i}a}, I. and {Gonz{\'a}lez-Vidal}, J.~J. and {Granvik}, M. and {Guillout}, P. and {Guiraud}, J. and {Guti{\'e}rrez-S{\'a}nchez}, R. and {Guy}, L.~P. and {Hatzidimitriou}, D. and {Hauser}, M. and {Haywood}, M. and {Helmer}, A. and {Helmi}, A. and {Sarmiento}, M.~H. and {Hidalgo}, S.~L. and {Hilger}, T. and {H{\l}adczuk}, N. and {Hobbs}, D. and {Holland}, G. and {Huckle}, H.~E. and {Jardine}, K. and {Jasniewicz}, G. and {Jean-Antoine Piccolo}, A. and {Jim{\'e}nez-Arranz}, {\'O}. and {Jorissen}, A. and {Juaristi Campillo}, J. and {Julbe}, F. and {Karbevska}, L. and {Kervella}, P. and {Khanna}, S. and {Kontizas}, M. and {Kordopatis}, G. and {Korn}, A.~J. and {K{\'o}sp{\'a}l}, {\'A}. and {Kostrzewa-Rutkowska}, Z. and {Kruszy{\'n}ska}, K. and {Kun}, M. and {Laizeau}, P. and {Lambert}, S. and {Lanza}, A.~F. and {Lasne}, Y. and {Le Campion}, J. -F. and {Lebreton}, Y. and {Lebzelter}, T. and {Leccia}, S. and {Leclerc}, N. and {Lecoeur-Taibi}, I. and {Liao}, S. and {Licata}, E.~L. and {Lindstr{\o}m}, H.~E.~P. and {Lister}, T.~A. and {Livanou}, E. and {Lobel}, A. and {Lorca}, A. and {Loup}, C. and {Madrero Pardo}, P. and {Magdaleno Romeo}, A. and {Managau}, S. and {Mann}, R.~G. and {Manteiga}, M. and {Marchant}, J.~M. and {Marconi}, M. and {Marcos}, J. and {Marcos Santos}, M.~M.~S. and {Mar{\'\i}n Pina}, D. and {Marinoni}, S. and {Marocco}, F. and {Marshall}, D.~J. and {Martin Polo}, L. and {Mart{\'\i}n-Fleitas}, J.~M. and {Marton}, G. and {Mary}, N. and {Masip}, A. and {Massari}, D. and {Mastrobuono-Battisti}, A. and {Mazeh}, T. and {McMillan}, P.~J. and {Messina}, S. and {Michalik}, D. and {Millar}, N.~R. and {Mints}, A. and {Molina}, D. and {Molinaro}, R. and {Moln{\'a}r}, L. and {Monari}, G. and {Mongui{\'o}}, M. and {Montegriffo}, P. and {Montero}, A. and {Mor}, R. and {Mora}, A. and {Morbidelli}, R. and {Morel}, T. and {Morris}, D. and {Muraveva}, T. and {Murphy}, C.~P. and {Musella}, I. and {Nagy}, Z. and {Noval}, L. and {Oca{\~n}a}, F. and {Ogden}, A. and {Ordenovic}, C. and {Osinde}, J.~O. and {Pagani}, C. and {Pagano}, I. and {Palaversa}, L. and {Palicio}, P.~A. and {Pallas-Quintela}, L. and {Panahi}, A. and {Payne-Wardenaar}, S. and {Pe{\~n}alosa Esteller}, X. and {Penttil{\"a}}, A. and {Pichon}, B. and {Piersimoni}, A.~M. and {Pineau}, F. -X. and {Plachy}, E. and {Plum}, G. and {Poggio}, E. and {Pr{\v{s}}a}, A. and {Pulone}, L. and {Racero}, E. and {Ragaini}, S. and {Rainer}, M. and {Raiteri}, C.~M. and {Rambaux}, N. and {Ramos}, P. and {Ramos-Lerate}, M. and {Re Fiorentin}, P. and {Regibo}, S. and {Richards}, P.~J. and {Rios Diaz}, C. and {Ripepi}, V. and {Riva}, A. and {Rix}, H. -W. and {Rixon}, G. and {Robichon}, N. and {Robin}, A.~C. and {Robin}, C. and {Roelens}, M. and {Rogues}, H.~R.~O. and {Rohrbasser}, L. and {Romero-G{\'o}mez}, M. and {Rowell}, N. and {Royer}, F. and {Ruz Mieres}, D. and {Rybicki}, K.~A. and {Sadowski}, G. and {S{\'a}ez N{\'u}{\~n}ez}, A. and {Sagrist{\`a} Sell{\'e}s}, A. and {Sahlmann}, J. and {Salguero}, E. and {Samaras}, N. and {Sanchez Gimenez}, V. and {Sanna}, N. and {Santove{\~n}a}, R. and {Sarasso}, M. and {Schultheis}, M. and {Sciacca}, E. and {Segol}, M. and {Segovia}, J.~C. and {S{\'e}gransan}, D. and {Semeux}, D. and {Shahaf}, S. and {Siddiqui}, H.~I. and {Siebert}, A. and {Siltala}, L. and {Silvelo}, A. and {Slezak}, E. and {Slezak}, I. and {Smart}, R.~L. and {Snaith}, O.~N. and {Solano}, E. and {Solitro}, F. and {Souami}, D. and {Souchay}, J. and {Spagna}, A. and {Spina}, L. and {Spoto}, F. and {Steele}, I.~A. and {Steidelm{\"u}ller}, H. and {Stephenson}, C.~A. and {S{\"u}veges}, M. and {Surdej}, J. and {Szabados}, L. and {Szegedi-Elek}, E. and {Taris}, F. and {Taylor}, M.~B. and {Teixeira}, R. and {Tolomei}, L. and {Tonello}, N. and {Torra}, F. and {Torra}, J. and {Torralba Elipe}, G. and {Trabucchi}, M. and {Tsounis}, A.~T. and {Turon}, C. and {Ulla}, A. and {Unger}, N. and {Vaillant}, M.~V. and {van Dillen}, E. and {van Reeven}, W. and {Vanel}, O. and {Vecchiato}, A. and {Viala}, Y. and {Vicente}, D. and {Voutsinas}, S. and {Weiler}, M. and {Wevers}, T. and {Wyrzykowski}, {\L}. and {Yoldas}, A. and {Yvard}, P. and {Zhao}, H. and {Zorec}, J. and {Zucker}, S. and {Zwitter}, T.},
        title = "{Gaia Data Release 3. Summary of the content and survey properties}",
      journal = {\aap},
         year = 2023,
        month = jun,
       volume = {674},
          eid = {A1},
        pages = {A1},
          doi = {10.1051/0004-6361/202243940},
archivePrefix = {arXiv},
       eprint = {2208.00211},
 primaryClass = {astro-ph.GA},
       adsurl = {https://ui.adsabs.harvard.edu/abs/2023A&A...674A...1G}
}

@ARTICLE{wilson1964,
       author = {{Wilson}, O.~C. and {Skumanich}, Andrew},
        title = "{Dependence of Chromospheric Activity upon Age in Main-Sequence Stars: Additional Evidence.}",
      journal = {\apj},
         year = 1964,
        month = nov,
       volume = {140},
        pages = {1401},
          doi = {10.1086/148046},
       adsurl = {https://ui.adsabs.harvard.edu/abs/1964ApJ...140.1401W}
}

@ARTICLE{fekel2002,
       author = {{Fekel}, Francis C. and {Henry}, Gregory W. and {Eaton}, Joel A. and {Sperauskas}, Julius and {Hall}, Douglas S.},
        title = "{Chromospherically Active Stars. XXI. The Giant, Single-lined Binaries HD 89546 And HD 113816}",
      journal = {\aj},
         year = 2002,
        month = aug,
       volume = {124},
       number = {2},
        pages = {1064-1076},
          doi = {10.1086/341612},
       adsurl = {https://ui.adsabs.harvard.edu/abs/2002AJ....124.1064F}
}

@ARTICLE{xing2021,
       author = {{Xing}, Li-Feng and {Li}, Yuan-Chao and {Chang}, Liang and {Wang}, Chuan-Jun and {Bai}, Jin-Ming},
        title = "{Lithium abundance in a sample of active stars: High-resolution spectrograph observation with the 1.8 m telescope}",
      journal = {\aap},
         year = 2021,
        month = sep,
       volume = {653},
          eid = {A28},
        pages = {A28},
          doi = {10.1051/0004-6361/202039203},
       adsurl = {https://ui.adsabs.harvard.edu/abs/2021A&A...653A..28X}
}

@ARTICLE{sneden2022,
       author = {{Sneden}, Christopher and {Af{\c{s}}ar}, Melike and {Bozkurt}, Zeynep and {Adam{\'o}w}, Monika and {Mallick}, Anohita and {Reddy}, Bacham E. and {Janowiecki}, Steven and {Mahadevan}, Suvrath and {Bowler}, Brendan P. and {Hawkins}, Keith and {Lind}, Karin and {Dupree}, Andrea K. and {Ninan}, Joe P. and {Nagarajan}, Neel and {Topcu}, Gamze B{\"o}cek and {Froning}, Cynthia S. and {Bender}, Chad F. and {Terrien}, Ryan and {Ramsey}, Lawrence W. and {Mace}, Gregory N.},
        title = "{The Active Chromospheres of Lithium-rich Red Giant Stars}",
      journal = {\apj},
         year = 2022,
        month = nov,
       volume = {940},
       number = {1},
          eid = {12},
        pages = {12},
          doi = {10.3847/1538-4357/ac922e},
archivePrefix = {arXiv},
       eprint = {2209.05941},
 primaryClass = {astro-ph.SR},
       adsurl = {https://ui.adsabs.harvard.edu/abs/2022ApJ...940...12S}
}

@ARTICLE{rebolo1988,
       author = {{Rebolo}, R. and {Beckman}, J.~E.},
        title = "{Lithium and rotation in the Hyades late F and G stars.}",
      journal = {\aap},
         year = 1988,
        month = aug,
       volume = {201},
        pages = {267-272},
       adsurl = {https://ui.adsabs.harvard.edu/abs/1988A&A...201..267R}
}

@ARTICLE{badry2018,
       author = {{El-Badry}, Kareem and {Rix}, Hans-Walter and {Weisz}, Daniel R.},
        title = "{An Empirical Measurement of the Initial-Final Mass Relation with Gaia White Dwarfs}",
      journal = {\apjl},
         year = 2018,
        month = jun,
       volume = {860},
       number = {2},
          eid = {L17},
        pages = {L17},
          doi = {10.3847/2041-8213/aaca9c},
archivePrefix = {arXiv},
       eprint = {1805.05849},
 primaryClass = {astro-ph.SR},
       adsurl = {https://ui.adsabs.harvard.edu/abs/2018ApJ...860L..17E}
}

@ARTICLE{flaulhabe2025,
       author = {{Flaulhabe}, T. and {Holanda}, N. and {Tautvai{\v{s}}ien{\.{e}}}, G. and {Katime Santrich}, O.~J. and {Maia}, F.~F.~S. and {Ferreira}, B.~P.~L. and {Corradi}, W.~J.~B. and {Pereira}, C.~B. and {Carlos}, M. and {Daflon}, S.},
        title = "{Astrometric and Spectroscopic Analysis of IC 2714: An Open Cluster Hosting a Lithium-Rich Giant}",
      journal = {arXiv e-prints},
         year = 2025,
        month = dec,
          eid = {arXiv:2512.16042},
        pages = {arXiv:2512.16042},
          doi = {10.48550/arXiv.2512.16042},
archivePrefix = {arXiv},
       eprint = {2512.16042},
 primaryClass = {astro-ph.GA},
       adsurl = {https://ui.adsabs.harvard.edu/abs/2025arXiv251216042F}
}

@ARTICLE{bossini2019,
       author = {{Bossini}, D. and {Vallenari}, A. and {Bragaglia}, A. and {Cantat-Gaudin}, T. and {Sordo}, R. and {Balaguer-N{\'u}{\~n}ez}, L. and {Jordi}, C. and {Moitinho}, A. and {Soubiran}, C. and {Casamiquela}, L. and {Carrera}, R. and {Heiter}, U.},
        title = "{Age determination for 269 Gaia DR2 open clusters}",
      journal = {\aap},
         year = 2019,
        month = mar,
       volume = {623},
          eid = {A108},
        pages = {A108},
          doi = {10.1051/0004-6361/201834693},
archivePrefix = {arXiv},
       eprint = {1901.04733},
 primaryClass = {astro-ph.SR},
       adsurl = {https://ui.adsabs.harvard.edu/abs/2019A&A...623A.108B}
}

@ARTICLE{swaelmen2017,
       author = {{Van der Swaelmen}, M. and {Boffin}, H.~M.~J. and {Jorissen}, A. and {Van Eck}, S.},
        title = "{The mass-ratio and eccentricity distributions of barium and S stars, and red giants in open clusters}",
      journal = {\aap},
         year = 2017,
        month = jan,
       volume = {597},
          eid = {A68},
        pages = {A68},
          doi = {10.1051/0004-6361/201628867},
archivePrefix = {arXiv},
       eprint = {1608.04949},
 primaryClass = {astro-ph.SR},
       adsurl = {https://ui.adsabs.harvard.edu/abs/2017A&A...597A..68V}
}

@ARTICLE{chaboyer1998,
       author = {{Chaboyer}, Brian and {Demarque}, P. and {Kernan}, Peter J. and {Krauss}, Lawrence M.},
        title = "{The Age of Globular Clusters in Light of Hipparcos: Resolving the Age Problem?}",
      journal = {\apj},
         year = 1998,
        month = feb,
       volume = {494},
       number = {1},
        pages = {96-110},
          doi = {10.1086/305201},
archivePrefix = {arXiv},
       eprint = {astro-ph/9706128},
 primaryClass = {astro-ph},
       adsurl = {https://ui.adsabs.harvard.edu/abs/1998ApJ...494...96C}
}

@ARTICLE{goncalvez2020,
       author = {{Gon{\c{c}}alves}, B.~F.~O. and {da Costa}, J.~S. and {de Almeida}, L. and {Castro}, M. and {do Nascimento}, Jr., J.-D.},
        title = "{Li-rich giant stars under scrutiny: binarity, magnetic activity, and the evolutionary status after Gaia DR2}",
      journal = {\mnras},
         year = 2020,
        month = oct,
       volume = {498},
       number = {2},
        pages = {2295-2308},
          doi = {10.1093/mnras/staa2408},
archivePrefix = {arXiv},
       eprint = {2008.02948},
 primaryClass = {astro-ph.SR},
       adsurl = {https://ui.adsabs.harvard.edu/abs/2020MNRAS.498.2295G}
}

@ARTICLE{rolo2024,
       author = {{Rolo}, In{\^e}s and {Delgado Mena}, Elisa and {Tsantaki}, Maria and {Gomes da Silva}, Jo{\~a}o},
        title = "{The enigma of Li-rich giants and its relation with temporal variations observed in radial velocity and stellar activity signals}",
      journal = {\aap},
         year = 2024,
        month = aug,
       volume = {688},
          eid = {A68},
        pages = {A68},
          doi = {10.1051/0004-6361/202449873},
archivePrefix = {arXiv},
       eprint = {2406.12711},
 primaryClass = {astro-ph.SR},
       adsurl = {https://ui.adsabs.harvard.edu/abs/2024A&A...688A..68R}
}

@ARTICLE{morel2004,
       author = {{Morel}, T. and {Micela}, G. and {Favata}, F. and {Katz}, D.},
        title = "{The photospheric abundances of active binaries. III. Abundance peculiarities at high activity levels}",
      journal = {\aap},
         year = 2004,
        month = nov,
       volume = {426},
        pages = {1007-1020},
          doi = {10.1051/0004-6361:20041047},
archivePrefix = {arXiv},
       eprint = {astro-ph/0406348},
 primaryClass = {astro-ph},
       adsurl = {https://ui.adsabs.harvard.edu/abs/2004A&A...426.1007M}
}

@ARTICLE{fekel2005,
       author = {{Fekel}, Francis C. and {Henry}, Gregory W.},
        title = "{Chromospherically Active Stars. XXIV. The Giant, Single-lined Binaries HD 37824, HD 181809, and HD 217188}",
      journal = {\aj},
         year = 2005,
        month = mar,
       volume = {129},
       number = {3},
        pages = {1669-1685},
          doi = {10.1086/427713},
       adsurl = {https://ui.adsabs.harvard.edu/abs/2005AJ....129.1669F}
}

@ARTICLE{skumanich1972,
       author = {{Skumanich}, A.},
        title = "{Time Scales for Ca II Emission Decay, Rotational Braking, and Lithium Depletion}",
      journal = {\apj},
         year = 1972,
        month = feb,
       volume = {171},
        pages = {565},
          doi = {10.1086/151310},
       adsurl = {https://ui.adsabs.harvard.edu/abs/1972ApJ...171..565S}
}

@ARTICLE{osorio2015,
       author = {{Osorio}, Y. and {Barklem}, P.~S. and {Lind}, K. and {Belyaev}, A.~K. and {Spielfiedel}, A. and {Guitou}, M. and {Feautrier}, N.},
        title = "{Mg line formation in late-type stellar atmospheres. I. The model atom}",
      journal = {\aap},
         year = 2015,
        month = jul,
       volume = {579},
          eid = {A53},
        pages = {A53},
          doi = {10.1051/0004-6361/201525846},
archivePrefix = {arXiv},
       eprint = {1504.07593},
 primaryClass = {astro-ph.SR},
       adsurl = {https://ui.adsabs.harvard.edu/abs/2015A&A...579A..53O}
}

@ARTICLE{nandakumar2024,
       author = {{Nandakumar}, G. and {Ryde}, N. and {Hartman}, H. and {Mace}, G.},
        title = "{M giants with IGRINS: IV. Identification and characterisation of a NIR line of the s-element barium}",
      journal = {\aap},
         year = 2024,
        month = oct,
       volume = {690},
          eid = {A226},
        pages = {A226},
          doi = {10.1051/0004-6361/202451255},
archivePrefix = {arXiv},
       eprint = {2408.12971},
 primaryClass = {astro-ph.GA},
       adsurl = {https://ui.adsabs.harvard.edu/abs/2024A&A...690A.226N}
}

@ARTICLE{nandakumar2025,
       author = {{Nandakumar}, Govind and {Ryde}, Nils and {Schultheis}, Mathias and {Rich}, R. Michael and {di Matteo}, Paola and {Thorsbro}, Brian and {Mace}, Gregory},
        title = "{The First Chemical Census of the Milky Way's Nuclear Star Cluster}",
      journal = {\apjl},
         year = 2025,
        month = mar,
       volume = {982},
       number = {1},
          eid = {L14},
        pages = {L14},
          doi = {10.3847/2041-8213/adbb6d},
archivePrefix = {arXiv},
       eprint = {2502.17756},
 primaryClass = {astro-ph.GA},
       adsurl = {https://ui.adsabs.harvard.edu/abs/2025ApJ...982L..14N}
}

@ARTICLE{roriz2024,
       author = {{Roriz}, M.~P. and {Holanda}, N. and {da Concei{\c{c}}{\~a}o}, L.~V. and {Junqueira}, S. and {Drake}, N.~A. and {Sonally}, A. and {Pereira}, C.~B.},
        title = "{High-resolution Spectroscopic Analysis of Four Unevolved Barium Stars}",
      journal = {\aj},
         year = 2024,
        month = apr,
       volume = {167},
       number = {4},
          eid = {184},
        pages = {184},
          doi = {10.3847/1538-3881/ad29f2},
archivePrefix = {arXiv},
       eprint = {2402.14709},
 primaryClass = {astro-ph.SR},
       adsurl = {https://ui.adsabs.harvard.edu/abs/2024AJ....167..184R}
}

@ARTICLE{cseh2022,
       author = {{Cseh}, B. and {Vil{\'a}gos}, B. and {Roriz}, M.~P. and {Pereira}, C.~B. and {D'Orazi}, V. and {Karakas}, A.~I. and {So{\'o}s}, B. and {Drake}, N.~A. and {Junqueira}, S. and {Lugaro}, M.},
        title = "{Barium stars as tracers of s-process nucleosynthesis in AGB stars. I. 28 stars with independently derived AGB mass}",
      journal = {\aap},
         year = 2022,
        month = apr,
       volume = {660},
          eid = {A128},
        pages = {A128},
          doi = {10.1051/0004-6361/202142468},
archivePrefix = {arXiv},
       eprint = {2201.13379},
 primaryClass = {astro-ph.SR},
       adsurl = {https://ui.adsabs.harvard.edu/abs/2022A&A...660A.128C}
}

@ARTICLE{battino2019,
       author = {{Battino}, U. and {Tattersall}, A. and {Lederer-Woods}, C. and {Herwig}, F. and {Denissenkov}, P. and {Hirschi}, R. and {Trappitsch}, R. and {den Hartogh}, J.~W. and {Pignatari}, M. and {NuGrid Collaboration}},
        title = "{NuGrid stellar data set - III. Updated low-mass AGB models and s-process nucleosynthesis with metallicities Z= 0.01, Z = 0.02, and Z = 0.03}",
      journal = {\mnras},
         year = 2019,
        month = oct,
       volume = {489},
       number = {1},
        pages = {1082-1098},
          doi = {10.1093/mnras/stz2158},
archivePrefix = {arXiv},
       eprint = {1906.01952},
 primaryClass = {astro-ph.SR},
       adsurl = {https://ui.adsabs.harvard.edu/abs/2019MNRAS.489.1082B}
}

@ARTICLE{bedding2011,
       author = {{Bedding}, Timothy R. and {Mosser}, Benoit and {Huber}, Daniel and {Montalb{\'a}n}, Josefina and {Beck}, Paul and {Christensen-Dalsgaard}, J{\o}rgen and {Elsworth}, Yvonne P. and {Garc{\'\i}a}, Rafael A. and {Miglio}, Andrea and {Stello}, Dennis and {White}, Timothy R. and {De Ridder}, Joris and {Hekker}, Saskia and {Aerts}, Conny and {Barban}, Caroline and {Belkacem}, Kevin and {Broomhall}, Anne-Marie and {Brown}, Timothy M. and {Buzasi}, Derek L. and {Carrier}, Fabien and {Chaplin}, William J. and {di Mauro}, Maria Pia and {Dupret}, Marc-Antoine and {Frandsen}, S{\o}ren and {Gilliland}, Ronald L. and {Goupil}, Marie-Jo and {Jenkins}, Jon M. and {Kallinger}, Thomas and {Kawaler}, Steven and {Kjeldsen}, Hans and {Mathur}, Savita and {Noels}, Arlette and {Silva Aguirre}, Victor and {Ventura}, Paolo},
        title = "{Gravity modes as a way to distinguish between hydrogen- and helium-burning red giant stars}",
      journal = {\nat},
         year = 2011,
        month = mar,
       volume = {471},
       number = {7340},
        pages = {608-611},
          doi = {10.1038/nature09935},
archivePrefix = {arXiv},
       eprint = {1103.5805},
 primaryClass = {astro-ph.SR},
       adsurl = {https://ui.adsabs.harvard.edu/abs/2011Natur.471..608B}
}

@ARTICLE{mosser2014,
       author = {{Mosser}, B. and {Benomar}, O. and {Belkacem}, K. and {Goupil}, M.~J. and {Lagarde}, N. and {Michel}, E. and {Lebreton}, Y. and {Stello}, D. and {Vrard}, M. and {Barban}, C. and {Bedding}, T.~R. and {Deheuvels}, S. and {Chaplin}, W.~J. and {De Ridder}, J. and {Elsworth}, Y. and {Montalban}, J. and {Noels}, A. and {Ouazzani}, R.~M. and {Samadi}, R. and {White}, T.~R. and {Kjeldsen}, H.},
        title = "{Mixed modes in red giants: a window on stellar evolution}",
      journal = {\aap},
         year = 2014,
        month = dec,
       volume = {572},
          eid = {L5},
        pages = {L5},
          doi = {10.1051/0004-6361/201425039},
archivePrefix = {arXiv},
       eprint = {1411.1082},
 primaryClass = {astro-ph.SR},
       adsurl = {https://ui.adsabs.harvard.edu/abs/2014A&A...572L...5M}
}

@ARTICLE{bressan2012,
       author = {{Bressan}, Alessandro and {Marigo}, Paola and {Girardi}, L{\'e}o. and {Salasnich}, Bernardo and {Dal Cero}, Claudia and {Rubele}, Stefano and {Nanni}, Ambra},
        title = "{PARSEC: stellar tracks and isochrones with the PAdova and TRieste Stellar Evolution Code}",
      journal = {\mnras},
         year = 2012,
        month = nov,
       volume = {427},
       number = {1},
        pages = {127-145},
          doi = {10.1111/j.1365-2966.2012.21948.x},
archivePrefix = {arXiv},
       eprint = {1208.4498},
 primaryClass = {astro-ph.SR},
       adsurl = {https://ui.adsabs.harvard.edu/abs/2012MNRAS.427..127B}
}

@ARTICLE{lucertini2022,
       author = {{Lucertini}, F. and {Monaco}, L. and {Caffau}, E. and {Bonifacio}, P. and {Mucciarelli}, A.},
        title = "{Sulfur abundances in the Galactic bulge and disk}",
      journal = {\aap},
         year = 2022,
        month = jan,
       volume = {657},
          eid = {A29},
        pages = {A29},
          doi = {10.1051/0004-6361/202140947},
archivePrefix = {arXiv},
       eprint = {2109.06216},
 primaryClass = {astro-ph.GA},
       adsurl = {https://ui.adsabs.harvard.edu/abs/2022A&A...657A..29L}
}

@ARTICLE{martell2021,
       author = {{Martell}, Sarah L. and {Simpson}, Jeffrey D. and {Balasubramaniam}, Adithya G. and {Buder}, Sven and {Sharma}, Sanjib and {Hon}, Marc and {Stello}, Dennis and {Ting}, Yuan-Sen and {Asplund}, Martin and {Bland-Hawthorn}, Joss and {De Silva}, Gayandhi M. and {Freeman}, Ken C. and {Hayden}, Michael and {Kos}, Janez and {Lewis}, Geraint F. and {Lind}, Karin and {Zucker}, Daniel B. and {Zwitter}, Toma{\v{z}} and {Campbell}, Simon W. and {{\v{C}}otar}, Klemen and {Horner}, Jonathan and {Montet}, Benjamin and {Wittenmyer}, Rob},
        title = "{The GALAH survey: a census of lithium-rich giant stars}",
      journal = {\mnras},
         year = 2021,
        month = aug,
       volume = {505},
       number = {4},
        pages = {5340-5355},
          doi = {10.1093/mnras/stab1356},
archivePrefix = {arXiv},
       eprint = {2006.02106},
 primaryClass = {astro-ph.SR},
       adsurl = {https://ui.adsabs.harvard.edu/abs/2021MNRAS.505.5340M}
}

@ARTICLE{schwab2020,
       author = {{Schwab}, Josiah},
        title = "{A Helium-flash-induced Mixing Event Can Explain the Lithium Abundances of Red Clump Stars}",
      journal = {\apjl},
         year = 2020,
        month = sep,
       volume = {901},
       number = {1},
          eid = {L18},
        pages = {L18},
          doi = {10.3847/2041-8213/abb45f},
archivePrefix = {arXiv},
       eprint = {2009.01248},
 primaryClass = {astro-ph.SR},
       adsurl = {https://ui.adsabs.harvard.edu/abs/2020ApJ...901L..18S}
}

@PHDTHESIS{sneden1973,
       author = {{Sneden}, Christopher Alan},
        title = "{Carbon and Nitrogen Abundances in Metal-Poor Stars.}",
       school = {THE UNIVERSITY OF TEXAS AT AUSTIN.},
         year = "1973",
        month = "Jan",
       adsurl = {https://ui.adsabs.harvard.edu/abs/1973PhDT.......180S}
}

@ARTICLE{asplund2009,
       author = {{Asplund}, Martin and {Grevesse}, Nicolas and {Sauval}, A. Jacques and
         {Scott}, Pat},
        title = "{The Chemical Composition of the Sun}",
      journal = {\araa},
         year = "2009",
        month = "Sep",
       volume = {47},
       number = {1},
        pages = {481-522},
          doi = {10.1146/annurev.astro.46.060407.145222},
archivePrefix = {arXiv},
       eprint = {0909.0948},
 primaryClass = {astro-ph.SR},
       adsurl = {https://ui.adsabs.harvard.edu/abs/2009ARA&A..47..481A}
}

@ARTICLE{fekel1997,
       author = {{Fekel}, F.~C.},
        title = "{Rotational Velocities of Late-Type Stars}",
      journal = {\pasp},
         year = "1997",
        month = "May",
       volume = {109},
        pages = {514-523},
          doi = {10.1086/133908},
       adsurl = {https://ui.adsabs.harvard.edu/abs/1997PASP..109..514F}
}

@ARTICLE{lambert1996,
       author = {{Lambert}, David L. and {Heath}, James E. and {Lemke}, Michael and
         {Drake}, Jeremy},
        title = "{The Chemical Composition of Field RR Lyrae Stars. I. Iron and Calcium}",
      journal = {\apjs},
         year = "1996",
        month = "Mar",
       volume = {103},
        pages = {183},
          doi = {10.1086/192274},
       adsurl = {https://ui.adsabs.harvard.edu/abs/1996ApJS..103..183L}
}

@ARTICLE{takeda2017,
       author = {{Takeda}, Yoichi and {Tajitsu}, Akito},
        title = "{On the observational characteristics of lithium-enhanced giant stars in comparison with normal red giants$^{{\textdagger}}$}",
      journal = {\pasj},
         year = 2017,
        month = aug,
       volume = {69},
       number = {4},
          eid = {74},
        pages = {74},
          doi = {10.1093/pasj/psx057},
archivePrefix = {arXiv},
       eprint = {1706.02273},
 primaryClass = {astro-ph.SR},
       adsurl = {https://ui.adsabs.harvard.edu/abs/2017PASJ...69...74T}
}

@ARTICLE{boothroyd1999,
       author = {{Boothroyd}, Arnold I. and {Sackmann}, I. -Juliana},
        title = "{The CNO Isotopes: Deep Circulation in Red Giants and First and Second Dredge-up}",
      journal = {\apj},
         year = 1999,
        month = jan,
       volume = {510},
       number = {1},
        pages = {232-250},
          doi = {10.1086/306546},
       adsurl = {https://ui.adsabs.harvard.edu/abs/1999ApJ...510..232B}
}

\bsp	
\label{lastpage}
\end{document}